\documentclass[fleqn,usenatbib]{mnras}

\usepackage{newtxtext,newtxmath}
\usepackage[T1]{fontenc}

\DeclareRobustCommand{\VAN}[3]{#2}
\let\VANthebibliography\thebibliography
\def\thebibliography{\DeclareRobustCommand{\VAN}[3]{##3}\VANthebibliography}

\usepackage{graphicx}	% Including figure files
\usepackage{amsmath}	% Advanced maths commands
\usepackage{hyperref}
\usepackage{longtable}
\usepackage{tabularx}
\usepackage{multirow}
\usepackage{rotating}
\usepackage{lscape}     % landscape mode
\usepackage[dvipsnames]{xcolor}
\usepackage[caption = false]{subfig}
\usepackage{float}
\usepackage{soul}
\usepackage[normalem]{ulem}
\usepackage{animate}    % Including animation
\newcommand{\mean}{\langle m \rangle}
\newcommand{\xmean}{\langle x \rangle}
\newcommand{\median}{\mathrm{Median}}
\newcommand{\sqd}{$\mathrm{deg}^2$}
\newcommand{\andro}{\textsc{M\,31}}
\newcommand{\deep}{\textsc{Deep}}
\newcommand{\disk}{\textsc{Disk}}

\title[Anomaly detection in ZTF DR3]{Anomaly detection in the Zwicky Transient Facility DR3}

\author[K. L. Malanchev et al.]{
K. L. Malanchev,$^{1,2}$\thanks{E-mail: kostya@illinois.edu}
M. V. Pruzhinskaya,$^{2}$
V. S. Korolev,$^{3,4}$
P. D. Aleo,$^{1,5}$
\newauthor M. V. Kornilov,$^{2,6}$
E. E. O. Ishida,$^{7}$
V. V. Krushinsky,$^{8}$
F. Mondon,$^{7}$
S. Sreejith,$^{7,11}$
\newauthor A. A. Volnova,$^{9}$
A. A. Belinski,$^{2}$
A. V. Dodin,$^{2}$
A. M. Tatarnikov$^{2}$
and S. G. Zheltoukhov$^{2,10}$
\newauthor (The SNAD Team) \\
$^{1}$Department of Astronomy, University of Illinois at Urbana-Champaign, 1002 West Green Street, Urbana, IL 61801, USA \\
$^{2}$Lomonosov Moscow State University, Sternberg Astronomical Institute, Universitetsky pr. 13, Moscow, 119234, Russia\\
$^{3}$Central Aerohydrodynamic Institute, 1 Zhukovsky st, Zhukovsky, Moscow Region, 140180, Russia\\
$^{4}$Moscow Institute of Physics and Technology , 9 Institutskiy per., Dolgoprudny, Moscow Region, 141701, Russia\\
$^{5}$Center for Astrophysical Surveys Fellow, National Center for Supercomputing Applications, USA\\
$^{6}$National Research University Higher School of Economics, 21/4 Staraya Basmannaya Ulitsa, Moscow, 105066, Russia\\
$^{7}$Universit\'e Clermont Auvergne, CNRS/IN2P3, LPC, F-63000 Clermont-Ferrand, France\\
$^{8}$Laboratory of Astrochemical Research, Ural Federal University, Ekaterinburg, Russia, ul. Mira d. 19, Yekaterinburg, Russia, 620002\\
$^{9}$Space Research Institute of the Russian Academy of Sciences (IKI), 84/32 Profsoyuznaya Street, Moscow, 117997, Russia\\
$^{10}$Faculty of Physics, Lomonosov Moscow State University, Leninskie Gory 1-2, 119991 Moscow, Russia\\
$^{11}$Physics Department, Brookhaven National Laboratory, Upton, NY 11973\\
}

\date{Accepted XXX. Received YYY; in original form ZZZ}

\pubyear{2015}

\begin{document}
\label{firstpage}
\pagerange{\pageref{firstpage}--\pageref{lastpage}}
\maketitle

% Abstract of the paper
\begin{abstract}
We present results from applying the SNAD anomaly detection pipeline to the third public data release of the Zwicky Transient Facility (ZTF DR3). The pipeline is composed of 3 stages: feature extraction, search of outliers with machine learning algorithms and anomaly identification with followup by human experts. Our analysis concentrates in three ZTF fields, comprising more than 2.25 million objects. A set of 4 automatic learning algorithms was used to identify 277 outliers, which were subsequently scrutinised by an expert. From these, 188 (68\%) were found to be bogus light curves -- including effects from the image subtraction pipeline as well as overlapping between a star and a known asteroid, 66 (24\%) were previously reported sources whereas 23 (8\%) correspond to non-catalogued objects, with the two latter cases of potential scientific interest (e.~g. 1 spectroscopically confirmed RS Canum Venaticorum star, 4 supernovae candidates, 1 red dwarf flare). Moreover, using results from the expert analysis, we were able to identify a simple bi-dimensional relation which can be used to aid filtering  potentially bogus light curves in future studies. We provide a complete list of objects with potential scientific application so they can be further scrutinised by the community. These results confirm the importance of combining automatic machine learning algorithms with domain knowledge in the construction of recommendation systems for astronomy. Our code is publicly available\footnotemark.
\end{abstract}

% Select between one and six entries from the list of approved keywords.
% Don't make up new ones.

\footnotetext{ \url{https://github.com/snad-space/zwad}}

%%%%%%%%%%%%%%%%%%%%%%%%%%%%%%%%%%%%%%%%%%%%%%%%%%

%%%%%%%%%%%%%%%%% BODY OF PAPER %%%%%%%%%%%%%%%%%%
\section{Introduction}

The meaning of astronomical discovery has changed throughout history, but it has always been highly correlated with the technological development of the era and the sociological construct which accompanies it. In astronomy, the discovery of new or unexpected astronomical sources was often serendipitous \citep{dick_2013}, for example, the discovery of gamma-ray bursts in the late 1960s by the Vela satellites~\citep{1973ApJ...182L..85K} or the discovery of the cosmic microwave background radiation by Robert Wilson and Arno Penzias~\citep{1965ApJ...142..419P}.
The advent of the CCD highly increased our ability to gather photometric data, and allowed programmed systematic searches for scientifically interesting objects  \citep{Perlmutter1992,Filippenko1992, Richmond1993}. Nevertheless, manual screening continued to play a central role in the discovery of new sources \citep[e.~g.][]{Cardamone2009,borisovcomet}.

The arrival of large scale sky surveys like the Sloan Digital Sky Survey (SDSS;~\citealt{2017AJ....154...28B}) and, more recently,  the Zwicky Transient Facility~\citep[ZTF;][]{Bellm2019, 2019PASP..131a8003M} pushed this paradigm even further. Confronted with data sets holding  observational data from around a billion  sources, the use of automated machine learning methods to search for new physics was unavoidable \citep{Ball2010}. This trend, shifting the astronomical data analysis towards data-driven approaches, is highly recognised in supervised learning tasks  \citep[e.g. photometric classification of transients, ][]{Ishida2019Nat}, but it is as crucial for unsupervised ones. Given that new astronomical surveys are always accompanied by technological developments which allow them to probe different epochs in the evolution of the Universe, or different regions within our own galaxy, every new instrument has a high probability of observing previously non-catalogued phenomena. At the same time, every new data set is more complex and bigger than its precedents. In the absence of appropriate unsupervised learning pipelines, we may miss the discovery opportunity which is in the root of all scientific endeavour.

This key aspect of unsupervised learning in astronomical research has been recognised by the research community with some extent, producing interesting illustrative examples: \cite{Rebbapragada2009} discovered periodic variable stars using a Phased $k$-means algorithm; \cite{Hoyle2015} isolated and removed problematic objects for photometric redshift estimation with an Elliptical Envelope routine; \cite{nun2016} selected five different algorithms for an ensemble method tested on three MACHO fields and confirmed some objects belonging to rare classes; \cite{Baron2017} identified anomalous galaxy spectra based on an Unsupervised Random Forest (URF); \cite{solarz2017} used One-Class Support Vector Machine (O-SVM) to find weird objects in photometric data, including a few misclassifications; \cite{reyes2020} used a geometric transformation-based model to highlight artefacts' properties from real objects in ZTF images; and \cite{soraisam2020,2021AJ....161...15S} identified novel events in variable source population by computing distributions of magnitude changes over time intervals for a given filter. 

The task of applying machine learning algorithms to real data also requires the data to be translated into an acceptable format for automatic learning applications, for example each light curve could be transformed to a vector of features. Modern astronomical data are frequently composed of a large number of correlated features which must be simplified and/or homogenised. Several works provide details on efforts in the dimensionality reduction stage. For example: \cite{Pruzhinskaya2019} located both novel and misclassified transient events light curves in the Open Supernova Catalog~\citep{2017ApJ...835...64G} after applying t-SNE to Gaussian-process approximated light curves, combined with an isolation forest (IF) and confirmed by expert analysis. \cite{Galarza2020} implemented tree-based algorithms (IF, URF) and two manifold-based algorithms (t-SNE, UMAP) to identify anomalies, using their \emph{bona fide} anomaly Boyajian's star \citep[KIC 8462852,][]{Boyajian2016} as reference, and investigated how an ensemble of such methods can be used to find objects of similar phenomenology or astrophysical characteristics. This work was built off of \cite{giles2019}, who used t-SNE to visually represent cluster membership designations of Kepler light curve data (including Boyajian's star as a ``ground truth'' anomaly) from  DBSCAN. The same authors applied the same technique to range \textit{Kepler} objects by their outlier score \citep{2020MNRAS.499..524G}. Similarly, \cite{webb2020} used t-SNE to visually represent HDBSCAN cluster membership designations of \textit{Deeper, Wider, Faster program} light curves combined with anomaly scores assigned by an IF algorithm in their interactive \textsc{Astronomaly} package \citep{Lochner2020}.

In this work we present detailed description of a complete anomaly detection pipeline and its results when applied to three fields from ZTF Data Release~3 (DR3)\footnote{\url{https://www.ztf.caltech.edu/page/dr3}}. The SNAD\footnote{\url{https://snad.space}} pipeline was built with the goal of exploiting the potential of modern astronomical data sets for discovery and under the hypothesis that, although automatic learning algorithms have a crucial role to play in this task, the scientific discovery is only completely realised when such systems are designed to boost the impact of domain knowledge experts. Our approach combines a diverse set of machine learning algorithms, tailored  feature extraction procedures and domain knowledge experts who validate the results of the machine learning pipeline   \citep{Pruzhinskaya2019, 2019mmag.conf..100K, Ishida2019, malanchev2020, aleo2020}. 

In what follows, Section~\ref{sec:data} describes the ZTF data and  the adopted preprocessing steps. Section~\ref{sec:methodology} details the outlier detection algorithms, followup observations and expert analysis employed in this work. Section~\ref{sec:results}  lists the outliers found by the complete pipeline and Section~\ref{sec:justification} reports numerical efficiency of our pipeline in identifying fake light curves. Section \ref{subsec:phase-space} provides a deeper analysis of our feature parameter space. Our conclusions are outlined  in Section~\ref{sec:conclusions}. We also provide further details of our analysis in the appendices: Appendix \ref{sec:appendix-features} mathematically defines the extracted light curve features, and Appendix 
\ref{sec:dr-viewer} describes the SNAD ZTF DR object web-viewer and cross-match tool, constructed to help the experts in their critical evaluation of each outlier. Appendix \ref{ap:lcmodels} illustrates results from the light curve fit of the supernova candidates and Appendix \ref{sec:appendix-outliers} lists properties of each anomaly candidate discovered from this work. 

Our code, {\tt zwad}, is publicly available at \url{https://github.com/snad-space/zwad}, and the SNAD team's ZTF DR3 object viewer can be found at \url{https://ztf.snad.space/}.

\section{ZTF Data}
\label{sec:data}

The Zwicky Transient Facility is a 48-inch Schmidt telescope on Mount Palomar, equipped with a 47\,\sqd{} camera, which allows rapid scanning of the entire north sky. The ZTF survey started on March 2018 and during its initial phase has observed around a billion objects~\citep{Bellm2019}. Beyond representing the current state of the art in rapid photometric observations, ZTF also has the crucial role of being the predecessor of the next generation of large scale surveys like the Vera Rubin Observatory Legacy Survey of Space and Time\footnote{\url{https://www.lsst.org/}} (LSST). Many of its data protocols are being used as test-bench for LSST systems.  

In this work, we analysed data from the first 9.4 months of the ZTF survey, between 17 March and 31 December 2018 $(58194 \leq \mathrm{MJD} \leq 58483)$. This period includes data from ZTF private survey, thus having a better cadence than the rest of DR3. We selected only perfectly clean extractions at every epoch ($\texttt{catflag} = 0$) obtained in passband $zr$, which encloses more than half of all objects. 

Light curves from ZTF DR3 are spread over 1020 fields, with objects within the same field being sampled with a similar cadence --- on average $\sim$ 1~day for the Galactic plane and $\sim$ 3~days for the Northern-equatorial sky. In order to minimise effects due to different cadences, we performed our analysis in three separate fields: ``\andro{}'' (a part of ZTF field 695),  which includes the Andromeda galaxy~\citep{Messier1781}, ``\deep{}'' (ZTF field 795), located far from the galactic plane, and ``\disk{}'' (ZTF field 805), located in the Galaxy plane.  The Andromeda galaxy region is a well-studied part of the sky, thus allowing further scrutiny of candidates with external information.  The Andromeda galaxy is fully imaged by only one CCD, thus the \andro{} field of view (FOV) is only $1/16$ of ZTF light-sensitive area ($\sim 3$\,\sqd{}), and does not contain as many objects as the two other fields. For \disk{}  and \deep{}, we chose fields with high declination, which allow easier follow-up observations by northern hemisphere facilities, and which maximise the number of field objects. 

We imposed a final selection cut by choosing light curves with at least 100 observations in \andro{} and \deep{}, and at least 500 observations in \disk{}. All data were downloaded from IRSA IPAC\footnote{\url{https://irsa.ipac.caltech.edu/data/ZTF/lc_dr3/}}. The number of light curves passed through the selection cuts are shown in  Table~\ref{table:fields} along with a few other properties of these fields.

\begin{table*}

\centering
\caption{Properties of three ZTF fields analysed in this work.}
\label{table:fields}
\begin{tabular}{l|l|c|c|c|c|r}
     Field name & ZTF field & Centre ($\alpha$, $\delta$)& Centre ($l$, $b$) & FOV & N & Object count \\ \hline
     \andro{} & Field 695, ccdid 11 & $10\fdg72, 41\fdg34$ & $121\fdg20, -21\fdg51$ & 3\,\sqd{} & $\geq 100$ & 57\,546 \\
     \deep{} & Field 795 & $246\fdg72, 54\fdg88$ & $83\fdg91, 42\fdg45$ & 55\,\sqd{} & $\geq 100$ & 406\,611 \\
     \disk{} & Field 807 & $19\fdg99, 62\fdg05$ & $126\fdg27, -0\fdg63$ & 55\,\sqd{} & $\geq 500$ & 1\,790\,565
\end{tabular}
\end{table*}

\section{Methodology}\label{sec:methodology}

Our anomaly detection pipeline is composed of 3 stages: feature extraction (Section \ref{subsec:features}), outlier detection using machine learning (Section \ref{subsec:ml}) and domain expert analysis (Section \ref{subsec:expert-analysis}).

We consider as outliers all the objects which result from the machine learning algorithms. Among the outliers we define two groups of objects, i.~e. ``bogus outliers'' and ``anomaly candidates''. To the former case, we assign all artefacts of image processing, bad columns on CCDs, tracks from satellites, diffraction spikes and other cases of non-astrophysical variability. To the latter case, we attribute the objects whose variability is proved to be astrophysically real and related either to the internal properties of the source or to the environmental effects, e.~g. present of a companion. All the anomaly candidates have a potential to be interesting for the experts in the corresponding domain. If after thorough expert-investigation an object is shown to behave unusually for its suspected astrophysical type, has unknown nature of variability, or represents a rare class of objects, then we deem it an ``anomaly''. It has to be stressed that the experts can miss some anomalies from the anomaly candidates due to insufficient knowledge in a particular domain of variable stars or transients or the lack of opportunity to perform the additional observations with required facilities. Therefore, presenting the results of this work in Table~\ref{sec:appendix-outliers}, we show not only anomalies confirmed by us, but the other anomaly candidates that have a potential to be proved as anomalies by other experts in the field.

All light-curve data sets were submitted to the same feature extraction procedure before being used as input to different machine learning algorithms.

\subsection{Feature extraction}
\label{subsec:features}

We extracted 42 features from each light curve. Some of them  probe magnitude properties (e.~g. Amplitude, Von Neumann $\eta$, Standard Deviation), while others are specific summary statistics for periodic signals (e.~g. Periodogram Amplitude, Periodogram $\eta$, Periodogram Standard Deviation). This ensemble of features describe different aspects of the light curve shape, with the tails of feature distributions indicating objects with less common light curves properties,  or potential outliers.  A detailed description of each feature is given in Appendix~\ref{sec:appendix-features}. Light-curve feature data within each field were standardised by shifting to zero mean and re-scaling to unity standard deviation. This ensures the results from outlier detection strategies will not depend on the units of the input variables.

\subsection{Outlier detection algorithms}
\label{subsec:ml}

This section describes the outlier detection algorithms  used in this study, which are aimed to select objects located in sparse regions of the feature space. Our pipeline profits from the Python  \texttt{scikit-learn}~\citep{scikit-learn} implementation for all strategies. 

\subsubsection{Isolation Forest}

Isolation Forest is an unsupervised outlier detection algorithm  first proposed by \citet{Liu_etal2008}. It considers that outliers are normally isolated from nominal data in the input parameter space, hence requiring less number of random partitions to be separated from it. Its scores are based on the inverse distance from root to leave node in an ensemble of decision trees built with random split points. Outliers are identified as objects with shorter path length from root to leaf node and the score is inversely proportional to the average path root-to-leaf among all trees in the forest. 

The two main parameters that describe an IF are: the number of trees and the sub-sample size used to train each tree. In what follows all results were obtained using sub-samples of 1000 elements and a forest containing 1000 trees. 

\subsubsection{Local Outlier Factor}

The Local Outlier Factor \citep[LOF, ][]{Breunig_etal2000} algorithm detects outliers in a data set based on comparative local density estimation. For each object in the data set, its local density is calculated using the distance between the object and its $k$-th nearest neighbour and the total number of neighbours it encloses ($k$). The average local density is calculated over all $k$ neighbours and the score is defined as the ratio between the individual and average local densities. Normal objects are expected to have similar local density to that of its neighbours, while outliers will show significant differences.

The behaviour of the LOF algorithm is dictated by two main parameters: the number of neighbours $k$ and the metric used to calculate the distance between instances. The number of neighbours, $k$, in particular is an important criterion, since too high a value of $k$ might cause the algorithm to miss outliers while too small a value results in a narrow focus which might be especially problematic in the case of noisy datasets. In what follows, all results were obtained using $k = 100$ with the distances given by the Euclidean metric.

\subsubsection{Gaussian Mixture Model}

A Gaussian Mixture Model \citep[GMM,][]{MCLA2000} is a parametric description of the data which assumes that multiple classes can be modelled as a superposition of multivariate Gaussian distributions. 
An object is considered to belong to a group according to its probability of being generated by one of the Gaussians within the model. Alternatively, an outlier is defined as an object for which all Gaussians within the mixture assign a low probability. The most important variable that defines the GMM is the number of components, or groups, to consider. In what follows, we used 10 isotropic Gaussian components.

\subsubsection{One-class Support Vector Machines }

Support Vector Machines \citep{CT2000,HTF2009} were originally formulated to tackle supervised binary classification scenarios. This is achieved by defining a hyperplane which separates points belonging to different classes. If the different classes are not linearly separable in a given feature space, they are projected into a higher dimensional feature space where they can be linearly separated using a kernel (`kernel trick'). \citet{schonesvm} proposed a modification to this method for novelty/outlier detection called One-class SVM. Analogously to its supervised learning counterpart, O-SVM identifies a decision boundary, by defining the smallest hypersphere  containing the bulk part of the data.  The model has two parameters: $\nu$, often called the margin which defines the probability of finding a new but normal point outside the decision boundary, and the kernel. In this work we used $\nu = 0.01$ and a Gaussian radial basis function kernel.

\subsection{Domain expert analysis}
\label{subsec:expert-analysis}

Outlier detection algorithms are, in general, able to identify statistical abnormalities within a large data set, e.~g. objects from sparse regions of the feature space or which do not conform with the general statistical description of the data. However, these may not be of astrophysical interest. Our approach to mining scientifically significant peculiarities admits that the light curve itself is typically not enough to determine the scientific content of a given source. Therefore, each outlier automatically identified  is subjected to analysis by a human expert with the goal of scrutinising its characteristics and determining the degree of its astrophysical content. In this context, all automatic learning algorithms can be seen as  recommendation systems whose goal is to perform a first triage which is subsequently confirmed by a human. The expert analysis includes literature and community search, cross-matching with known databases and catalogues, analysis of compatibility with theoretical models and, when possible, additional photometric and/or spectroscopic observations.
The final aim of the last stage is identification of anomalies (scientifically interesting objects confirmed by the expert) from outliers (candidates identified with high scores by the machine learning algorithms).

\subsubsection{The ZTF viewer}

In order to provide a smooth experience for the experts in charge of analysing outliers, we constructed a specially designed web-interface which allows smooth visualisation of several light curve characteristics: the SNAD ZTF viewer\footnote{\url{https://ztf.snad.space/}} (Fig.~\ref{fig:viewer}). It enables easy access to the individual exposure images; to the Aladin Sky Atlas~\citep{2000A&AS..143...33B,2014ASPC..485..277B} and to various catalogues of variable stars and transients, including the General Catalogue of Variable Stars (GCVS,~\citealt{2017ARep...61...80S}), the American Association of Variable Star Observers' Variable Star Index (AAVSO VSX,~\citealt{2006SASS...25...47W}), the Asteroid Terrestrial-impact Last Alert System (ATLAS,~\citealt{2018AJ....156..241H}), the ZTF Catalog of Periodic Variable Stars~\citep{2020ApJS..249...18C}, astrocats\footnote{\url{https://astrocats.space/}}, the OGLE-III On-line Catalog of Variable Stars~\citep{Soszynski2008}, and the  \textsc{SIMBAD} database~\citep{2000A&AS..143....9W}.

In ZTF DR3 each OID corresponds to an object in a particular field and passband, therefore the same source can have several OIDs. Our viewer allows the user to perform fast coordinate cross-match to associate a given OID with others from different fields and passbands, under a user-defined cross-match radius. Full description of the SNAD ZTF viewer is given in Appendix~\ref{sec:dr-viewer}.

\subsubsection{Additional observations}
\label{sec:observations}

For a few anomaly candidates we also performed additional observations with the telescopes at the Caucasus Mountain Observatory which belongs to the Sternberg Astronomical Institute, Lomonosov Moscow State University (CMO~SAI~MSU, \citealt{2020arXiv201010850S}). 

Photometric observations were carried with the 60-cm Ritchie-Chretien telescope in $g',r',i'$ passbands~\citep{1996AJ....111.1748F}, in remote control mode (RC600,~\citealt{2020ARep...64..310B}). Photometric reductions were performed using standard methods of dark frames and twilight sky flat-fields. Fluxes were extracted with the aperture photometry technique using fixed aperture radius $\approx$2.5 full width at half maximum (FWHM) 
of the stellar point spread function of the frame. We used an ensemble of 400 nearby stars from Pan-STARRS DR1 catalogue~\citep{2016Pan-starrs1,2016Pan-starrs2} to derive the linear solution between instrumental and Pan-STARRS DR1 magnitudes. Some stars were ejected from ensemble due to  $3\sigma$-clipping. The final number of comparison stars in each  ensemble was $\sim$\,200--300 for different objects.

Spectra were obtained with the Transient Double-beam Spectrograph (TDS) of the 2.5-meter telescope~\citep{2017ARep...61..715P}. The general characteristics of the TDS and the data reduction methods are described in ~\citet{Potanin2020}. The slit was oriented vertically in order to reduce wavelength dependent slit losses caused by atmospheric dispersion. The wavelength calibration was performed with a Ne-Al-Si lamp and corrected by using night-sky emission lines that allowed us to achieve an accuracy of $\lesssim10$\,km\,s$^{-1}$. The spectra were extracted with an aperture of 4.5\,arcsec. The flux calibration was performed by dividing the extracted spectra by the response curve, calculated with the spectrophotometric standard BD+28d4211. However, the photometric accuracy was lost due to the narrow slit, which was used to achieve a higher spectral resolution. Barycentric radial velocity corrections were applied.
\section{Results}
\label{sec:results}

We applied the outlier detection algorithms separately to each  field. \andro{} and \deep{} fields were scrutinised by all four algorithms, while for \disk{} we used only IF, GMM and O-SVM (LOF had a prohibitively high  computation cost given the number of objects in this field).  For all machine learning algorithms, the 40 objects with largest values of outlier score were submitted to the expert analysis. Taking into account that a few objects were assigned high anomaly scores by more than one algorithm, the final list of unique outliers contained 277 objects: 101 in \andro{}, 113 in \deep{} and 63 in \disk{}.
Each one of the 277 was subjected to the expert analysis using the utilities described in Section~\ref{subsec:expert-analysis}. 
A summary of the properties describing anomaly candidates is given in Table~\ref{tab:disk_deep_m31}. The first column is the object identifier from ZTF DR3. The second column contains alternative nomenclature by which the object is known. %the other names corresponding to the known objects. 
The equatorial coordinates ($\alpha$, $\delta$) in degrees are presented in the 3rd column. The 4th and 5th columns show the minimum and maximum $zr$ magnitudes derived from the entire public light curve of ZTF DR3. The line-of-sight reddening in our galaxy $E(B - V)$ is given in column 6~\citep{2011ApJ...737..103S}. The 7th column indicates the distance to the object ($D$) 
in parsec according to~\cite{2018AJ....156...58B}; for objects belonging to M\,31 we adopt a distance of $\sim780$~kpc~\citep{2014A&A...570A..13M}. The 8th column gives the approximate absolute $zr$ magnitude derived as $M_{r, {\rm max}} = m_{r, {\rm max}} - 5\log_{10}\,D + 5 - A_r$, where $A_r$ is the Milky Way foreground absorption in $zr$ passband. Column 9 contains the best period $P_0$, in days,  extracted from either the Lomb–Scargle periodogram (see Appendix \ref{sec:appendix-features}) or one of the catalogues listed in the ZTF-viewer or determined by us. In case of previously catalogued objects, their types and the source of classification are listed in columns 10 and 11, respectively.

\subsection{\andro{}}

Among the 101 outliers automatically identified in the \andro{} field there are 80 bogus light curves and 21 objects of astrophysical nature, 7 of which are not listed in  known catalogues and/or databases of variable sources. As expected, a large part of anomaly candidates in this field belongs to the M\,31 galaxy. Further information about these objects is given in Table~\ref{tab:disk_deep_m31}.

The known variables from our list are distributed by types as follows:  3 classical Cepheids (M\,31), 2 red supergiants (RSG, M\,31), 1 eclipsing binary (EB, MW), 2 possible novae (PNV, M\,31), 1 RS Canum Venaticorum-type binary system (RSCVN, MW), and 5 objects of unknown nature of variability.

\subsubsection{{\tt 695211200019653} --- RSCVN}

\begin{figure*}
    \includegraphics[width=\columnwidth]{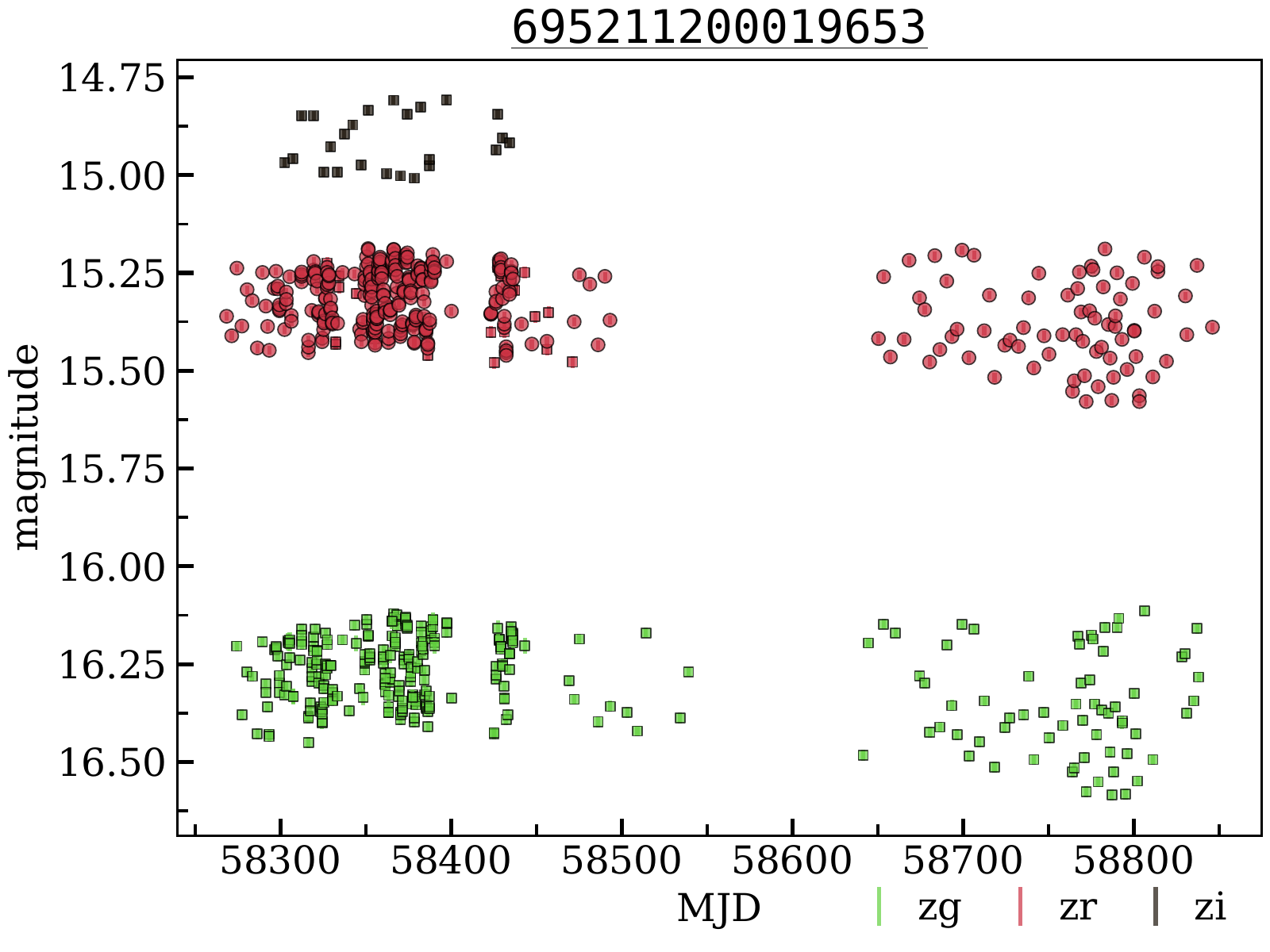}
    \includegraphics[width=\columnwidth]{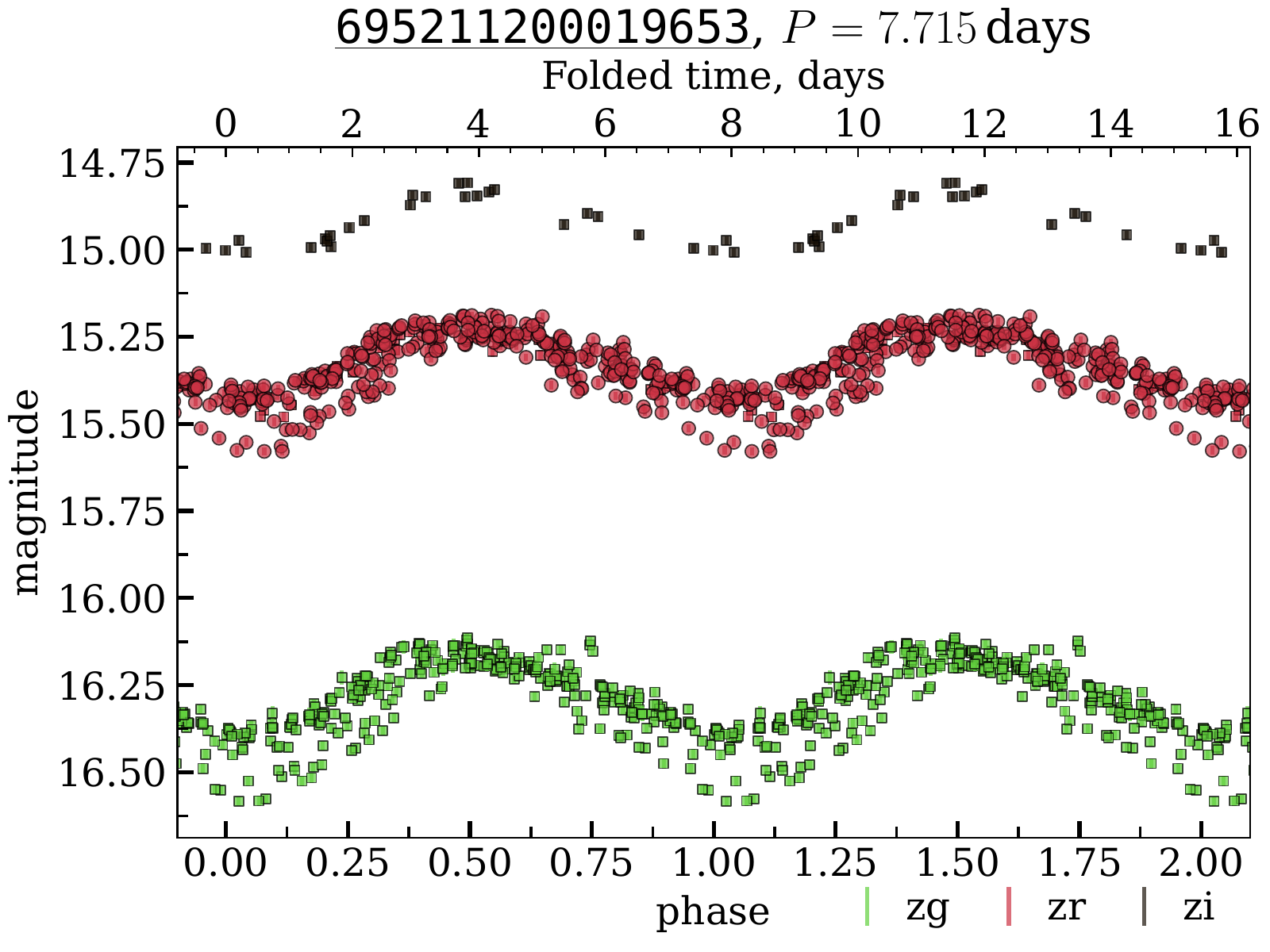}
    \caption{The light curve of RS Canum Venaticorum-type variable. \textbf{Left}: observed ZTF light curve.  \textbf{Right}: folded light curve using $P=7.715$\,d  obtained using our feature extraction code (see Appendix~\ref{sec:appendix-features}). Note the minimum depth change after MJD$\gtrsim 58750$ (left) and the two outlier $zg$ observations in the folded light curve (right). OIDs: {\tt 695111200008124}, {\tt 695211200019653}, {\tt 695311200027644}, {\tt 1735101200013615}, {\tt 1735201200018614}. Different colours denote the different ZTF passbands. Red circles correspond to the outlier OID, squares show data from other OIDs corresponding to the same source but found in different passbands and ZTF fields.}
    \label{fig:695211200019653}
\end{figure*}

The object {\tt 695211200019653} (Fig. \ref{fig:695211200019653}) was previously classified as a star with sine-like variability and period $P=7.696$\,d in the ATLAS catalog of variable stars~\citep{2018AJ....156..241H}. The ASAS-SN Catalog of Variable Stars~\citep{2020MNRAS.493.4186J} classifies it as a rotation variable with $P=7.709$\,d and the ZTF Catalog of Periodic Variable Stars~\citep{2020ApJS..249...18C} considers it a RSCVN with period $P=7.734$\,d based on an automatic DBSCAN classifier. Moreover, this object is also  marked as X-ray and UV source by {\em ROSAT}, {\em XMM-Newton} and {\em Swift/UVOT}. According to the Gaia DR2~\citep{2018A&A...616A...1G}, it is a Milky Way object at the distance $D\approx1640$~pc, $R\approx2.34$\,R$_{\odot}$, $T_{\rm eff}\approx4500$\,K, and $L\approx2$\,L$_{\odot}$.

Its ZTF DR3 light curves are characterised by amplitude variability, e.~g. around $\rm MJD\simeq 58750-58850$ when the minimums become deeper. Moreover, its $zg$ light curve shows two observations at $\rm MJD\simeq 58368.4$ which could be a flare with an amplitude of $\gtrsim 0.1$\,mag. The Transiting Exoplanet Survey Satellite  \citep[{\em TESS}, ][]{2014SPIE.9143E..20R} observed {\tt 695211200019653} continuously during $\sim3$ periods. Its light curve shows an  asymmetric sine-like variability with amplitude of $\approx3$\%, significant inter-period changing (typical for stars with spot activity) and $\sim0.05$\,mag flare near $\rm MJD\simeq58779.5$. The estimated period is 7.662\,d. The difference in periods estimation from different surveys can be explained by the low signal to noise ratio of ASAS-SN data, irregular sampling of ZTF and ATLAS data, and by the short observation sequence of {\em TESS}. Alternatively, the difference could also be explained by %An alternate explanation is 
a real fluctuation in %of light curve 
period due to changing of positions and temperatures of stellar spots.

\begin{figure*}
  \includegraphics[width=2\columnwidth]{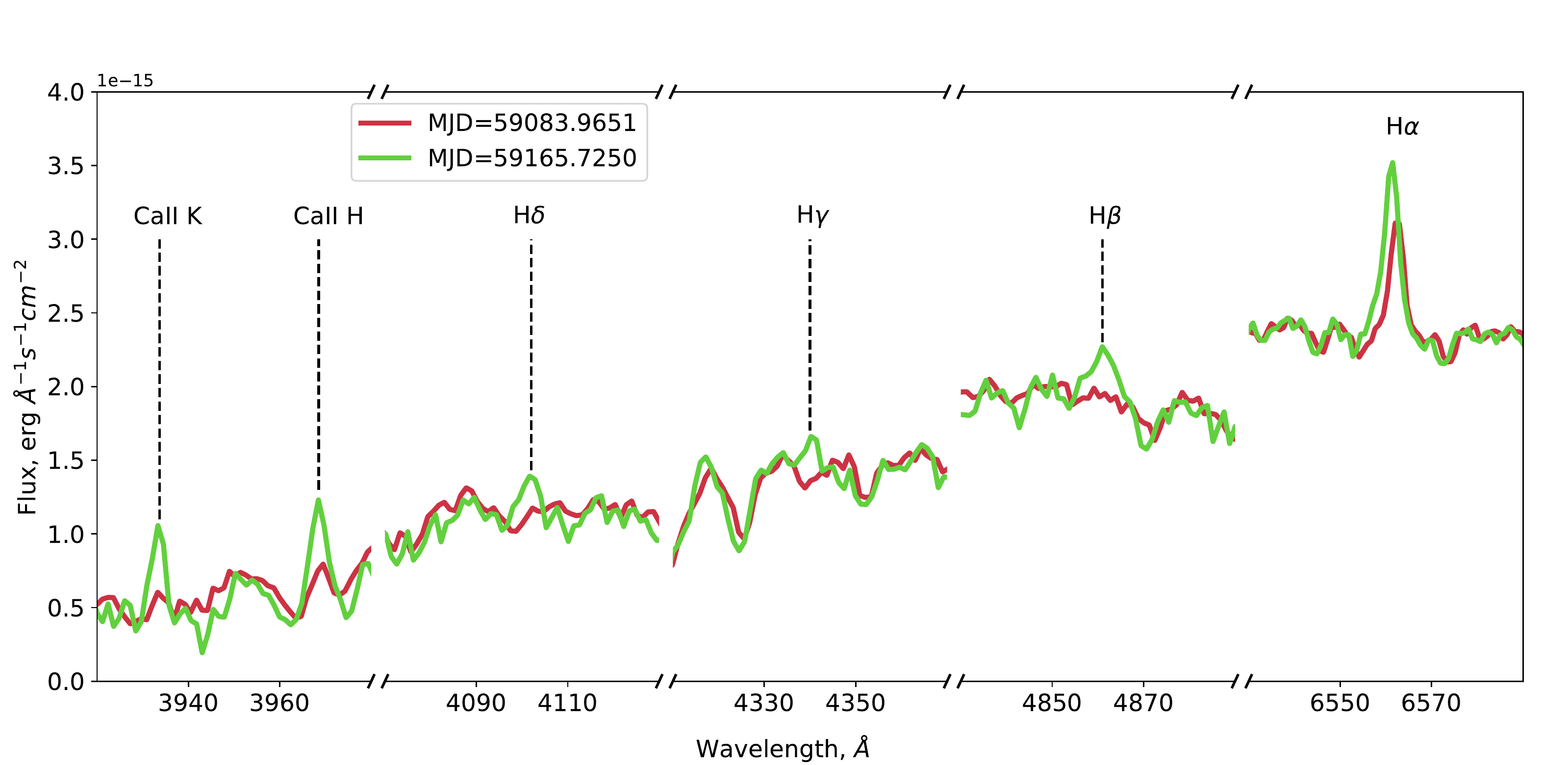}
  \caption{Ca\,\textsc{II} H-K and Balmer series emissions on two spectra of {\tt 695211200019653} obtained with the 2.5-meter telescope of CMO SAI MSU. }
    \label{fig:Spec_695211200019653}
\end{figure*}

We obtained two spectra of {\tt 6952112007019653} with a resolution of $\sim1500$ and a signal to noise ratio of $\sim50$ on 2020 August 22 and November 12 (MJD = 59083.9651 and MJD = 59165.7250) with a 1\,arcsec slit and total exposure times of 1500\,s and 1800\,s, correspondingly (see Fig.~\ref{fig:Spec_695211200019653}).
Spectral classification was done by comparing with the spectral library~\citep{Valdes}. We determined the spectral class as K3. Strong variable Balmer and Ca\,\textsc{II} H-K emissions are present in all spectra, indicating chromospheric activity (\citealt{1968ApJ...153..221W}). Velocities of the emission lines correspond to the absorption lines. The radial velocity shift between two spectra is $46\pm10$\,km\,s$^{-1}$, that points out to the presence of a second companion. However, we did not detect any obvious spectral signs of a companion in the spectra. Based on overall analysis, distance estimation,  photometric and spectral data we definitely classified {\tt 695211200019653} as RS Canum Venaticorum-type binary system \citep{2005LRSP....2....8B}.

\subsubsection{{\tt 695211200075348} --- Unclassified Variable}

The object M31N 2013-11b was first discovered by~\cite{2013ATel.5569....1O} on 2013 Nov. 7.780 UT (MJD = 56603.780) with $R \simeq 20.3$~mag and classified as probable nova in M31. Later on, 2013 Nov.-Dec., \citet{2013ATel.5640....1H} observed the object using the 0.65-m telescope at Ondrejov and the Danish 1.54-m telescope at La Silla. The observations revealed a  significant red colour which is not typical for a classical nova unless the object is highly reddened, which is not expected for its line of sight. \citet{2013ATel.5640....1H} concluded that it is more likely a red long-period variable which is supported also by its very slow brightening. 

\begin{figure}
\begin{center}
\includegraphics[width=\columnwidth]{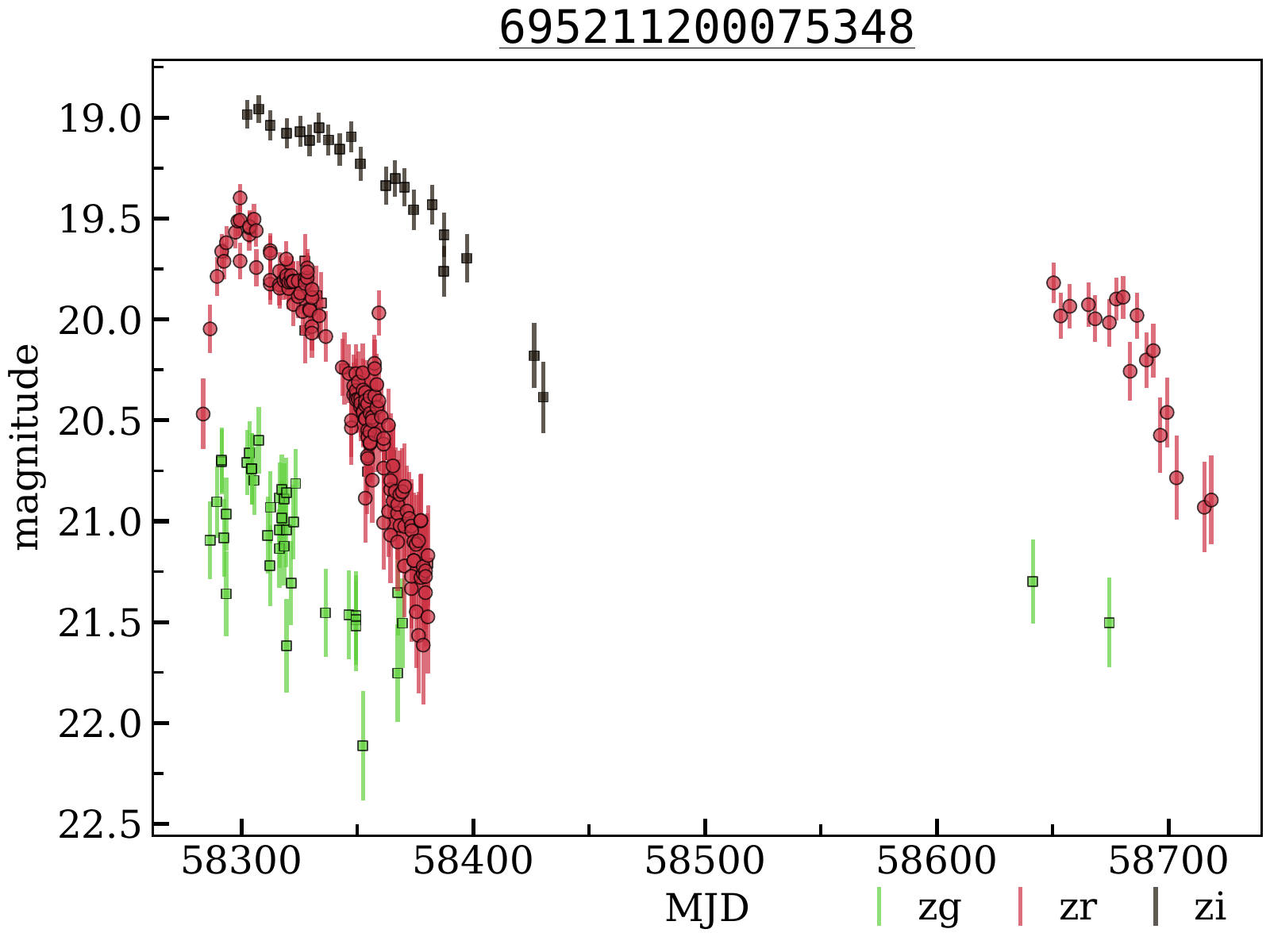}
\caption{The light curves of M31N 2013-11b/MASTER OT J004126.22+414350.0 --- unclassified variable object from the \andro{} field. OIDs: {\tt 695111200015523}, {\tt 695211200075348}, {\tt 695311200020148}, {\tt 1735101200009309}, {\tt 1735201200040725}.}
\label{fig:695211200075348}
\end{center}
\end{figure}

On MJD = 57633.1234 MASTER-IAC auto-detection system discovered the optical transient source, MASTER OT J004126.22+414350.0, at the same position with unfiltered magnitude $\sim19.7$~mag~\citep{2016ATel.9470....1S}. \citet{2016ATel.9554....1W} performed spectroscopic and additional photometric observations with the 2-m Liverpool Telescope on 2016 Sep 9 
UT. The spectrum revealed no obvious emission or absorption lines, but the continuum was clearly detected. Spectroscopic and photometric observations of the transient implied it is unlikely to be a recurrent nova eruption in M\,31. The colour of the transient also suggests it is unlikely to be a Galactic dwarf nova outburst. 
The ZTF object light curve is given in Fig.~\ref{fig:695211200075348}.

\subsubsection{{\tt 695211100022045} --- Nova Candidate}

\begin{figure}
\begin{center}
\includegraphics[width=\columnwidth]{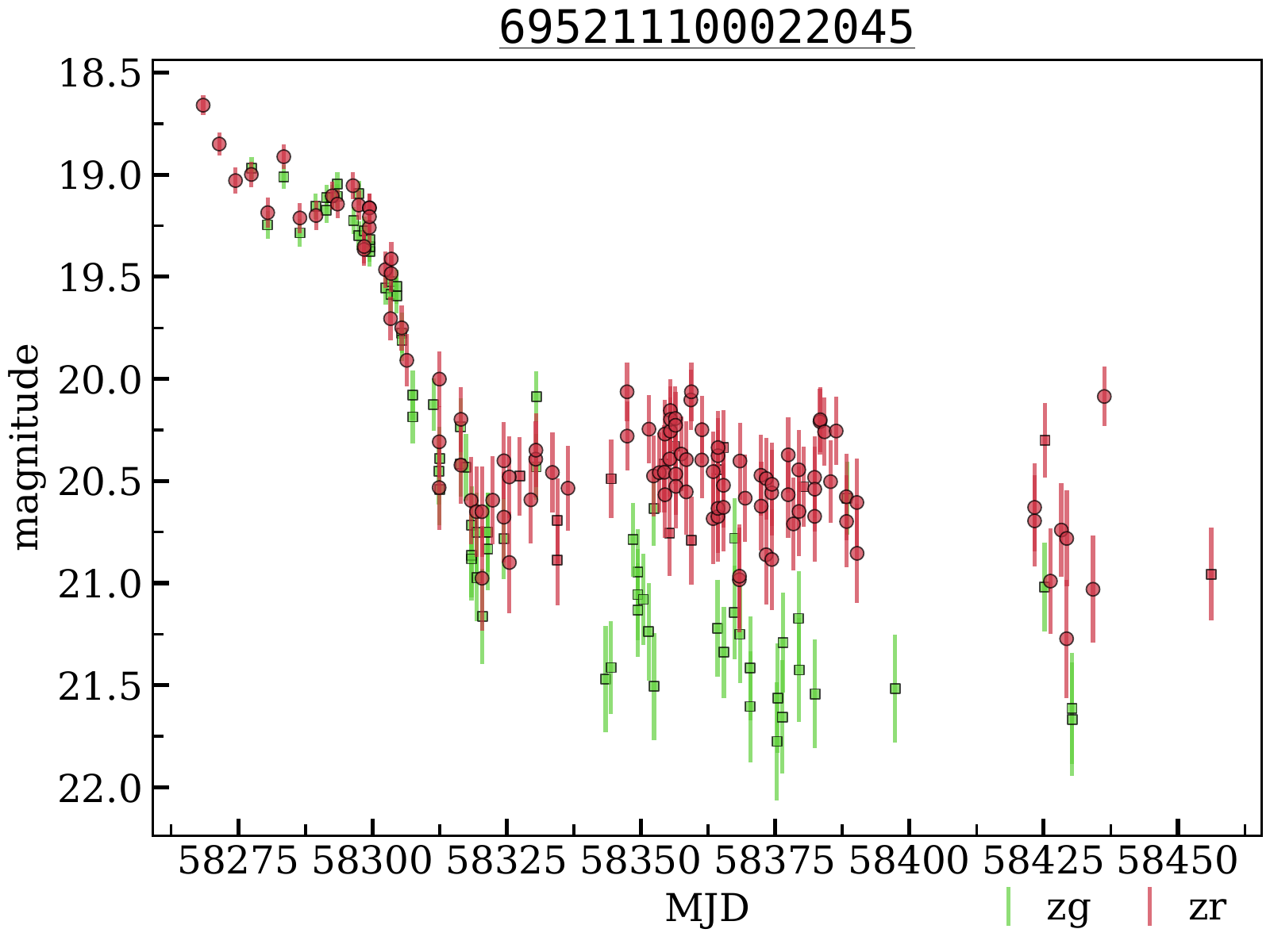}
\caption{The light curves of AT 2017ixs/MASTER OT J004355.89+413209.9 --- nova candidate from the \andro{} field. OIDs: {\tt 695111100031492}, {\tt 695211100022045}, {\tt 1735101200034700}, {\tt 1735201200049286}.}
\label{fig:695211100022045}
\end{center}
\end{figure}

According to the Transient Name Server\footnote{\url{https://www.wis-tns.org}} (TNS), the object {\tt 695211100022045} was first seen on 2017-10-29 (MJD = 58055) as AT 2017ixs. It was detected a second time on 2017-12-15 (MJD = 58102) with $19.5$~mag in Clear filter and classified as a possible nova \citep{2017TNSTR1418....1C} in M\,31. Six months latter, on 2018-06-20 (MJD = 58289),  MASTER-Kislovodsk auto-detection system discovered MASTER OT J004355.89+413209.9 with an unfiltered magnitude of $19.0$~mag at AT 2017ixs position~\citep{2018ATel11755....1B}. The behaviour of its light curve is not typical for a  dwarf nova or a cataclysmic variable, therefore, AT 2017ixs is the interesting anomaly for the further study. The ZTF object light curve is given in Fig.~\ref{fig:695211100022045}.

\subsubsection{{\tt 695211100131796} --- LBV Candidate}

\begin{figure}
\begin{center}
\includegraphics[width=\columnwidth]{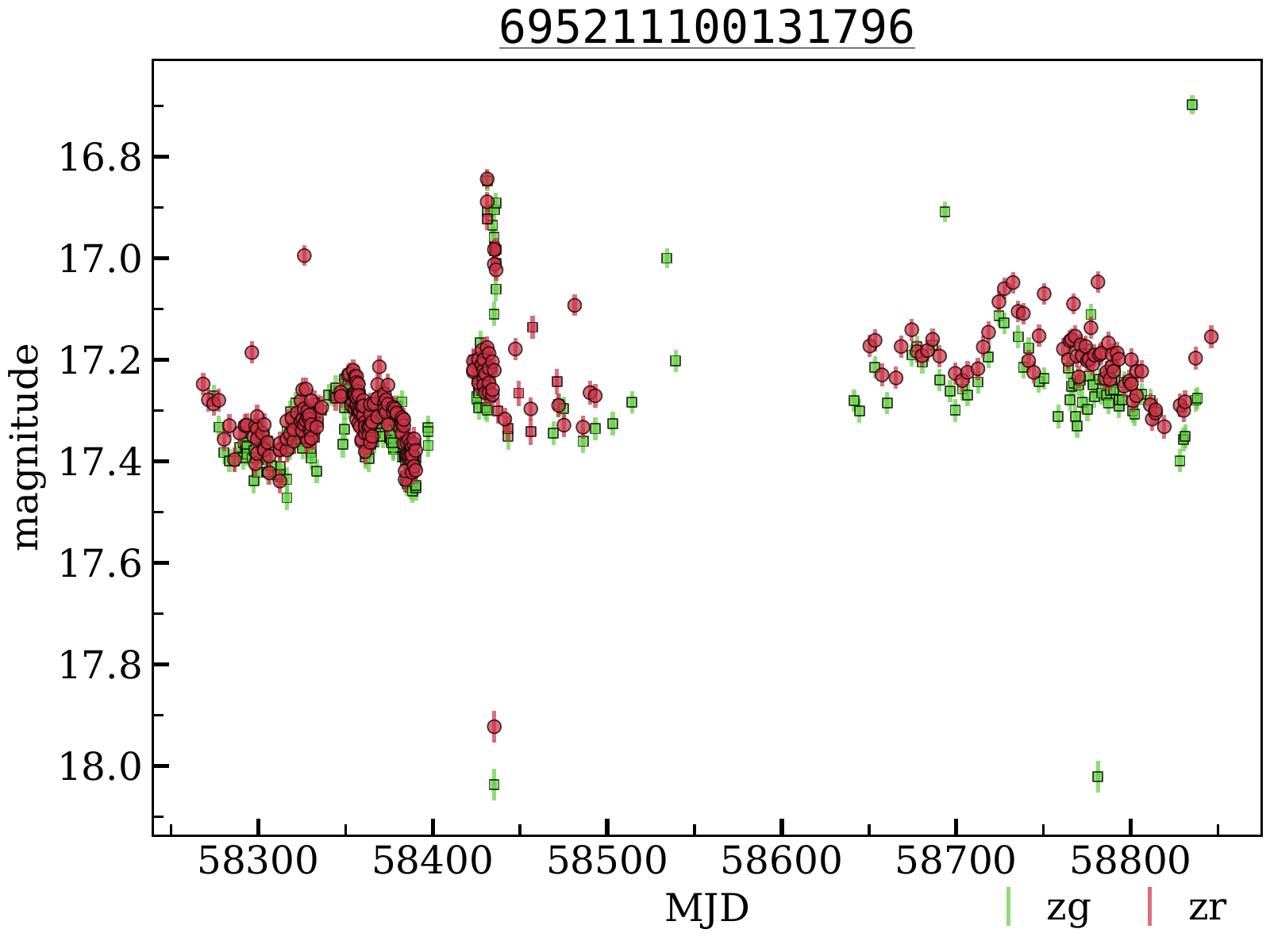}
\caption{The light curves of  PSO J011.0457+41.5548 --- luminous blue variable candidate from the \andro{} field. OIDs: {\tt 695111100014264}, {\tt 695211100131796}, {\tt 1735101200058759}, {\tt 1735201200018887}.}
\label{fig:695211100131796}
\end{center}
\end{figure}

The object {\tt 695211100131796} (Fig.~\ref{fig:695211100131796}) is located near the ionised hydrogen region [AMB2011]~HII~2692~\citep{2011AJ....142..139A}. It was previously detected as an object of unknown nature PSO J011.0457+41.5548~\citep{2014ApJ...785...11L}.  Based on the spectra of PSO J011.0457+41.5548, which is turned to be typical of B- and A-type supergiants, \citet{2017ApJ...836...64H} concluded that the available information is  insufficient  to confirm  it as a luminous blue variable (LBV). We consider it a luminous blue variable or  variable of the S~Doradus type (SDOR) candidate.

\subsection{\textsc{DEEP}}

Among the 113 outliers in the \deep{} field there are 95 bogus light curves and 18 objects of astrophysical nature, 8 of which are not listed in known catalogues and/or databases of variable sources. The information about these objects is given in Table~\ref{tab:disk_deep_m31}. 

The known variables from our list are distributed by types as follows: 2 eclipsing binaries, 2 semi-regular variables, 1 Mira variable, 1 RR Lyrae variable with asymmetric light curve (steep ascending branches), 1 polar, 1 Type Ia Supernova (SN), 2 supernova candidates.

\subsubsection{{\tt 795214300016648} --- Red Dwarf Flare}

\begin{figure}
\begin{center}
\includegraphics[width=\columnwidth]{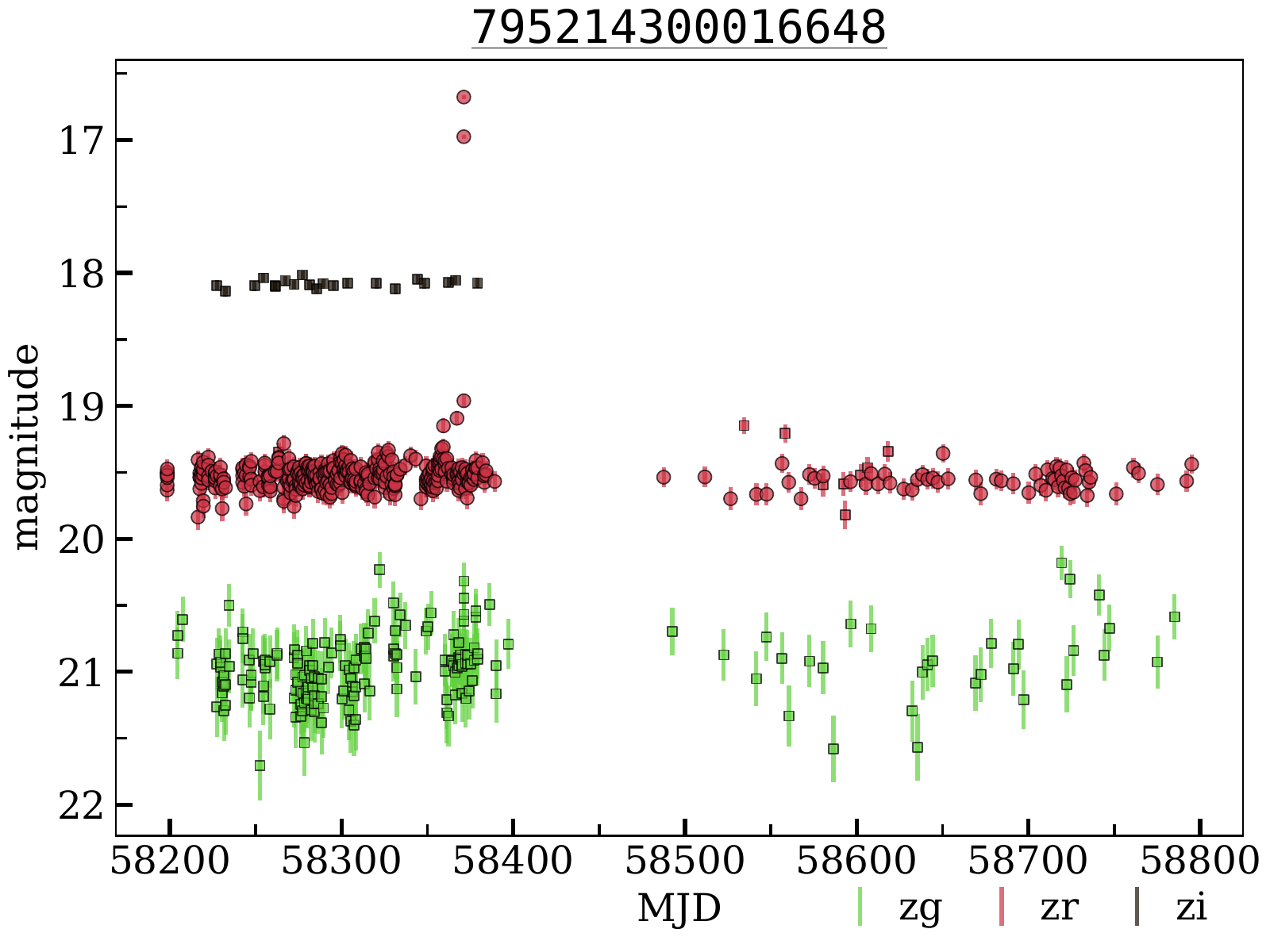}
\caption{The light curves of {\tt 795214300016648} from the \deep{} field classified by us as red dwarf flare. OIDs: {\tt 795114300008349}, {\tt 795214300016648}, {\tt 795314300011499}, {\tt 1829206200019338}.}
\label{fig:795214300016648}
\end{center}
\end{figure}

Among the unclassified objects we would like to mention, {\tt 795214300016648} is a possible red dwarf flare (Fig.~\ref{fig:795214300016648}). It is a relatively nearby object at a distance of $\sim400$~pc~\citep{2018AJ....156...58B}, with a high  proper motion $\rm PM_{R.A.} = 8.122\pm0.502$~mas/yr,	
$\rm PM_{Dec.} = -29.269\pm0.596$~mas/yr~\citep{2018A&A...616A...1G}. Despite the fact that such objects are quite common in our galaxy, their detection is rare because of the small duration and low luminosity of its flares. Their study is of great scientific interest due to the potential habitability associated to the planets it hosts (e.~g.~\citealt{2010AsBio..10..751S,2013ApJ...763..149F}).

\subsubsection{Supernova Candidates}

In the outlier list from \deep{} there are 6 objects that are possibly of extragalactic origin --- {\tt 795202100005941}, {\tt 795204100013041}, {\tt 795205100007271}, {\tt 795209200003484}, {\tt 795212100007964}, {\tt 795213200000671}. Two of them appeared in the databases as known supernova candidates: {\tt 795213200000671} (AT 2018afr/Gaia18apj\footnote{\url{http://gsaweb.ast.cam.ac.uk/alerts/home}}) and {\tt 795202100005941} (MLS180307:163438+521642;~\citealt{2009ApJ...696..870D}). Considering the photometric redshift of the possible host galaxy, {\tt 795202100005941} absolute magnitude is $\sim-22.4$, making it a potential superluminous supernova (SLSN) candidate. The four remaining objects are not catalogued. 

We use the \textsc{Python} library \textsc{sncosmo}\footnote{\url{https://sncosmo.readthedocs.io/en/stable/}} to fit their light curves with several supernova models: Peter Nugent's spectral templates\footnote{\url{https://c3.lbl.gov/nugent/nugent_templates.html}} which cover the main supernova types (Ia, Ib/c, IIP, IIL, IIn) and \textsc{SALT2} model of Type Ia Supernovae~\citep{Guy07}. Each model is characterised by a set of parameters. Nugent's models are the simple spectral time series that can be scaled up and down. The parameters of the models are the redshift $z$, the observer-frame time corresponding to the zero source's phase, $t_0$, and the amplitude. The zero phase is defined relative to the explosion moment and the observed time $t$ is related to phase via $t = t_0 + {\rm phase} \times (1 + z)$.

The \textsc{SALT2} model is more sophisticated and contains the parameters that affect the shape of the spectrum at each phase. In addition to the redshift, $t_0$, and amplitude, the light curves are also characterised by $x_1$ (stretch) and $c$ (colour) parameters. The $x_1$ parameter describes the time-stretching of the light curve. The $c$ parameter corresponds to the colour offset with respect to the average at the date of maximum luminosity in $B$-band, i.~e. $c = (B - V )_{max} - \langle B - V\rangle$. In \textsc{SALT2} models the zero phase is defined relative to the maximum in $B$-band.

For each object we extracted photometry in $zg, zr, zi$ passbands from \deep{} field only. In a few cases, due to insufficient number of photometrical points in $zi$, we added to the $zi$-light curve observations from other fields. 
Then, we subtracted the reference magnitude from ZTF light curves to roughly account for the host galaxy contamination. In order to perform the fit, we determined the redshift bounds for each supernova candidate. For three objects (see Table~\ref{tab:sn_cand}) there are known SDSS galaxies at the source position with measured photometric redshift with errors, which we used for the redshift bounds. For the remaining ones, we adopted $[-15; -22]$ as an acceptable region for the supernovae absolute magnitude \citep{2014AJ....147..118R} and, then, using the apparent maximum magnitude, roughly transform it to the possible redshift range. 

\begin{table*}
% \centering
\caption{Results of the light curve fit with the \textsc{SALT2} model for supernova candidates from the \deep{} field.}
\begin{scriptsize}
\begin{tabular}{lccccccc}
\hline
OID & Host galaxy$^*$ & $z_{\rm ph}$ & $z$ & $t_0$ & $x_1$ & $c$ & Comments$^\dagger$\\
\hline
{\tt 795202100005941}/ZTF18aanbnjh & SDSS~J163437.92+521642.2 & $0.424\pm0.103$ & --- & --- & ---  & --- &  Blazar\\
{\tt 795204100013041}/ZTF18abgvctp &  SDSS~J160913.83+521251.3 & $0.375\pm0.138$ & $\sim$0.24 & 58320.9336$\pm$0.4389 & 1.71$\pm$0.51 & $-$0.044$\pm$0.035 & --- \\
{\tt 795205100007271}/ZTF18aayatjf & --- & --- & $\sim$0.20 & 58285.8334$\pm$0.1810 & $-$0.54$\pm$0.18 & $-$0.075$\pm$0.021 & SN Ia\\
{\tt 795209200003484}/ZTF18abbpebf & --- & --- & $\sim$0.11 & 58299.7269$\pm$0.0008 & 0.60$\pm$0.12 & $-$0.013$\pm$0.012 & SN Ia\\
{\tt 795212100007964}/ZTF18aanbksg & SDSS~J161144.90+555740.7 & $  0.288\pm0.122$& $\sim$0.18 & 58214.4470$\pm$0.0002 & 0.40$\pm$0.20 & $-$0.282$\pm$0.020& Blazar\\
{\tt 795213200000671}/ZTF18aaincjv & --- & --- & --- & --- & --- & --- & AGN-I \\
\hline
\end{tabular}
\label{tab:sn_cand}\\
\begin{flushleft}
$^*$
If available, candidate host galaxies from SDSS DR16~\citep{2020ApJS..249....3A} and their corresponding photometric redshifts ($z_{\rm ph}$, obtained via the KD-tree method).\\
$^\dagger$
According to the light curve classifier of the ALeRCE broker \citep{Forster2020}.
\end{flushleft}
\end{scriptsize}
\end{table*}

Since dust in the Galaxy also affects the shape of an observed spectrum, we accounted for it as an additional effect during the fitting procedure. We used \citet{1989ApJ...345..245C} extinction model and the individual object's colour excess $E(B-V)$ (see Table~\ref{tab:disk_deep_m31}).

Results from the light curve fit are presented in Fig.~\ref{fig:795209200003484},~\ref{fig:795212100007964},~\ref{fig:795205100007271},~\ref{fig:795204100013041}. We do not show the fitted light curves for {\tt 795202100005941}, {\tt 795213200000671} since both supernova candidates were discovered after the maximum light and only descending part of the light curves is available. The analysis of four non-catalogued supernova candidates revealed that their light curves are similar to those of Type Ia Supernovae.
The determined $x_1$ and $c$ parameters are typical for SNe~Ia. We summarised the main fit parameters using the \textsc{SALT2} model in Table~\ref{tab:sn_cand}.

\subsection{\disk{}}

Among the 63 outliers in the \disk{} field there are 13 bogus light curves and 50 variable objects of astrophysical nature, 8 of which are not listed in known catalogues and/or databases of variable sources. The information about these objects is given in Table~\ref{tab:disk_deep_m31}.

\begin{figure}
\begin{center}
\includegraphics[width=\columnwidth]{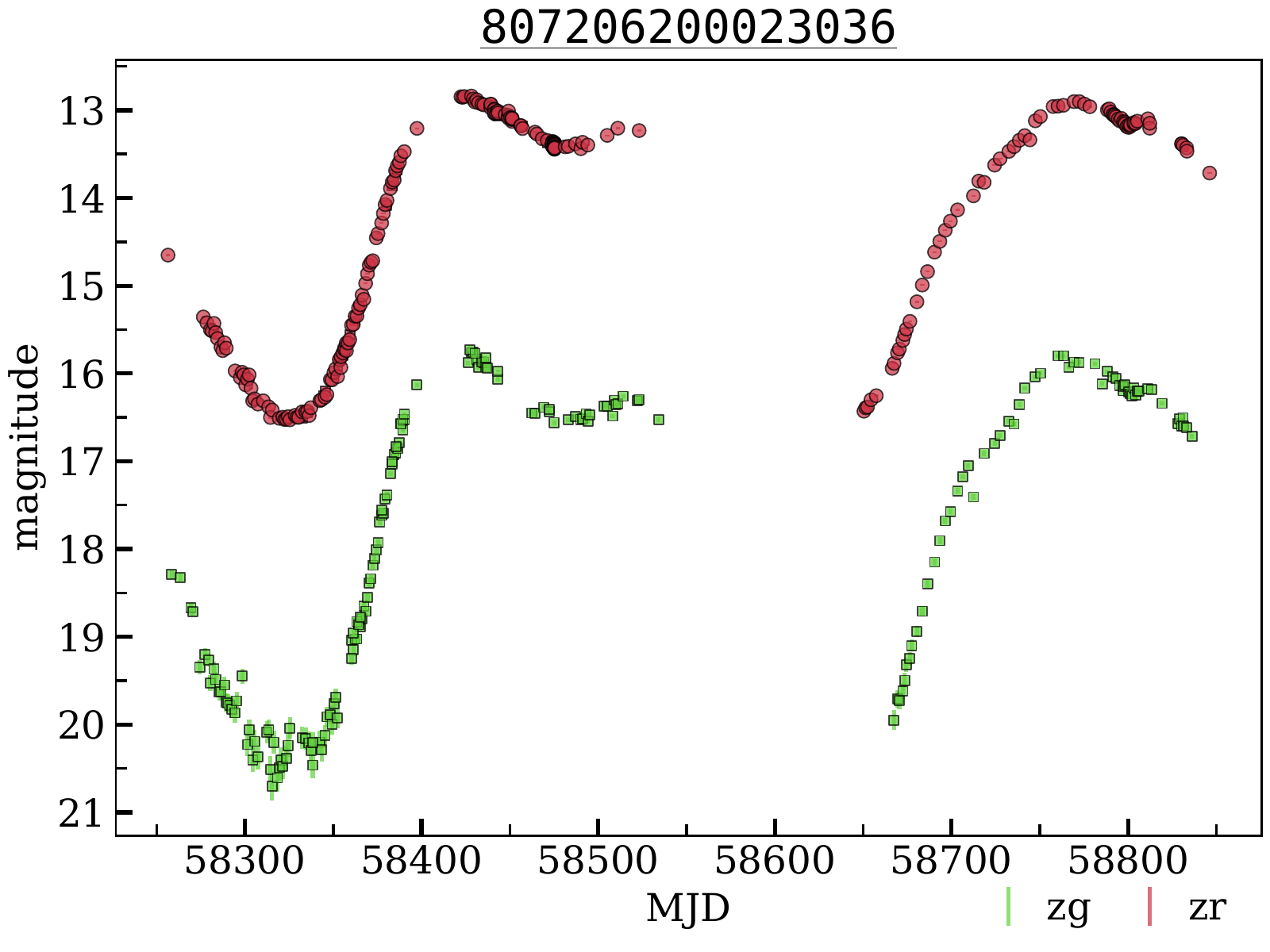}
\caption{The light  curves  of NSVS J0122238+611352 --- Mira variable from the \disk{} field. OIDs: {\tt 807106200016307}, {\tt 807206200023036}, {\tt 1810114300003360}, {\tt 1810214300006290}.}
\label{fig:807206200023036}
\end{center}
\end{figure}

Among the known variables from our list there are 3 dwarf novae, 5 eclipsing systems, 1 Orion variable with rapid light variations, 28 Mira variables, 1 other long-period variable of unspecified type and 4 candidates to pre-main-sequence (PMS) stars. These objects are considered anomaly candidates, and therefore, interesting sources to study more carefully. For example, some of these Mira variables have  asymmetric light curves which may indicate the presence of a companion (e.~g.~{\tt 807206200023036}, Fig.~\ref{fig:807206200023036}).

\subsubsection{Pre-main-sequence candidates}

\begin{figure}
\begin{center}
\includegraphics[width=\columnwidth]{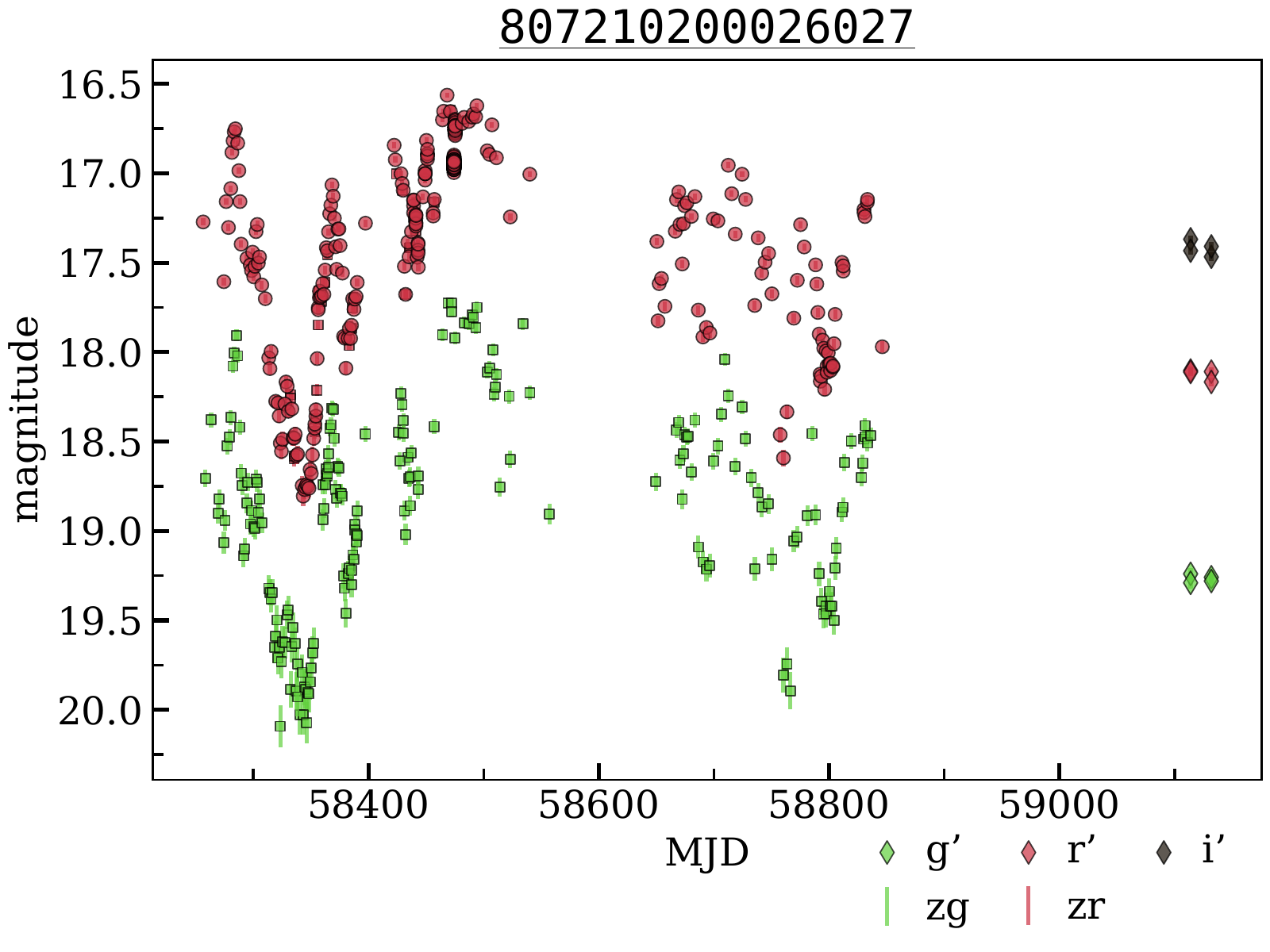}
\caption{The light  curves  of  poorly classified variable object --- pre-main-sequence star or slow red irregular variable candidate from the \disk{} field. OIDs: {\tt 807110200025843}, {\tt 807210200026027}, {\tt 1840103100026721}, {\tt 1840203100059097}. Diamonds show the additional $g',r',i'$ observations performed with RC600 telescope of CMO SAI MSU.}
\label{fig:807210200026027}
\end{center}
\end{figure}

We identified four objects ({\tt 807206200003542}, {\tt 807206200004116}, {\tt 807206200014645}, {\tt 807210200026027}) previously classified as pre-main-sequence candidates~\citep{2020AA...638A..21V}. For these objects, we gathered additional observations with the RC600 telescope of CMO SAI MSU  in $g',r',i'$ passbands (Table~\ref{tab:photometry_1}) and confirmed that the variability is still present. Three of them are located at a distance of  800--1000\,pc, and the last one ({\tt 807210200026027}) is significantly further, at 1700\,pc~\citep{2018AJ....156...58B}. In particular, {\tt 807210200026027} can be also found among the AGN candidates reported by ~\citet{2012ApJ...751...52E}. In order to identify AGN candidates, \citet{2012ApJ...751...52E} used the colour information from the Wide-field Infrared Survey Explorer (WISE;~\citealt{Wright2010}), the  Two  Micron All-Sky  Survey  (2MASS;~\citealt{Skrutskie2006}), and checked their presence in X-rays data with ROSAT All-Sky Survey  (RASS;~\citealt{1999AA...349..389V}). However, {\tt 807210200026027} was assigned as an AGN solely based on infrared observations, making it a weak candidate, and in contradiction with {\em Gaia} parallax~\citep{2018A&A...616A...1G}. Based on ML techniques, the object was also classified as a young stellar object (YSO;~\citealt{2016MNRAS.458.3479M}). However, the light curve seems to be typical for slow red irregular variables, LB type, according to the General Catalogue of Variable Stars classification system (Fig.~\ref{fig:807210200026027}), which does not contradict the {\em Gaia} parallax.

\begin{table}
\centering
\caption{Photometric observations of the candidates to pre-main-sequence sources with RC600 telescope of the Caucasus Mountain Observatory, SAI MSU.}
\begin{tabular}{lccc}
\hline
MJD & Filter & Magnitude & Error \\
\hline
\multicolumn{4}{c}{{\tt 807206200003542}}\\
\hline
59114.06219	&	$g'$	&	16.916	&	0.010	\\
59114.06499	&	$g'$	&	16.930	&	0.010	\\
59114.06780	&	$g'$	&	16.911	&	0.011	\\
59132.84609	&	$g'$	&	16.046	&	0.006	\\
59132.84720	&	$g'$	&	16.052	&	0.006	\\
59114.06319	&	$r'$	&	15.592	&	0.006	\\
59114.06600	&	$r'$	&	15.597	&	0.006	\\
59114.06881	&	$r'$	&	15.588	&	0.006	\\
59132.84437	&	$r'$	&	14.900	&	0.004	\\
59132.84514	&	$r'$	&	14.901	&	0.004	\\
59114.06403	&	$i'$	&	14.833	&	0.005	\\
59114.06683	&	$i'$	&	14.823	&	0.005	\\
59114.06964	&	$i'$	&	14.830	&	0.005	\\
59132.84283	&	$i'$	&	14.249	&	0.004	\\
59132.84359	&	$i'$	&	14.252	&	0.004	\\
\hline
\multicolumn{4}{c}{{\tt 807206200004116}}\\
\hline
59114.07549	&	$g'$	&	20.150	&	0.180	\\
59132.82983	&	$g'$	&	20.587	&	0.160	\\
59132.83456	&	$g'$	&	20.480	&	0.096	\\
59132.83812	&	$g'$	&	20.310	&	0.087	\\
59114.07953	&	$r'$	&	18.534	&	0.047	\\
59114.08100	&	$r'$	&	18.546	&	0.055	\\
59132.82603	&	$r'$	&	18.949	&	0.059	\\
59132.82783	&	$r'$	&	18.887	&	0.054	\\
59114.08235	&	$i'$	&	17.587	&	0.047	\\
59114.08347	&	$i'$	&	17.529	&	0.049	\\
59132.82239	&	$i'$	&	18.369	&	0.106	\\
59132.82420	&	$i'$	&	18.195	&	0.057	\\
\hline
\multicolumn{4}{c}{{\tt 807206200014645}}\\
\hline
59114.03133	&	$g'$	&	17.780	&	0.016	\\
59114.03279	&	$g'$	&	17.782	&	0.016	\\
59114.03426	&	$g'$	&	17.805	&	0.016	\\
59132.81479	&	$g'$	&	18.379	&	0.058	\\
59132.81625	&	$g'$	&	18.285	&	0.066	\\
59132.81771	&	$g'$	&	18.358	&	0.075	\\
59114.03544	&	$r'$	&	16.997	&	0.013	\\
59114.03621	&	$r'$	&	17.031	&	0.014	\\
59114.03699	&	$r'$	&	16.992	&	0.013	\\
59132.81127	&	$r'$	&	17.535	&	0.034	\\
59132.81238	&	$r'$	&	17.532	&	0.034	\\
59132.81349	&	$r'$	&	17.506	&	0.040	\\
59114.03781	&	$i'$	&	16.386	&	0.013	\\
59114.03859	&	$i'$	&	16.391	&	0.013	\\
59114.03936	&	$i'$	&	16.384	&	0.012	\\
59132.80879	&	$i'$	&	16.832	&	0.034	\\
59132.80955	&	$i'$	&	16.829	&	0.037	\\
59132.81031	&	$i'$	&	16.905	&	0.038	\\
\hline
\multicolumn{4}{c}{{\tt 807210200026027}}\\
\hline
59113.8449916	&	$g'$	&	19.240	&	0.033	\\
59113.8485605	&	$g'$	&	19.290	&	0.034	\\
59131.8365898	&	$g'$	&	19.259	&	0.039	\\
59131.8401575	&	$g'$	&	19.280	&	0.039	\\
59113.8513134	&	$r'$	&	18.103	&	0.022	\\
59113.8531301	&	$r'$	&	18.111	&	0.022	\\
59131.8430829	&	$r'$	&	18.109	&	0.023	\\
59131.8452469	&	$r'$	&	18.167	&	0.024	\\
59113.8549981	&	$i'$	&	17.369	&	0.019	\\
59113.8568122	&	$i'$	&	17.433	&	0.020	\\
59131.8472884	&	$i'$	&	17.467	&	0.024	\\
59131.8491064	&	$i'$	&	17.409	&	0.025	\\
\hline
\end{tabular}
\label{tab:photometry_1}
\end{table}

\subsubsection{Eclipsing binaries}

\begin{figure}
\begin{center}
\includegraphics[width=\columnwidth]{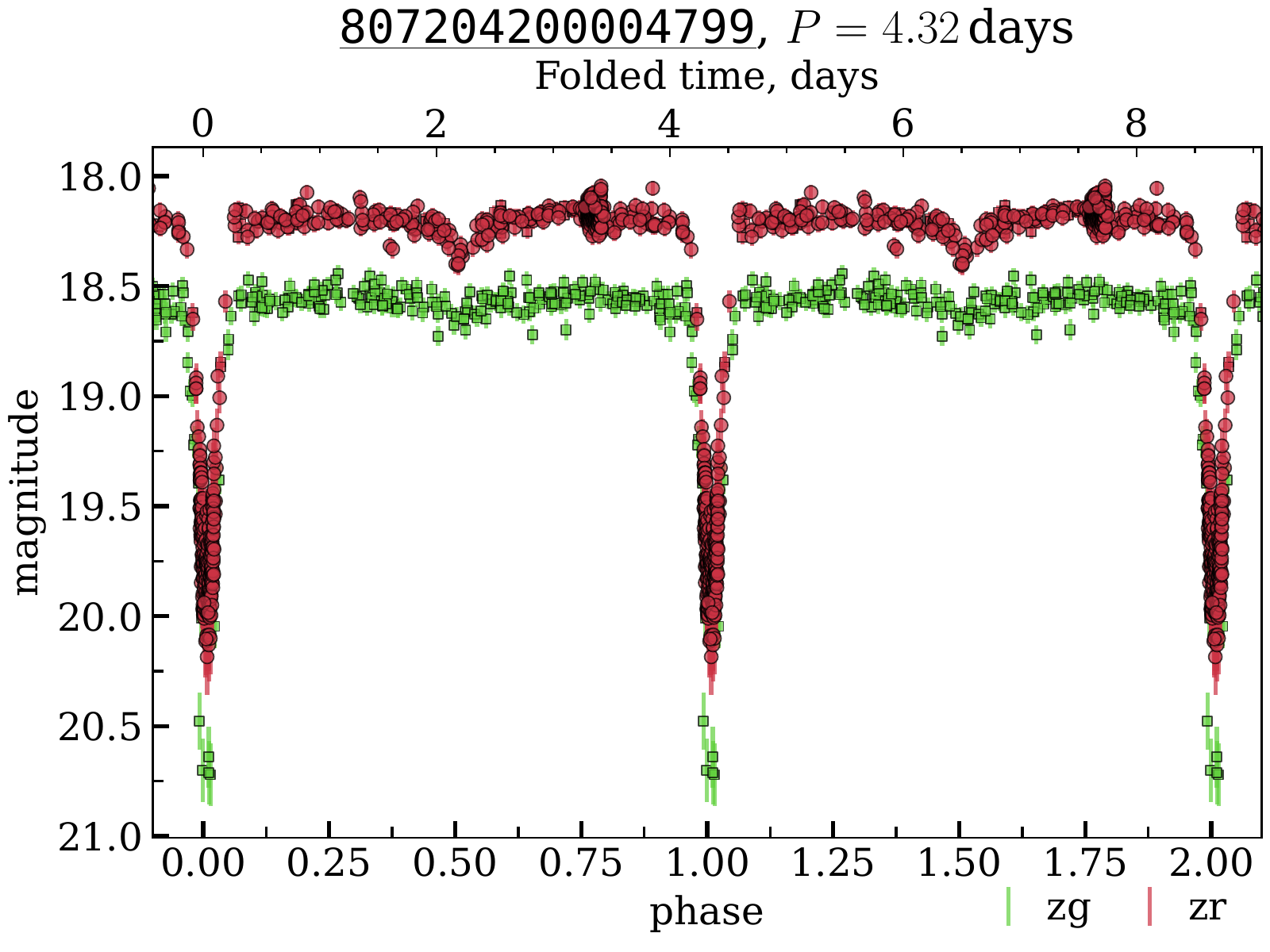}
\caption{The folded light curves of {\tt 807204200004799} from the \disk{} field classified by us as an eclipsing binary. OIDs: {\tt 807104200003543}, {\tt 807204200004799},  {\tt 1809109100009148}, {\tt 1809209100012225}, {\tt 1810112100013288},  {\tt 1810212100046961}.}
\label{fig:807204200004799}
\end{center}
\end{figure}

Among the unclassified objects we identified six new eclipsing binaries ({\tt 807203200013118}, {\tt 807204200004799}, {\tt 807204400014494}, {\tt 807208400036953}, {\tt 807211400009493}, {\tt 807216400013229}). As an example, Figure~\ref{fig:807204200004799} shows the folded light curve of the object {\tt 807204200004799} in $zr$, $zg$ passbands.

\subsubsection{Unclassified Variables}

\begin{figure}
\begin{center}
\includegraphics[width=\columnwidth]{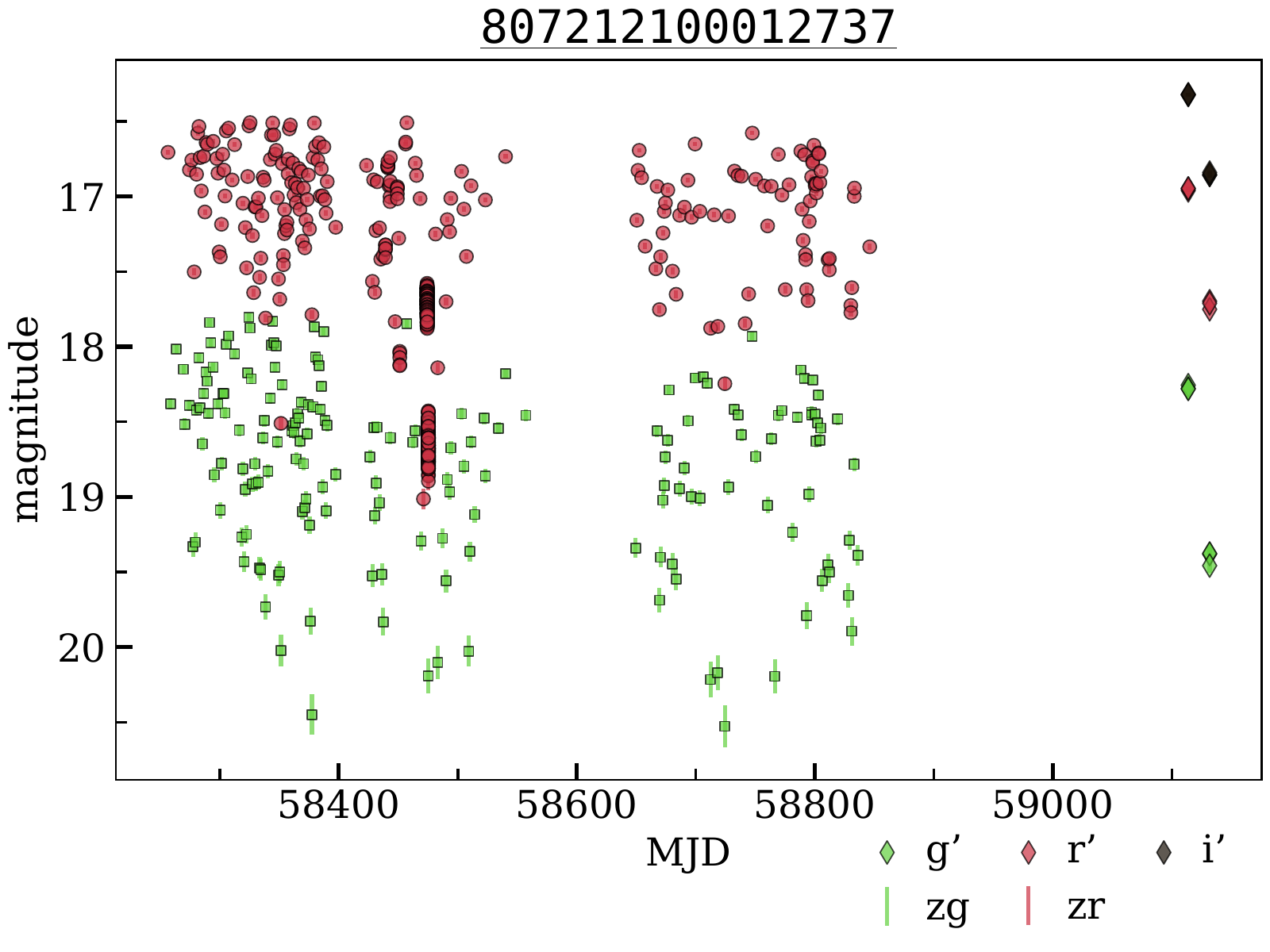}
\caption{The light curves of unclassified variable object from the \disk{} field. OIDs: {\tt 807112100008284}, {\tt 807212100012737}. Diamonds show the additional $g',r',i'$ observations performed with RC600 telescope of CMO SAI MSU.}
\label{fig:807212100012737}
\end{center}
\end{figure}

\begin{figure}
\begin{center}
\includegraphics[width=\columnwidth]{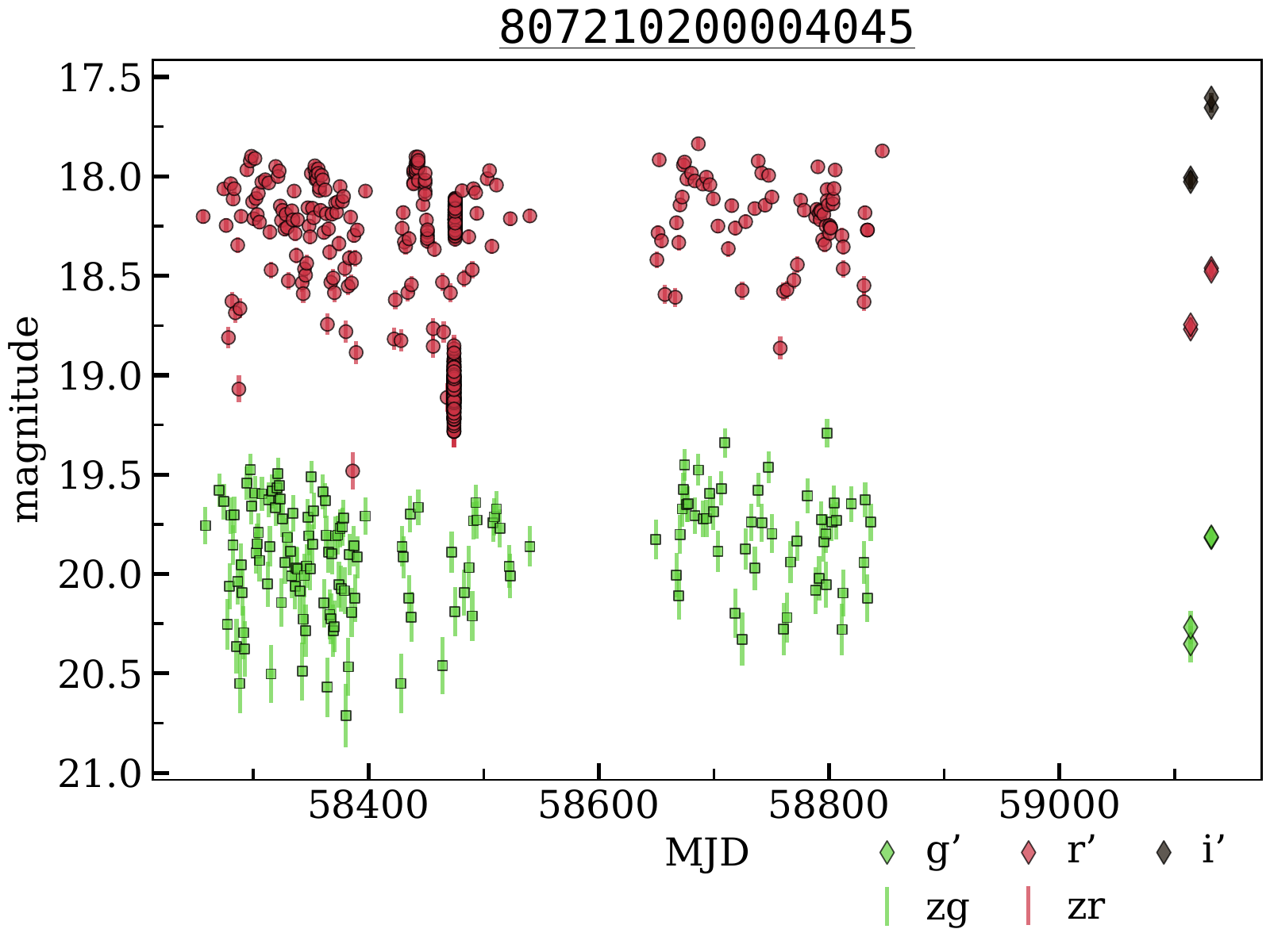}
\caption{The light curves of unclassified variable object from the \disk{} field. OIDs: {\tt 807110200019782}, {\tt 807210200004045}. Diamonds show the additional $g',r',i'$ observations performed with RC600 telescope of CMO SAI MSU.} 
\label{fig:807210200004045}
\end{center}
\end{figure}

Two other unclassified objects --- {\tt 807212100012737}, {\tt 807210200004045} --- show non-periodic variability with an amplitude $>1$~mag. Their light curves are shown in Figs.~\ref{fig:807212100012737} and \ref{fig:807210200004045}. We also obtained the observations with the RC600 telescope of CMO SAI MSU in $g',r',i'$ passbands (Table~\ref{tab:photometry_2}). The variability is still present. Both objects do not have a significant IR excess. The distance to  {\tt 807212100012737} is $\sim1000$\,pc. Based on this distance, brightness and colours we can make a dubious assumption that {\tt 807212100012737} is a red dwarf with strong spot activity. The distance to {\tt 807210200004045} is poorly defined; the {\em Gaia} parallax to error ratio is 1.14. Therefore, we do not make any assumption about its nature.

\begin{table}
\centering
\caption{Photometric observations of unclassified variable objects with RC600 telescope of the Caucasus Mountain Observatory, SAI MSU.}
\begin{tabular}{lccc}
\hline
MJD & Filter & Magnitude & Error \\
\hline
\multicolumn{4}{c}{{\tt 807210200004045}}\\
\hline
59113.82463	&	$g'$	&	20.352	&	0.091	\\
59113.82820	&	$g'$	&	20.268	&	0.084	\\
59131.79666	&	$g'$	&	19.818	&	0.048	\\
59131.80022	&	$g'$	&	19.813	&	0.049	\\
59113.83096	&	$r'$	&	18.769	&	0.039	\\
59113.83277	&	$r'$	&	18.746	&	0.038	\\
59131.80315	&	$r'$	&	18.462	&	0.027	\\
59131.80531	&	$r'$	&	18.478	&	0.028	\\
59113.83464	&	$i'$	&	18.007	&	0.034	\\
59113.83645	&	$i'$	&	18.027	&	0.035	\\
59131.80736	&	$i'$	&	17.654	&	0.027	\\
59131.80917	&	$i'$	&	17.605	&	0.026	\\
\hline
\multicolumn{4}{c}{{\tt 807212100012737}}\\
\hline
59113.79655	&	$g'$	&	18.256	&	0.026	\\
59113.79802	&	$g'$	&	18.278	&	0.026	\\
59113.79948	&	$g'$	&	18.281	&	0.026	\\
59131.77430	&	$g'$	&	19.380	&	0.058	\\
59131.77697	&	$g'$	&	19.376	&	0.043	\\
59131.77937	&	$g'$	&	19.457	&	0.045	\\
59113.80083	&	$r'$	&	16.960	&	0.012	\\
59113.80194	&	$r'$	&	16.948	&	0.012	\\
59113.80305	&	$r'$	&	16.946	&	0.012	\\
59131.78118	&	$r'$	&	17.697	&	0.020	\\
59131.78230	&	$r'$	&	17.751	&	0.022	\\
59131.78342	&	$r'$	&	17.711	&	0.021	\\
59113.80422	&	$i'$	&	16.325	&	0.011	\\
59113.80533	&	$i'$	&	16.321	&	0.011	\\
59113.80644	&	$i'$	&	16.320	&	0.011	\\
59131.78459	&	$i'$	&	16.857	&	0.017	\\
59131.78571	&	$i'$	&	16.856	&	0.017	\\
59131.78683	&	$i'$	&	16.841	&	0.016	\\
\hline
\end{tabular}
\label{tab:photometry_2}
\end{table}

\subsection{Bogus}
\label{sec:artefacts}

\begin{figure*}
\begin{center}
\includegraphics[width=2\columnwidth]{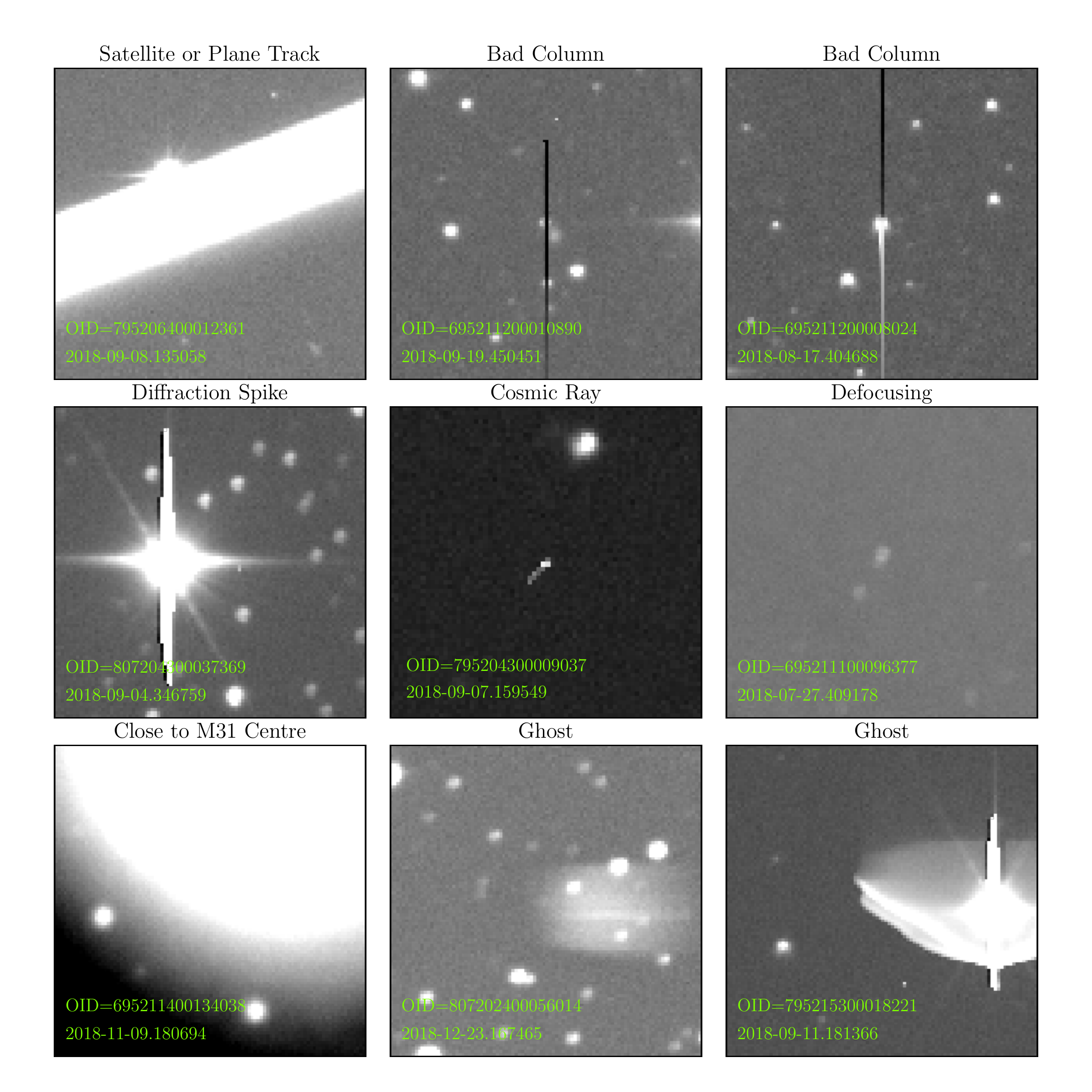}
\caption{Cases of bogus detections found among the outliers. The outlier is located in the image centre. The size of the cutout showing a cosmic ray is 70$\times$70 CCD pixels, for the others the image size is  100$\times$100 CCD pixels.}
\label{fig:art}
\end{center}
\end{figure*}

The ZTF data processing pipeline includes a procedure to separate the astrophysical events from bogus, i.~e. false positive detections \citep{2019PASP..131a8003M}. However, among the outliers we encountered a significant fraction of bogus light curves (80/101, 95/113, 13/63 for the \andro{}, \deep{}, \disk{} fields, respectively). A few examples are:
\begin{enumerate}
    \item a point sharply dropping up or down by several magnitudes --- possibly due to satellite/plane tracks, double star in one aperture due to defocusing, bad columns on CCDs, or cosmic particles;
    \item a random spread within several magnitudes due to ghosts, diffraction spikes, bright stars halos, cosmic rays or wrong background subtraction (close to M\,31 centre);
    \item a combination of (i) and (ii).
\end{enumerate}
Bogus examples with suggested classification are given in Fig.~\ref{fig:art}. We discuss below some interesting cases. 

\subsubsection{{\tt 795205400022890} and its echoes}

\begin{figure}
\begin{center}
\includegraphics[width=\columnwidth]{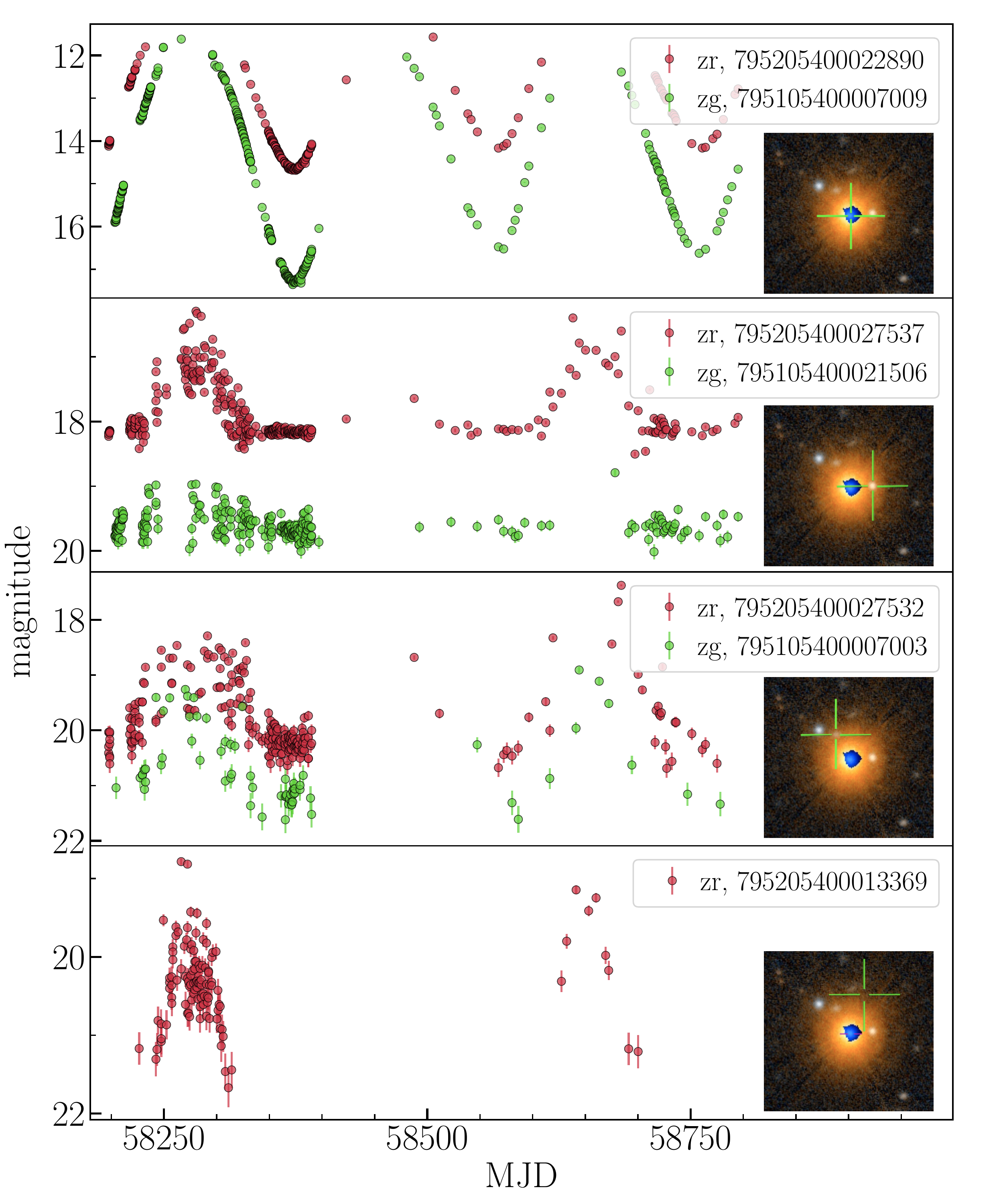}
\caption{IW~Dra and its echoes. SDSS image shows the position of IW~Dra and three ZTF sources showing the correlated variability.}
\label{fig:mira_art}
\end{center}
\end{figure}

In our outlier list there are four variables with the same phase and period. We found that one of them, known as IW Dra, is a typical Mira variable, while three others are located in $\sim20\arcsec$ region around it. It turned out that these three neighbours are artefacts of the automatic ZTF photometry arising from incorrect background subtraction of the halo of a bright variable star which overlaps with the light from the nearby objects. The light curves of IW Dra and its echoes are shown in Fig.~\ref{fig:mira_art}.

\subsubsection{{\tt 807203300039547} --- overlap of star and known asteroid}

One short transient was identified as a conjunction of asteroid 945 (Barcelona) and a  weak star in the \disk{} field (Fig.~\ref{fig:barcelona.pdf}). The identification of the asteroid was performed with SkyBot \citep{2006ASPC..351..367B}.

\begin{figure}
    \centering
    \animategraphics[autopause=false,autoplay=true,poster=last,width=\columnwidth,alttext=none,loop=false]{2}{images/barcelona/barcelona}{0}{8}
    \caption{Asteroid Barcelona overlapping object {\tt 807203300039547}. The blue crosses mark the location of the ZTF transient detection. An  animated version is available in the electronic version.}
    \label{fig:barcelona.pdf}
\end{figure}

\subsubsection{{\tt 695211200077906} --- double star artefact}

\begin{figure*}
    \centering
    \includegraphics[scale=0.45]{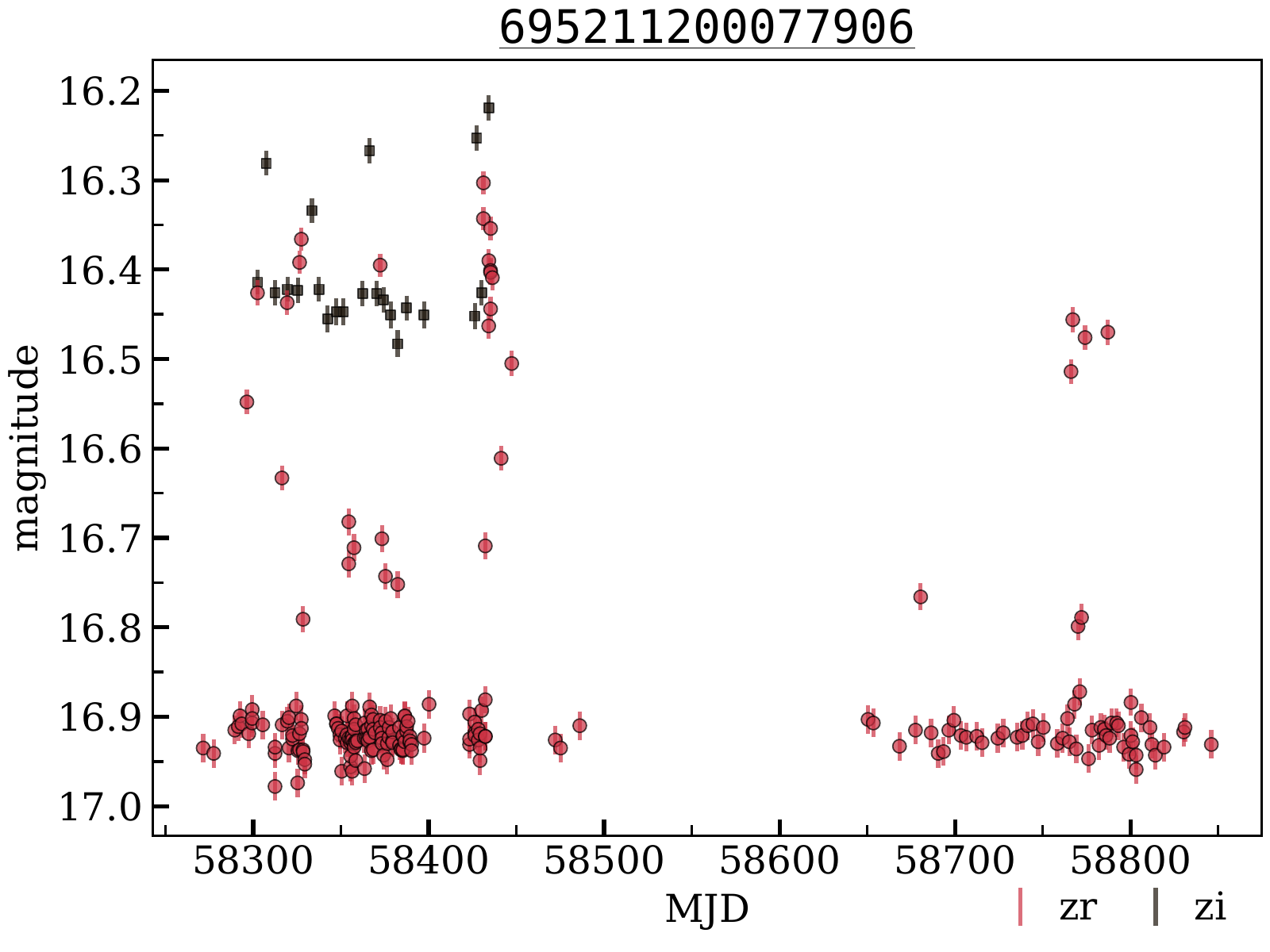}
    \includegraphics[scale=0.45]{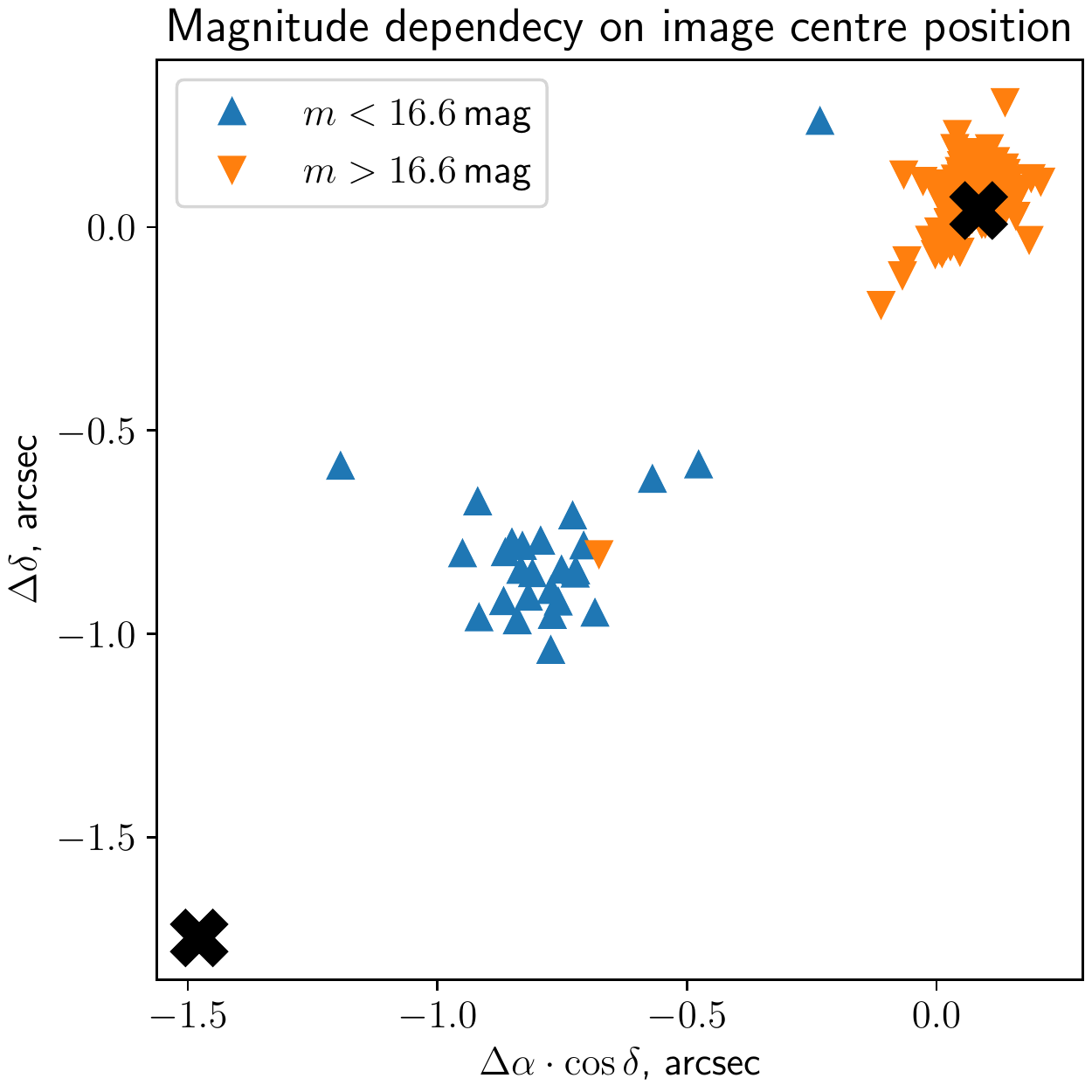}
    \caption{The light curves and the image centre positions of {\tt 695211200077906}.
    \textbf{Left}: light curves, OIDs: {\tt 695211200077906} ($zr$, red circles), {\tt 695311200089204} ($zi$, black squares).
    \textbf{Right}: spatial distribution of observations relative to the average object position, black crosses show Gaia~DR2 positions of the double star components.
    Note that the small brightness ($m > 16.6$\,mag) of the object corresponds to the position of one of the components, while the large brightness ($m < 16.6$\,mag) corresponds to the position somewhere between the components.
    We assume that the reason of such correlation between magnitude and position is in different FWHM of point spread function in different observations: for the first case FWHM is $\sim 2\arcsec$ and photometry counts signal from one component only, but for the other case FWHM is $\sim 4\arcsec$ and both components are being described by the common integrating area which makes measured magnitude closer to the integral magnitude of the double star and position to be near the double star centre.
    }
    \label{fig:695211200077906}
\end{figure*}

We found several objects that show bogus variability, and identified them as double stars with $\sim 2\arcsec$ separation. We assume that bogus measurements occur due defocusing when the FWHM becomes significantly larger than the separation between the components of the double star, which leads to overstatement of the object brightness.
The example of such a bogus light curve is {\tt 695211200077906}, see Fig.~\ref{fig:695211200077906}, where positions are given by the ZTF Lightcurve API\footnote{In this particular case we used the API because the individual observations coordinates are missed in the bulk-downloadable files. API description page is \url{https://irsa.ipac.caltech.edu/docs/program_interface/ztf_lightcurve_api.html}}.

\section{Pipeline validation}
\label{sec:justification}

In order to access the efficiency of our pipeline in identifying light curves with significantly different properties than those already present in the bulk of ZTF DR3 data, we created validation test-sets adding a few artificially generated light curves to the real data. This set of enclosed light curves is inspired by potentially interesting  astrophysical sources, as well as example cases of non-existing objects. The last case includes a perfectly plateau ``Flat Without Noise'' light curve, ``Flat With Noise'' which represents non-variable source, ``Point Up'', ``Point Down'', ``Step Up'', and ``Step Down'' mimicking a rapid magnitude change or bad column.
The astrophysically motivated fakes were built from
1) confirmed binary microlensing events Gaia16aye \citep[three different parts of its $r$ light curve,][]{2016ATel.9507....1W}, and candidates ZTF18ablruzq and ZTF18acskgwu \citep[$zr$ passband,][]{2020RNAAS...4...13M},
2) single microlensing events ZTF18aaztjyd and ZTF18abaqxrt \citep[$zr$ passband,][]{2020RNAAS...4...13M},
3) $r$ light curve of kilonova 170817 \citep{2017ApJ...851L..21V},
4) $V$ light curve of cepheid in eclipsing binary OGLE-LMC-CEP-0227 \citep{Soszynski2008}.

\begin{figure*}
\begin{center}
\includegraphics[width=0.95\linewidth]{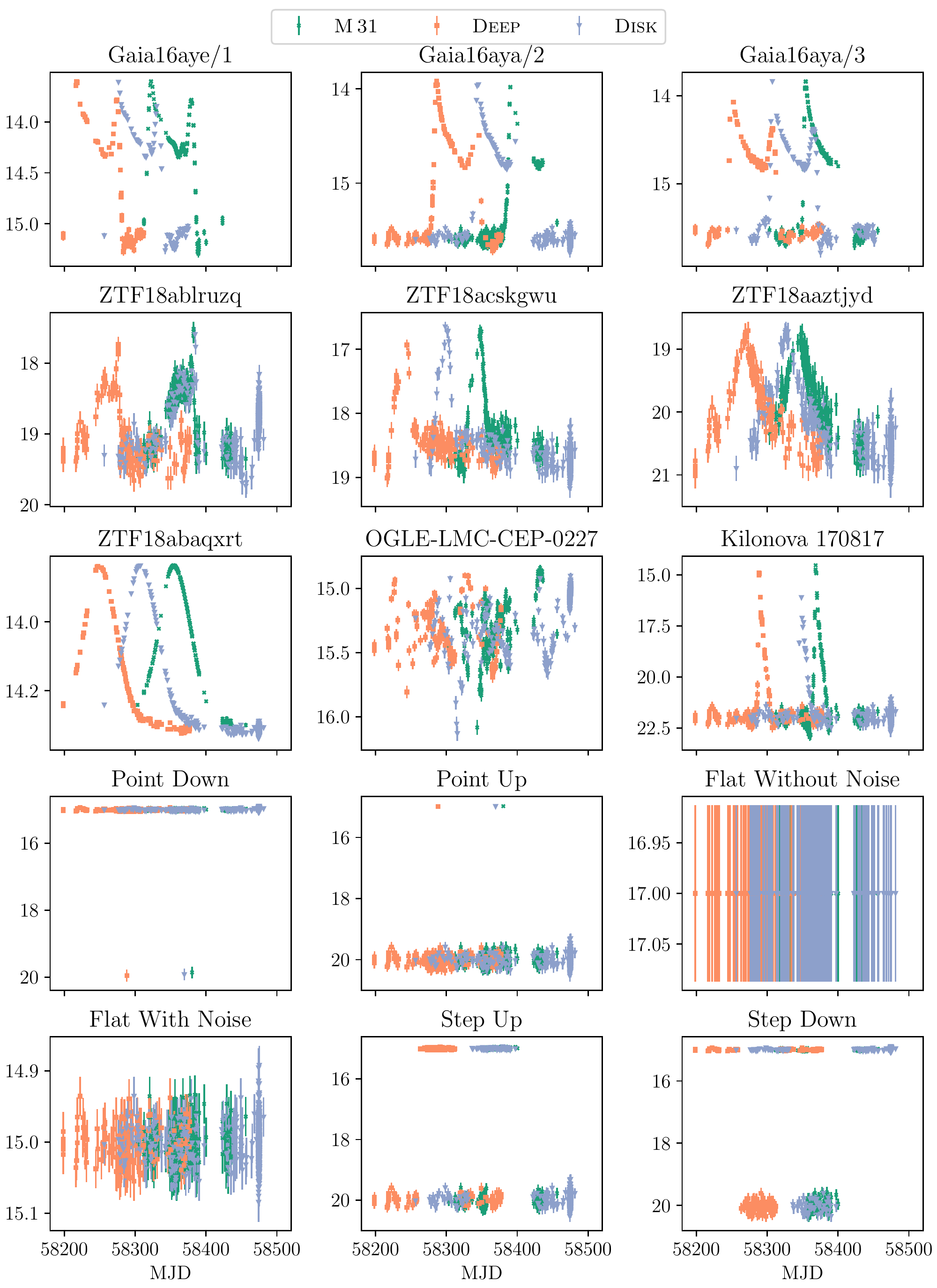}
\caption{Fake $zr$ light curves. Time-shift between different fields is caused by different initial observational date for each field. We show the light curves following cadences of \andro{} (green), \deep{} (orange) and \disk{} (blue) fields.}
\label{fig:fales_lc}
\end{center}
\end{figure*}

Original light curves were modified to conform with ZTF DR3 cadence and noise levels. We used a cubic smoothing spline fit for approximation of real objects and sampled the resulting function according to the cadence of real ZTF DR3 objects: \texttt{695211200035023} (for \andro{} field), \texttt{795216100016711} (for \deep{} field) and \texttt{807201300060502} (for \disk{} field).
The uncertainty of each observation was assigned using a linear relation between observed magnitude and its error, but with a lower bound of 0.001\,mag.
To derive this linear relation we took all $zr$ observations in the \disk{} field  with $\texttt{catflag} = 0$, averaged their magnitudes and uncertainties in intervals of 0.01\,mag, and fitted these binned uncertainties with a linear function of magnitude, resulting in:
\begin{equation*}
    \delta = \max(0.001; 0.0297\,m - 0.4179).
    \label{eq:delta}
\end{equation*}
Fit parameters for the other fields agree with this relation within 20\% margin. The final light curves for all fakes are shown in Fig.~\ref{fig:fales_lc}.

\begin{figure}
    \centering
    \includegraphics[width=0.49\textwidth]{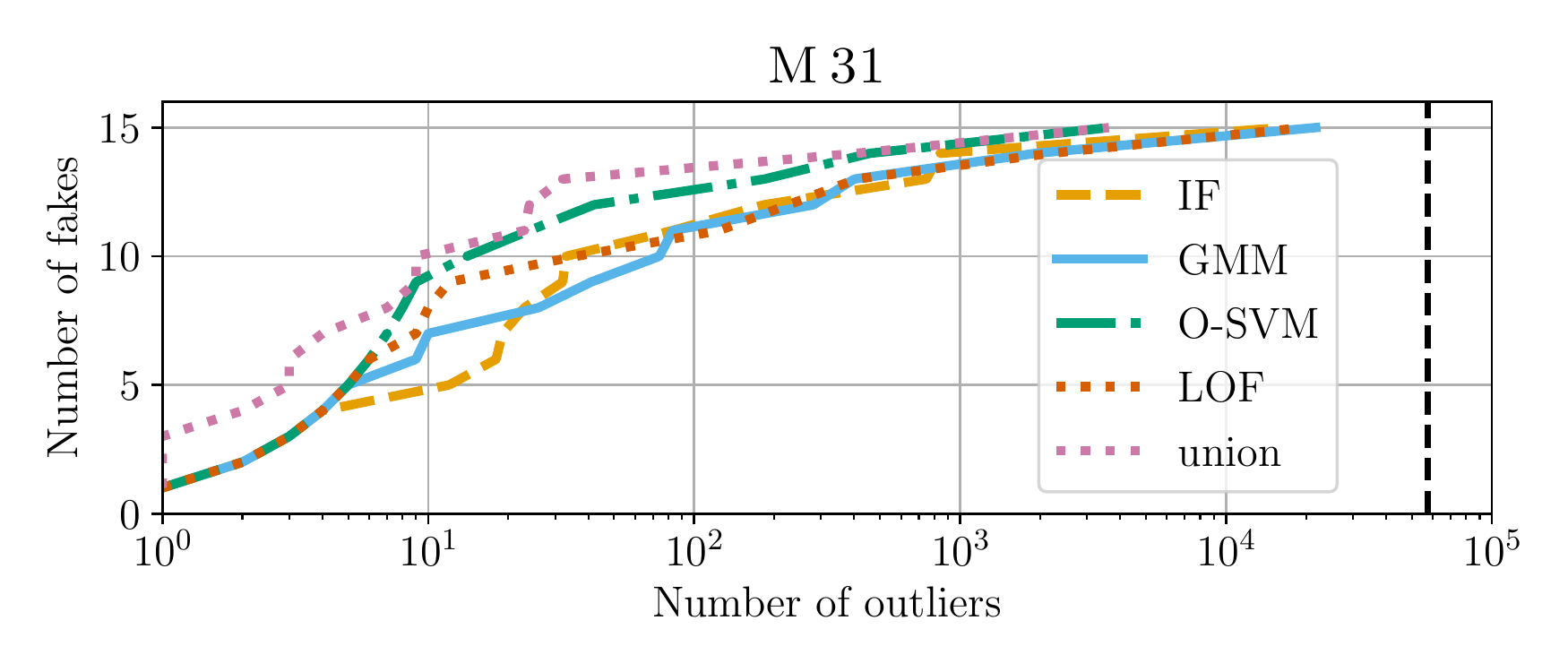}
    \includegraphics[width=0.49\textwidth]{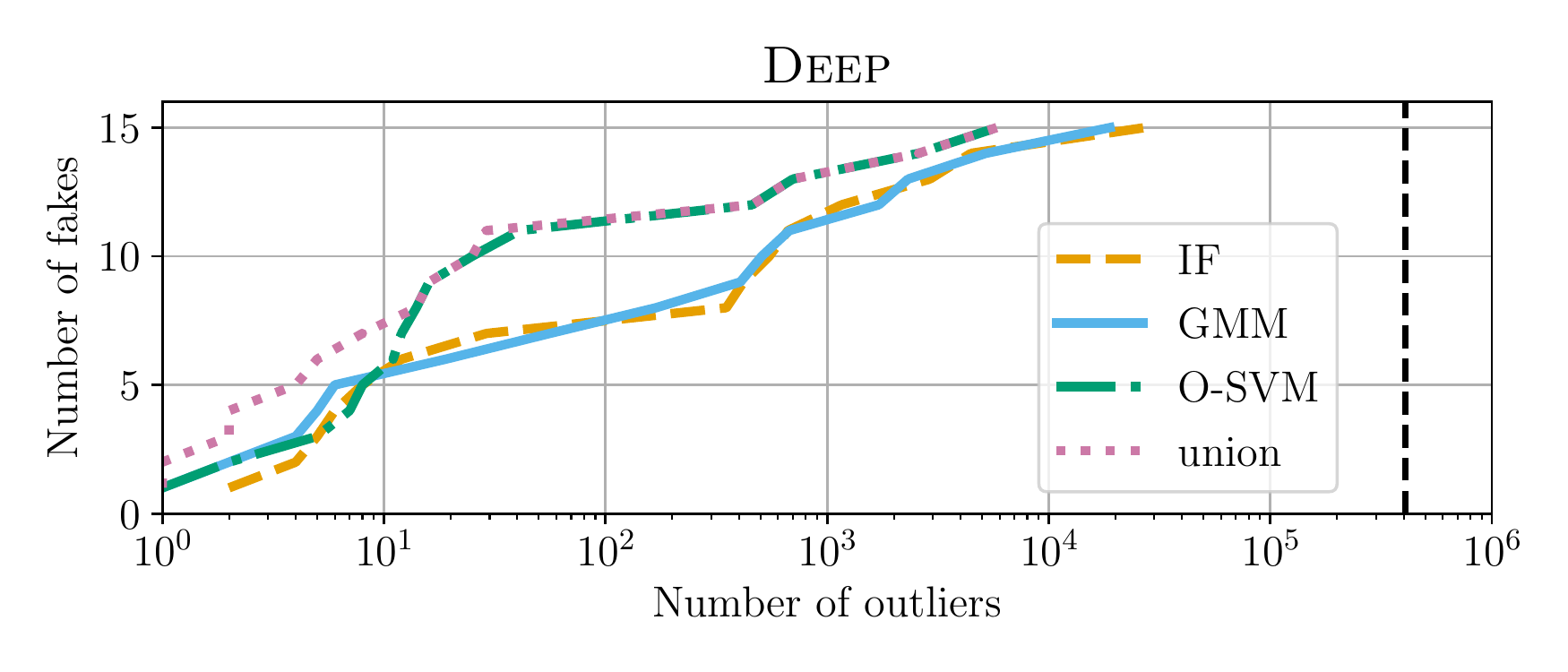}
    \includegraphics[width=0.49\textwidth]{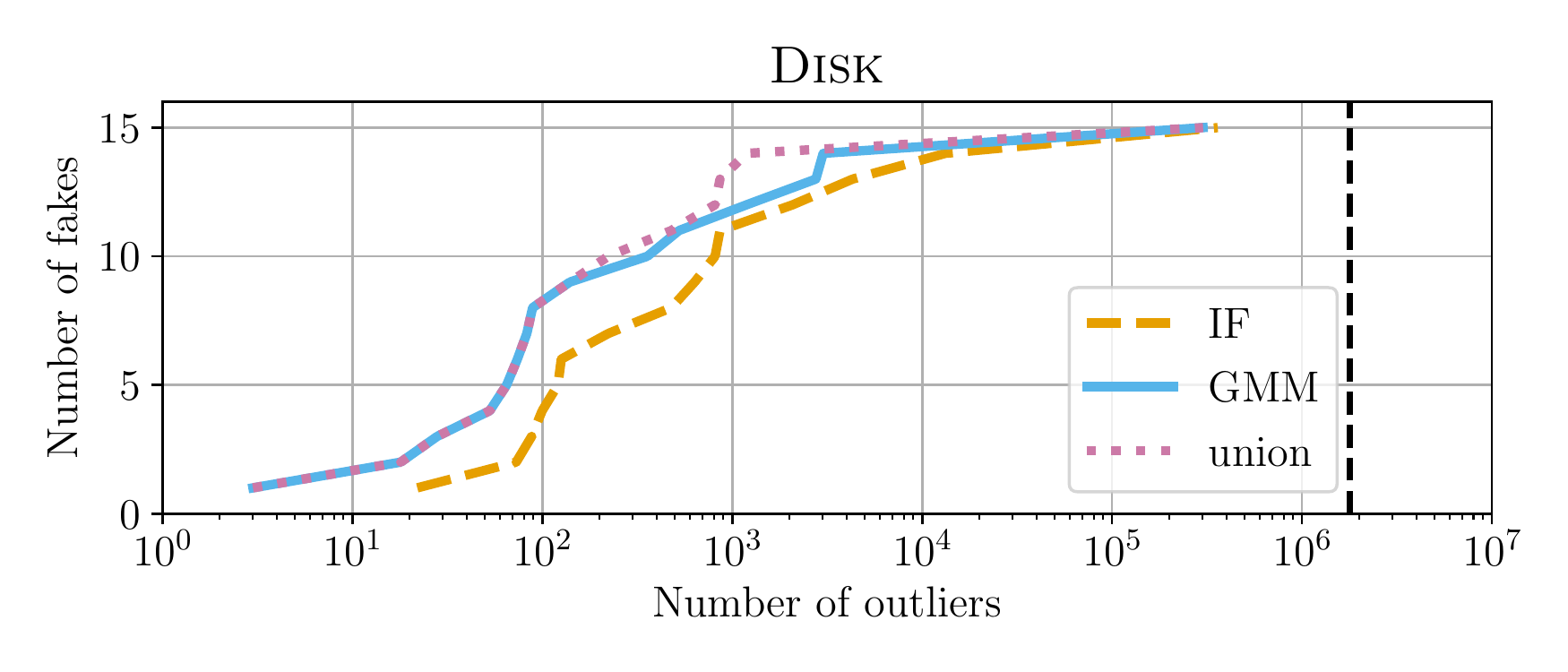}
    \caption{Fake detection rate. ``Union'' shows the total performance of all algorithms being run in parallel. Vertical dashed line shows the total number of light curves.}
    \label{fig:fake-detection}
\end{figure}

We built three validation data sets adding fifteen fake anomalous light curves to each of them and applied the same outlier detection algorithms as used for the ZTF data.
The resulting detection rates are shown in Fig.~\ref{fig:fake-detection}, where ``union'' denotes the total rate of all used algorithms run in parallel. \andro{} field is the smallest one, so the percent of outliers the expert can examine is the largest. This is why the very first 32 outliers contain 13 of 15 fake anomalies, while OGLE-LMC-CEP-0227 is the 408$^{\rm th}$ and ``Flat With Noise'' the 3630$^{\rm th}$ outlier. We use only three algorithms (IF, GMM \& O-SVM) for \deep{} field, but the fake detection rate is still very high, probably because its light curves are less diverse and the fake anomalies are more different. The first 30 outliers contain 11 fakes, while ZTF18ablruzq and ZTF18acskgwu are found within  the first 700 outliers, OGLE-LMC-CEP-0227 and ``Flat With Noise'' are within the first 6000 objects. The \disk{} field contains $\sim 1.8$ million objects and we used IF \& GMM only, but the fake performance is quite acceptable: first 8 fakes are found within the first 90 outliers, 14 of 15 fakes are found within the first 1200 outliers, and only ``Flat With Noise'' was not detected until the 400 000$^{\rm th}$ outlier. These results show that the outlier detection pipeline shows good anomaly detection rate and could detect physically interesting events. The ``Flat With Noise'' object is the hardest fake light curve to detect by all algorithms in all fields. This is possibly due to its location in a dense region of feature space which is also populated with low-variable sources and variable sources below ZTF’s threshold of detectability, or even stochastic variance stemming from a small subset of objects marked as variable from within a large distribution of non-variable objects.

%%%%%%%%%%%%%%%%%%%%%

\section{Exploratory data analysis}
\label{subsec:phase-space}

Once the experts finished their investigation we can now use the accumulated knowledge to identify simple strategies which would allow us to quickly separate interesting anomaly candidates from bogus light curves. We start with an exploratory data analysis (EDA) of our feature space. Figure~\ref{fig:all_features_histogram} shows histograms for all features and fields. In what follows we give an example EDA performed over the incremented parameter space, which includes features as well as the expert judgement. 

\begin{figure}
\begin{center}
\includegraphics[width=\columnwidth]{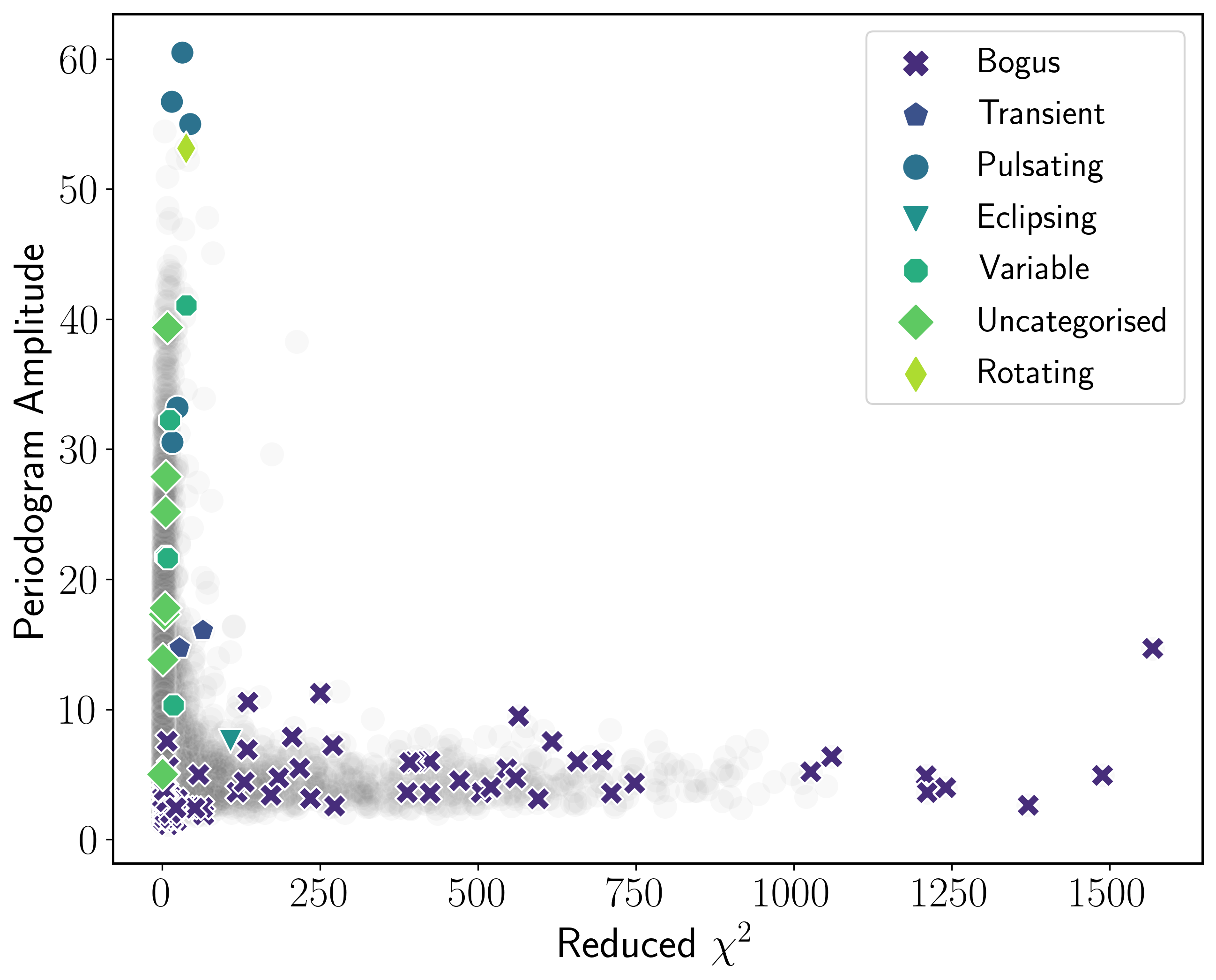}
\caption{A scatter plot of the 101 \andro{} expert-tagged outliers in a Periodogram Amplitude---Reduced~$\chi^2$ phase-space. The bogus objects can be well separated from anomaly candidates. All 57\,546 \andro{} objects are shown as grey circles.}
\label{fig:chi2_periodamp_phase}
\end{center}
\end{figure}

Figure~\ref{fig:artefacts_vs_anom_hist} shows the separate feature distribution for bogus objects and anomaly candidates in the \andro{} field. From this, it is evident that Periodogram Amplitude and Reduced~$\chi^2$ possess complementary dissimilar distributions between the two classes. Analysing the relationship between these two features  (Fig.~\ref{fig:chi2_periodamp_phase}) we see that it is possible to identify bogus light curves as those presenting low Periodogram Amplitude (frequently caused by chaotic signals) and high Reduced~$\chi^2$ values (when one or more images of some real source is overlapping with another source, such as a bad column or a glare, which mimics a constant light curve with bright outbursts). 

We emphasise that the expert-tagging of objects was performed prior to the EDA, and the overall near-orthogonal distribution of objects in the \andro{} field did not influence the expert's analysis. \par
Although we do not expect a 2-dimensional parameter space to completely enclose the complexity of a real/bogus separation pipeline, it is reasonable to assume that, for each data set, some low-dimensional representation will enable quick identification of these 2 classes --- even if we do not expect the final classification to be perfect. This is a direct  example of how we can harvest expert knowledge to optimise further automatic identifications. Similar approaches may be used to construct powerful alert stream filters in the context of real-time alert brokers like ANTARES \citep{Narayan18,2020arXiv201112385M}, ALeRCE \citep{Forster2020}, and Fink \citep{Moller2020}. Further exploration of feature phase-space anomaly detection is currently a main focus in Aleo \textit{et al.} (in prep.). \par

%%%%%%%%%%%%%%%%%%%%

\subsection{Principal Components}
\label{subsec:pca}

Principal Component Analysis \citep[PCA,][]{jolliffe2013principal} is a  dimensionality-reduction technique whereby multidimensional data are decomposed into an  orthogonal basis following the directions of maximum data variance.  We implemented the PCA module using the \texttt{sklearn} Python package \citep{scikit-learn}, and normalised the feature values across all objects using the \texttt{StandardScaler} module. Subsequently, we calculated 42 Principal Components (PCs), ordered by those which explain the most variance. For the \andro{} field, the first two PCs together explain 47.1\% of the variance, with the first 15 explaining 90.4\%. Likewise, for the \deep{} and \disk{} fields the first two PCs together explain 49.7\%, and 50.2\% of the variance, respectively.

\begin{figure*}
 \begin{minipage}[c]{1.0\linewidth}
  \centering
  \begin{center}
  \includegraphics[width=0.75\textwidth]{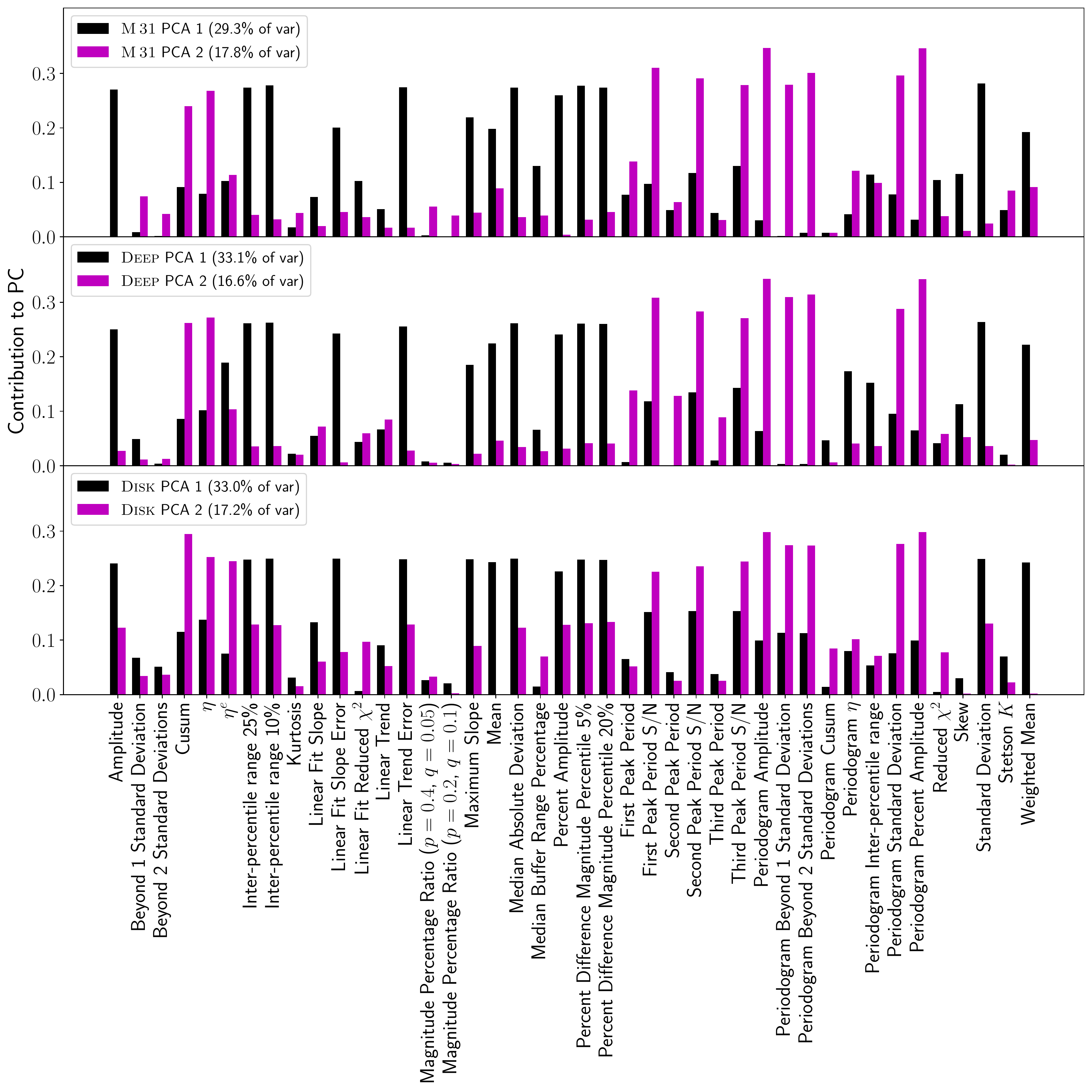}
  \caption{The first two principal components with ascribed feature loadings plotted for the \andro{} field (top panel, 47.1\% of variance), \deep{} field (middle panel, 49.7\% of variance), and \disk{} field (bottom panel, 50.2\% of variance). Notice that one principal component favours magnitude-based features (black), and the other favours periodogram and temporally-based features (magenta).}
  \label{fig:pca_barcharts}
  \end{center}
  \end{minipage}
\end{figure*}
The contribution to the first 2 PCs (loadings) from each of the 42 features are shown in Figure~\ref{fig:pca_barcharts}. For PC 1, shown in black, Standard Deviation, Inter Percentile Range 10\%, and Percent Difference Magnitude Percentile 5\% are the highest contributors. For PC 2, shown in magenta, Periodogram Amplitude, Periodogram Percent Amplitude, and Periodogram Beyond 2 Standard Deviations are those which contribute the most. From this, we note that PC 1 favours magnitude-based differences, and PC 2 favours periodogram-based features across all three fields. This is expected, since the two categories of information are complementary. 

This complementarity is also evident in the 
correlation matrix of normalised feature values (Fig.~\ref{fig:features_corr}). We see clearly two separate sets of features highly correlated between themselves and weakly correlated with members from the other group. This may be indicative that some features are degenerate. In fact, in Figure~\ref{fig:all_features_histogram}, those features with similar correlation vectors share similar distributions, albeit different values. However, since no singular feature had a consistently negligible contribution across all PCs and given the fluctuation in contribution by each features to successive PCs (\ref{fig:pca_barcharts}), we decided to keep all 42 features as input in our anomaly detection algorithm pipeline.  \par

\begin{figure*}
\begin{center}
\includegraphics[width=\textwidth]{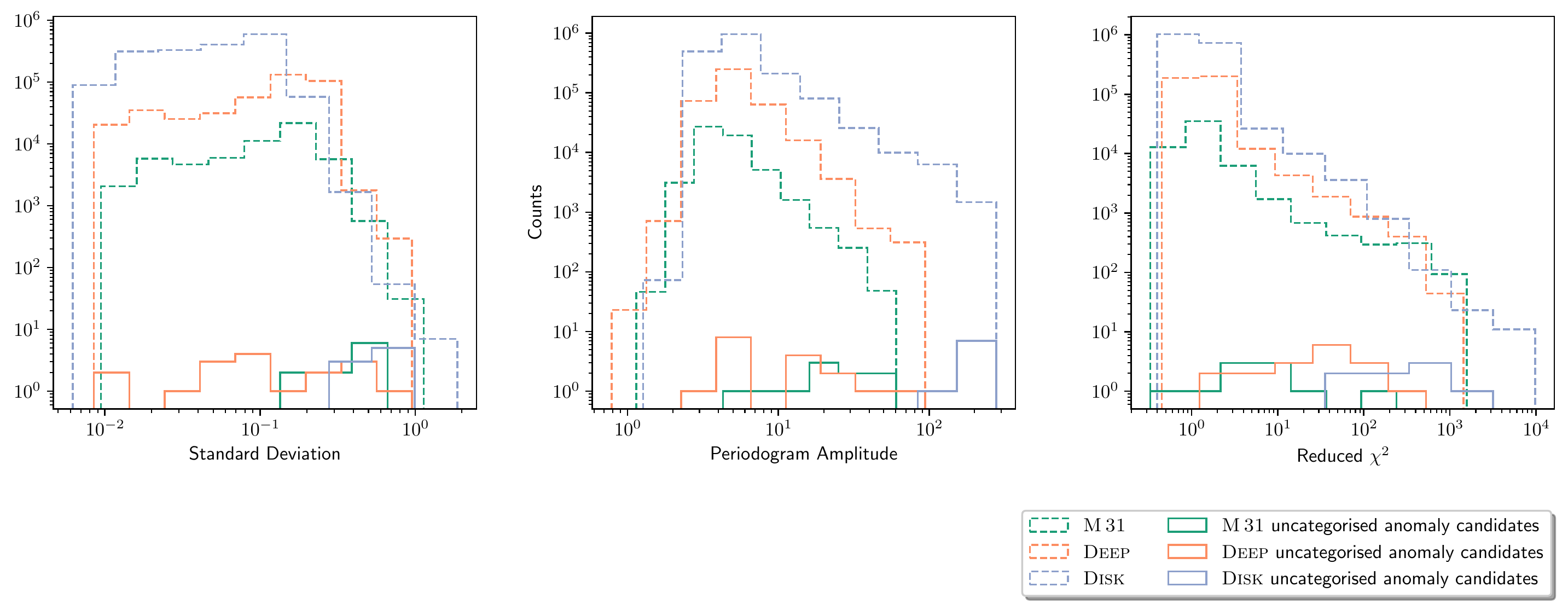}
\caption{Distribution of features with largest contribution to the first 3 PCs for objects in \andro{} (dashed green), \deep{} (dashed orange), and \disk{} (dashed blue) fields, as well as expert-identified uncategorised anomaly candidates (solid contours).}
\label{fig:three_features_hist_uncat_anom}
\end{center}
\end{figure*}

We show in Figure~\ref{fig:three_features_hist_uncat_anom} distributions for the features with largest contribution to the first \emph{three} PCs (PC 1---Standard Deviation, PC 2---Periodogram Amplitude, PC 3---Reduced~$\chi^2$) across all fields considering only objects identified as uncategorised anomaly candidates. All uncategorised anomaly candidates across the three fields are contained within, but do not constitute the edges of, their parent field distribution. That is, they are not in the tail of the distribution where the feature value is greatest, and are sometimes dwarfed in number by other objects in their field which make up their parent distribution. This could explain why such anomaly candidates were not yet categorised. These interesting, uncategorised candidates are high priority targets for future follow-up. See Section~\ref{sec:results} and Table~\ref{tab:disk_deep_m31} for their listings. \par

\section{Conclusions}
\label{sec:conclusions}

Despite all the expected scientific results which inspired the construction of modern astronomical observatories, there is little doubt about the potential of the resulting data sets for new discoveries. In the era of systematic large scale sky surveys, a telescope which only fulfils its science goals cannot be considered successful \citep{norris2017}. In this context, the use of automatic machine learning tools is unavoidable. They provide important insights into the statistical properties and limits of the data set at hand, and can be optimised to be good recommendation systems. Nevertheless, the discovery itself will always be a profoundly human experience. In this context, the identification of scientifically interesting sources is a product of the combination of data-driven machine learning and human-acquired domain knowledge.  The SNAD anomaly detection pipeline, presented in this work,  and its accompany results, are a concrete example of how powerful such a system can be when applied to a rich data set as the ZTF DR3. 

The pipeline consists of 3 separate stages: feature extraction, automatic outlier detection based on machine learning and anomaly confirmation by human experts. The infrastructure put in place to enable such detailed analysis includes not only the software pipeline\footnote{\url{https://github.com/snad-space/zwad}} for feature extraction and outlier detection, but also a  web-viewer\footnote{\url{https://ztf.snad.space/}} allowing cross-match with multiple catalogues and external data source --- specifically designed to help the expert in forming a global view of specific candidates. We apply the complete anomaly detection pipeline  to three ZTF fields, corresponding to more than 2.25 million objects. Four outlier detection algorithms (Isolation Forest, Gaussian Mixture Model, One-Class Support Vector Machine and Local Outlier Factor) were used to identify the top 277  outliers. From these, the expert analysis concluded that 188 (68\%) were bogus light curves and 89 (32\%) real astrophysical sources -- 66 of them being previously reported sources and 23 corresponding to non-catalogued objects. 

For a few of the most interesting anomaly candidates, the expert analysis included gathering additional observational data. This allowed us to spectroscopically classify one anomaly candidate as a RS Canum Venaticorum star --- a close detached binary whose variability outside the eclipse is due to stellar spots. Such objects are rare in ZTF DR3, therefore, we consider it an anomaly. Other found anomalies that belong to the rare types of objects are polar, supernovae and red dwarf flare. The interesting anomalies we found also include a few unclassified variable stars with controversial reports in the literature, among them the objects for which we performed the photometric observations. Nevertheless, additional studies are still required. We also found the sources that behave unusually for their suspected astrophysical type, e.~g. the burst frequency of  AT 2017ixs --- possible nova in M\,31. Even among the bogus light curves we were able to identify interesting events, like the IW Dra star and its three companion echoes or the interaction of an asteroid with a background star. 

Given the high incidence of bogus light curves, we searched for a simple rule which could allow us to easy estimate the likelihood of a given object being bogus. A quick exploratory data analysis,  over the data accumulated by the expert, showed that bogus light curves are more likely to have a high reduced~$\chi^2$ and low period amplitude (at least for data in the \andro{} field). At this point, this result should be seen only as an indication that a careful exploitation of the parameter space by the expert can potentially lead to useful relations. If confirmed, such a relation could be used in the construction of alert stream filters for future telescopes. 

Despite the encouraging results reported in this work, the incidence of bogus among the automatic identified outliers is still relatively high (68\%). In order to use similar techniques in the context of future surveys like the Vera Rubin Observatory Legacy Survey of Space and Time or the Nancy G. Roman Space Telescope\footnote{\url{https://www.stsci.edu/roman}}, it is necessary to reinforce and optimise the interaction between the domain expert and the machine learning algorithm. We are currently working on adaptive strategies which are proven to perform well in such scenarios \citep{Ishida2019} and intend to add similar capabilities to the SNAD pipeline. Nevertheless, this experiment confirms the potential of machine learning  techniques in pointing non-standard elements within a large data set. The list of anomaly candidates provided here may be of use for scientists working in a broad range of domains and consists of a concrete example of the crucial role to be played by automatic pipelines in the future of astronomical discovery.

\section*{Acknowledgements}
We are grateful to L.~N.~Berdnikov and S.~V.~Antipin for the helpful discussion on the physical nature of anomaly candidates.

M.V.K., V.S.K., K.L.M., A.A.V. and M.V.P. are supported by RFBR grant according to the research project 20-02-00779 for preparing ZTF data, implementation the anomaly detection algorithm, and analysis of outliers. We used the equipment funded by the Lomonosov Moscow State University Program of Development. The authors acknowledge the support by the Interdisciplinary Scientific and Educational School of Moscow University “Fundamental and Applied Space Research”. P.D.A. is supported by the Center for Astrophysical Surveys (CAPS) at the National Center for Supercomputing Applications (NCSA) as an Illinois Survey Science Graduate Fellow. V.V.K. is supported by the Ministry of science and higher education of Russian Federation, topic № FEUZ-2020-0038. E.~E.~O.~Ishida and S.~Sreejith acknowledge support from CNRS 2017 MOMENTUM grant under project \textit{Active Learning for Large Scale Sky Surveys}.
Observations with TDS and data reduction (S.G.Z., A.V.D.) are supported by
the Ministry of science and higher education of Russian Federation
under the contract 075-15-2020-778 in the framework of the Large
scientific projects program within the national project ``Science''. 

This research has made use of the International Variable Star Index (VSX) database, operated at AAVSO, Cambridge, Massachusetts, USA. This research has made use of data and/or software provided by the High Energy Astrophysics Science Archive Research Center (HEASARC), which is a service of the Astrophysics Science Division at NASA/GSFC. This research has made use of ``Aladin sky atlas'' developed at CDS, Strasbourg Observatory, France. This research has made use of the SIMBAD data base, operated at CDS, Strasbourg, France. We acknowledge the usage of the HyperLeda database\footnote{\url{http://leda.univ-lyon1.fr}}. This research has made use of NASA’s Astrophysics Data System Bibliographic Services and following Python software packages: {\sc NumPy}~\citep{numpy}, {\sc Matplotlib}~\citep{matplotlib}, {\sc SciPy}~\citep{scipy}, {\sc pandas}~\citep{reback2020pandas,mckinney-proc-scipy-2010}, {\sc scikit-learn}~\citep{scikit-learn}, {\sc astropy}~\citep{astropy:2013,astropy:2018}, and {\sc astroquery}~\citep{2019AJ....157...98G}.

\section*{Data availability}
The ZTF light-curve data underlying this article are available in NASA/IPAC Infrared Science Archive\footnote{\url{https://irsa.ipac.caltech.edu/}}.
Light-curve feature set is available in Zenodo, at \url{https://doi.org/10.5281/zenodo.4318700}.
Other data underlying this article are available in the source code GitHub repository at \url{https://github.com/snad-space/zwad}.

\appendix
\section{Features}\label{sec:appendix-features}

We extracted 42 features from every light curve.
For this purpose the library on Rust programming language was developed\footnote{\url{https://docs.rs/light-curve-feature/0.1.18/}}.

\subsection{Auxiliary Definitions}
\paragraph{Light Curve}
We define light curve as a list of triples of magnitude $m_i$, its error $\delta_i$ and observation time $t_i$, where $i = 0..N-1$ and $N$ is the number of observations.

\paragraph{Mean}\label{sec:def-mean}
We define mean of the sample $\{x_i\}$ as
\begin{equation}\label{eq:mean}
    \xmean \equiv \frac1{N} \sum_i x_i.
\end{equation}

\paragraph{Weighted Mean}\label{sec:def-weighed-mean}
We define weighted mean of the sample $\{x_i\}$ with corresponding observation errors $\{\Delta_i\}$ as
\begin{equation}
    \bar{x} \equiv \frac{\sum_i x_i / \Delta_i^2}{\sum_i 1 / \Delta_i^2}.
\end{equation}

\paragraph{Standard Deviation}\label{sec:def-standard-deviation}
We define standard deviation of the sample $\{x_i\}$ as
\begin{equation}
    \sigma_x \equiv \frac{\sum_i \left(x_i - \mean\right)^2}{N - 1}.
\end{equation}

\paragraph{Indicator Function}
The indicator function of a set $A$ $I_{A}(x)$ is a function that equals unity when $x \in A$ and equals zero when $x \notin A$.
For instance, $I_{x > 0}$ equals unity for positive numbers and zero for non-positive.

\paragraph{Distribution Quantile}
We define $Q_x(p)$ to be the R-5\footnote{See Type-5 at \url{https://stat.ethz.ch/R-manual/R-devel/library/stats/html/quantile.html}} $p$th quantile ($100\cdot p$ percentile) of $\{x_i\}$.

\paragraph{Median}
We define $\median(x) \equiv Q_x(0.5)$.

\paragraph{Periodogram}
We use Lomb--Scargle~\citep{Lomb1976,Scargle1982} periodogram as an estimation of spectral power of the light curve.
Our periodogram implementation is based on "fast" algorithm~\citep{Press_etal1992}, an estimation of Lomb--Scargle periodogram which can be evaluated in $O(N_\omega \log N_\omega)$, where $N_\omega$ is a number of considered angular frequencies.
This method requires interpolation of light curve to an evenly separated time grid, we used linear interpolation for this purpose~\citep[\texttt{MACC} $=1$ in \texttt{spread} function,][]{Press_etal1992}.
We found periodogram values $P_k = P(\omega_k)$ on an evenly angular frequency grid $\{\omega_k\}$:
\begin{equation}
\begin{gathered}
    \omega_k = k \Delta\omega, ~~k=1..N_\omega,\\
    \Delta\omega = \frac{2\piup (N - 1)}{N R (t_{N-1} - t_0)},\\
    N_\omega = 2^{ \lceil \log_2 \frac{M \pi}{\delta t \Delta\omega} \rceil},
\end{gathered}
\end{equation}
where $\lceil \cdot \rceil$ denotes the ceiling function, $\delta t$ is a typical time interval between observations, in our case it equals $\median(\Delta t_i)$, median time interval between consequential observations $\Delta t_i \equiv t_{i+1} - t_i$. $R$ and $M$ are coefficients which expand the grid to lower and higher frequencies respectively.
We used $R = 10$ and $M = 2$ in our analysis.

\subsection{Feature Description}

We follow feature names from the original papers when possible.

\paragraph{Amplitude}\label{sec:amplitude}
The half amplitude of the light curve:
\begin{equation}\label{eq.amplitude}
    \frac{\max(m) - \min(m)}{2}.
\end{equation}

\paragraph{Beyond $n$ Std}\label{sec:beyond-n-std}
\cite{DIsanto_etal2016}
The fraction of observations beyond $\mean \pm n\,\sigma_m$:
\begin{equation}
    \frac{\sum_i I_{|m - \mean| > n\sigma_m}(m_i)}{N}.
\end{equation}
We use $n = 1$ and $n = 2$ in our analysis.

\paragraph{Cusum}\label{sec:cusum}
\cite{Kim_etal2014}
The range of magnitude cumulative sums:
\begin{equation}\label{eq:cusum}
    \max_j(S_j) - \min_j(S_j),
\end{equation}
where
\begin{equation}
    S_j \equiv \frac1{N\,\sigma_m} \sum_{i=0}^j{\left(m_i - \mean\right)}.
\end{equation}

\paragraph{Von Neummann $\eta$}\label{sec:eta}
\cite{Kim_etal2014}
\begin{equation}\label{eq:eta}
    \eta \equiv \frac1{(N - 1)\,\sigma_m^2} \sum_{i=0}^{N-2}(m_{i+1} - m_i)^2.
\end{equation}

\paragraph{$\eta^e$}\label{}
\cite{Kim_etal2014}
Generalisation of $\eta$~\ref{sec:eta} for unevenly time series.
\begin{equation}\label{eq:eta_e}
    \eta^e \equiv (t_{N-1} - t_0)^2 \frac{\sum_{i=0}^{N-2} \left(\frac{m_{i+1} - m_i}{t_{i+1} - t_i}\right)^2}{(N - 1)\,\sigma_m^2}.
\end{equation}
Our version of this feature differs from \cite{Kim_etal2014}.

\paragraph{Inter-percentile range}\label{sec:inter-percentile-range}
\begin{equation}
    Q_m(1 - p) - Q_m(p).
\end{equation}
We use $p = 0.1$ and $p = 0.25$ in our analysis.

\paragraph{Kurtosis}
The kurtosis of the magnitude distribution
\begin{equation}
    G_2 \equiv \frac{N\,(N + 1)}{(N - 1)(N - 2)(N - 3)} \frac{\sum_i(m_i - \mean)^4}{\sigma_m^4} - \frac{3(N+1)^2}{(N-2)(N-3)}.
\end{equation}

\paragraph{Linear Trend}
The slope of the light curve and its standard deviation.
Least squares fit of the linear stochastic model with unknown constant Gaussian noise magnitude $\Sigma$ assuming observation errors to be zero:
\begin{equation}
    m_i = c + \mathrm{slope}\,t_i + \Sigma \varepsilon_i,
\end{equation}
where $c$ and $\Sigma$ are constants, $\{\varepsilon_i\}$ are standard distributed random variables.

\paragraph{Linear Fit}
The slope, its standard deviation and the reduced $\chi^2$ of the light curve linear fit.
Least squares fit of the linear stochastic model with Gaussian noise described by observation errors $\{\delta_i\}$:
\begin{equation}
    m_i = c + \mathrm{slope}\,t_i + \delta_i \varepsilon_i,
\end{equation}
where $c$ is a constant and $\{\varepsilon_i\}$ are standard distributed random variables.

\paragraph{Magnitude Percentage Ratio}
\cite{DIsanto_etal2016}
The ratio of $p$th and $q$th inter-percentile ranges~(see \ref{sec:inter-percentile-range}):
\begin{equation}
    \frac{Q_m(1-p) - Q_m(p)}{Q_m(1-q) - Q_m(q)}.
\end{equation}
We used $p = 0.4, q = 0.05$ and $p = 0.2, q = 0.1$ in our analysis.

\paragraph{Maximum Slope}
\cite{DIsanto_etal2016}
Maximum slope between two sequential observations:
\begin{equation}
    \max_{i=0..N-2} \left|\frac{m_{i+1} - m_i}{t_{i+1} - t_i}\right|.
\end{equation}

\paragraph{Mean}
See~\ref{sec:def-mean}.

\paragraph{Median Absolute Deviation}
\cite{DIsanto_etal2016}
Median of the absolute value of the difference between magnitudes and their median:
\begin{equation}
    \median\left(|m_i - \median(m)|\right).
\end{equation}

\paragraph{Median Buffer Range Percentage}
\cite{DIsanto_etal2016}
Fraction of observations inside $\mathrm{Median}(m) \pm n \times \mathrm{Median}(m)$ interval:
\begin{equation}
    \frac{\sum_i I_{|m - \median(m)| < n \,\median(m)}(m_i)}{N}.
\end{equation}
We used $n = 0.05$ in our analysis.

\paragraph{Percent Amplitude}\label{sec:percent-amplitude}
\cite{DIsanto_etal2016}
Maximum deviation of magnitude from its median:
\begin{equation}
    \max_i\left|m_i - \median(m)\right|.
\end{equation}

\paragraph{Percent Difference Magnitude Percentile}
\citet{DIsanto_etal2016}
Ratio of $p$th inter-percentile range (see~\ref{sec:inter-percentile-range}) to the median:
\begin{equation}
    \frac{Q_m(1-p) - Q_m(p)}{\mathrm{Median}(m)}.
\end{equation}
We used $p = 0.05$ and $p = 0.2$ in our analysis.

\paragraph{Periodogram Amplitude}
Same as Amplitude (see~\ref{sec:amplitude}) but for $\{P_k\}$.

\paragraph{Periodogram Beyond $n$ Std}
Same as Beyond $n$ Std (see~\ref{sec:beyond-n-std}) but for $\{P_k\}$.
We used $n = 1$ and $n = 2$ in our analysis.

\paragraph{Periodogram Cusum}
Same as Cusum (see~\ref{sec:cusum}) but for $\{P_k\}$.

\paragraph{Periodogram $\eta$}
Same as Von Neummann $\eta$ (see~\ref{sec:eta}) but for $\{P_k\}$.

\paragraph{Periodogram Inter-Percentile Range}
Same as Inter-Percentile range (see~\ref{sec:inter-percentile-range}) but for $\{P_k\}$.
We used $p = 0.25$ in our analysis.

\paragraph{Periodogram Peaks}
Three highest periodogram peaks' period and $P_\mathrm{peak} / \sigma_P$ (``signal-to-noise'' ratio, \citet{Kim_etal2014}), where $\sigma_P$ is periodogram values standard deviation (see~\ref{sec:periodogram-standard-deviation}).

\paragraph{Periodogram Standard Deviation}\label{sec:periodogram-standard-deviation}
Same as Standard Deviation (see~\ref{sec:standard-deviation}) but for $\{P_k\}$.

\paragraph{Peridogram Percent Amplitude}
Same as Percent Amplitude (see~\ref{sec:cusum}) but for $\{P_k\}$.

\paragraph{Reduced $\chi^2$}
Reduced $\chi^2$ of the plateau fit:
\begin{equation}
     \frac1{N-1} \sum_i\left(\frac{m_i - \bar{m}}{\delta_i}\right)^2.
\end{equation}

\paragraph{Skew}
The skewness of magnitude distribution:
\begin{equation}
    G_1 \equiv \frac{N}{(N - 1)(N - 2)} \frac{\sum_i(m_i - \mean)^3}{\sigma_m^3}.
\end{equation}

\paragraph{Standard Deviation}\label{sec:standard-deviation}
$\sigma_m$, see~\ref{sec:def-standard-deviation}.

\paragraph{Stetson $K$}
\cite{Stetson1996}
\begin{equation}
    K \equiv \frac{\sum_i\left|\frac{m_i - \bar{m}}{\delta_i}\right|}{\sqrt{N \sum_i\left(\frac{m_i - \bar{m}}{\delta_i}\right)^2}}.
\end{equation}

\paragraph{Weighted mean}\label{sec:weighted-mean}
$\bar{m}$, see~\ref{sec:def-weighed-mean}.

\begin{figure*}
\begin{center}
\includegraphics[width=0.9\textwidth]{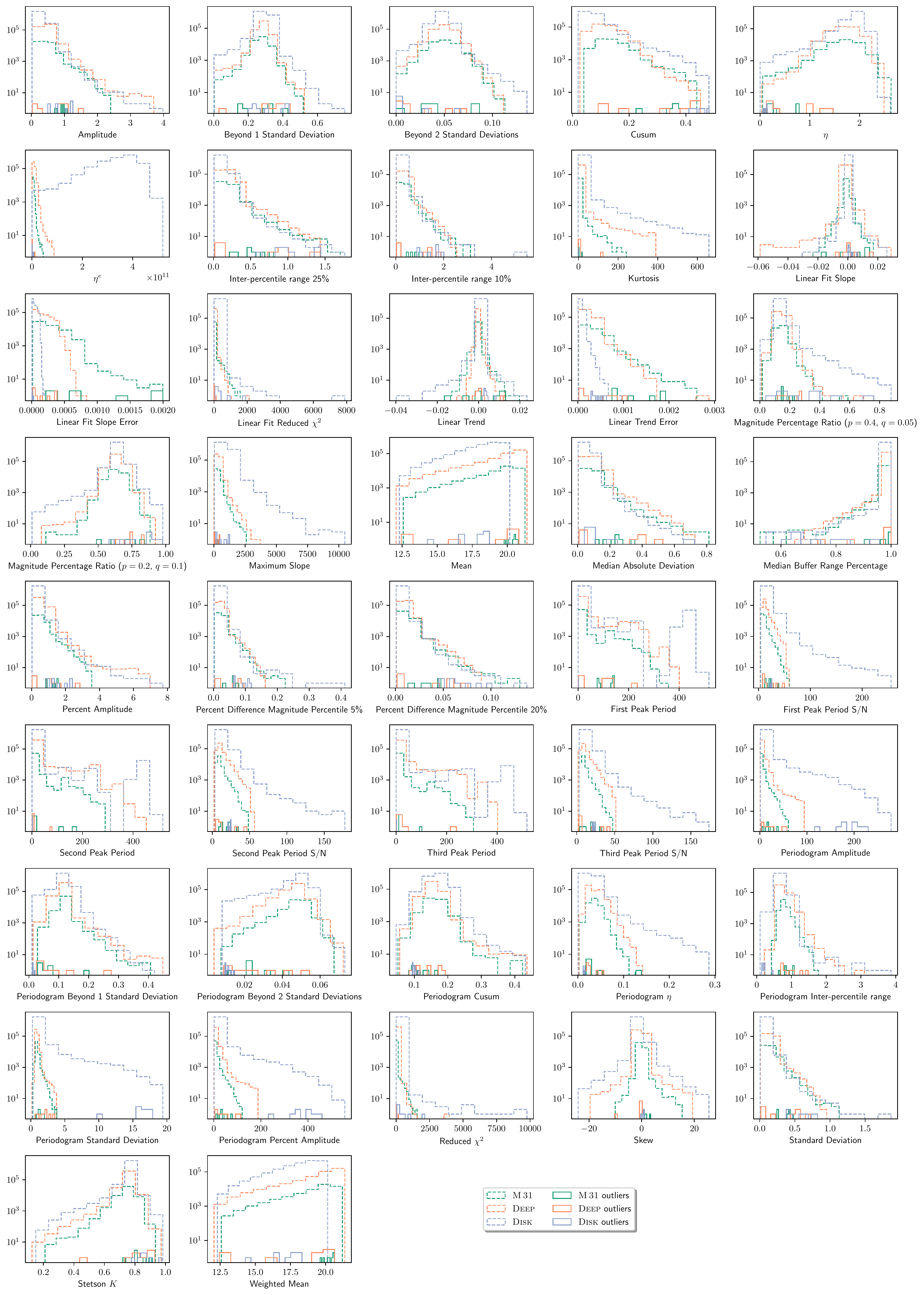}
\caption{Histogram of all 42 light curve features for objects in \andro{} (dashed green), \deep{} (dashed orange), and \disk{} (dashed blue), as well as outliers in \andro{} (solid green), \deep{} (solid orange), and \disk{} (solid blue) fields.}
\label{fig:all_features_histogram}
\end{center}
\end{figure*}

\begin{figure*}
\begin{center}
\includegraphics[width=0.9\textwidth]{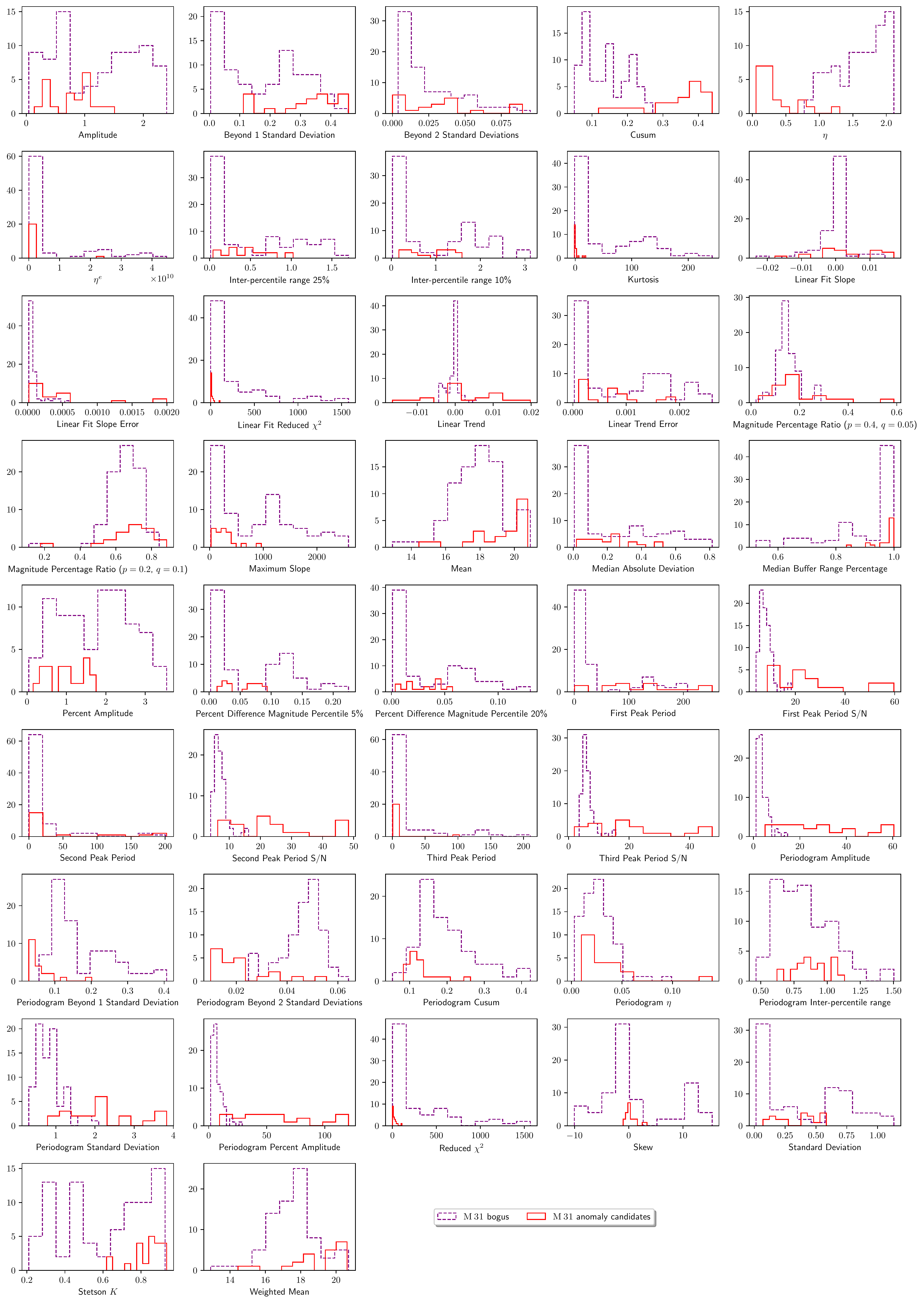}
\caption{Histogram of all 42 light curve features in \andro{} field for both bogus (dashed purple) and anomaly candidates (solid red).}
\label{fig:artefacts_vs_anom_hist}
\end{center}
\end{figure*}

% CORRELATION MATRIX 
\begin{figure*}
\begin{center}
\includegraphics[scale=0.5]{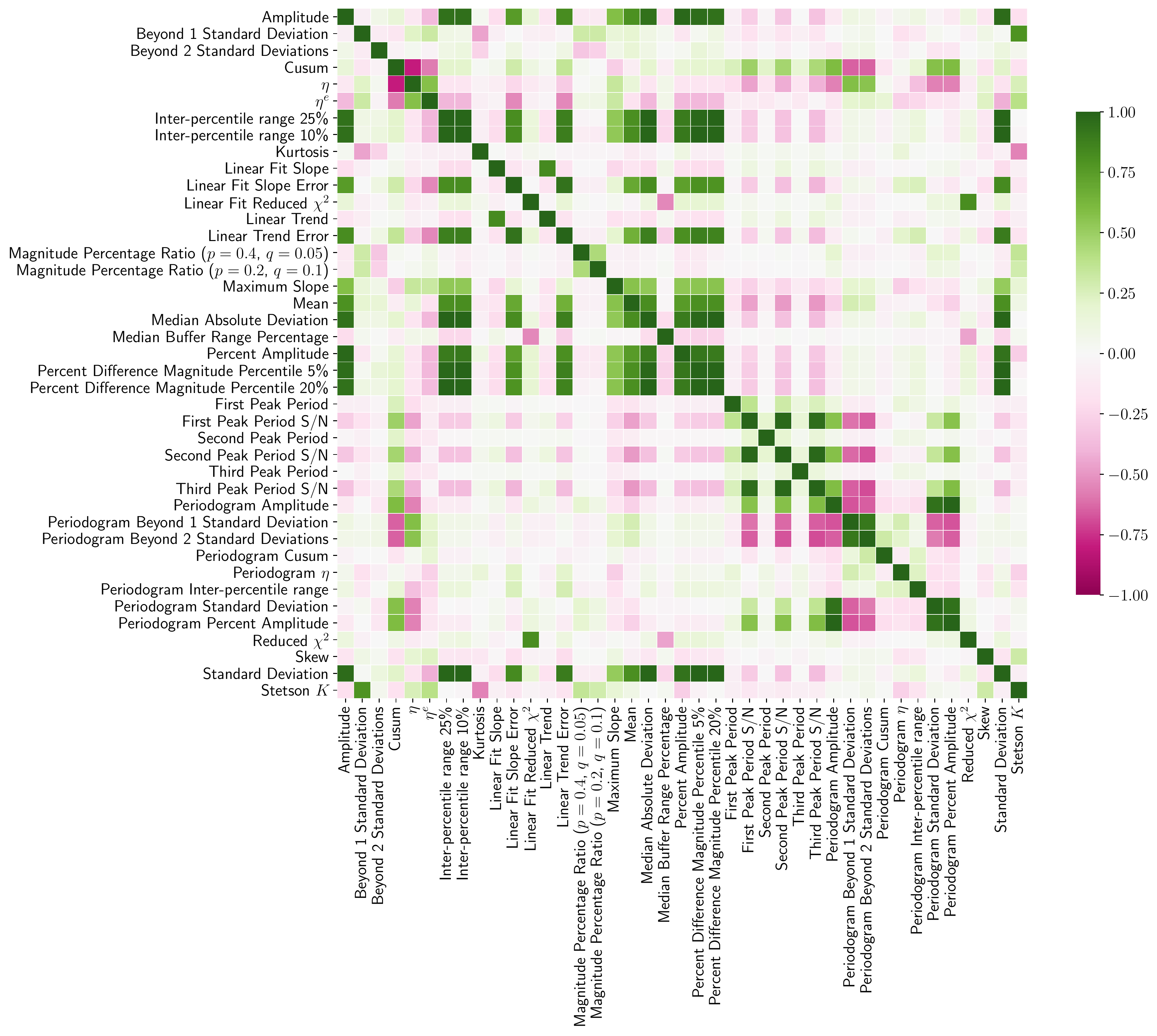}
\caption{A correlation matrix of all 42 light curve features on a combined data set of the \andro{}, \deep{}, and \disk{} fields.}
\label{fig:features_corr}
\end{center}
\end{figure*}
\section{ZTF DR objects web-viewer}
\label{sec:dr-viewer}

\begin{figure*}
    \centering
    \includegraphics[width=1.12\textwidth]{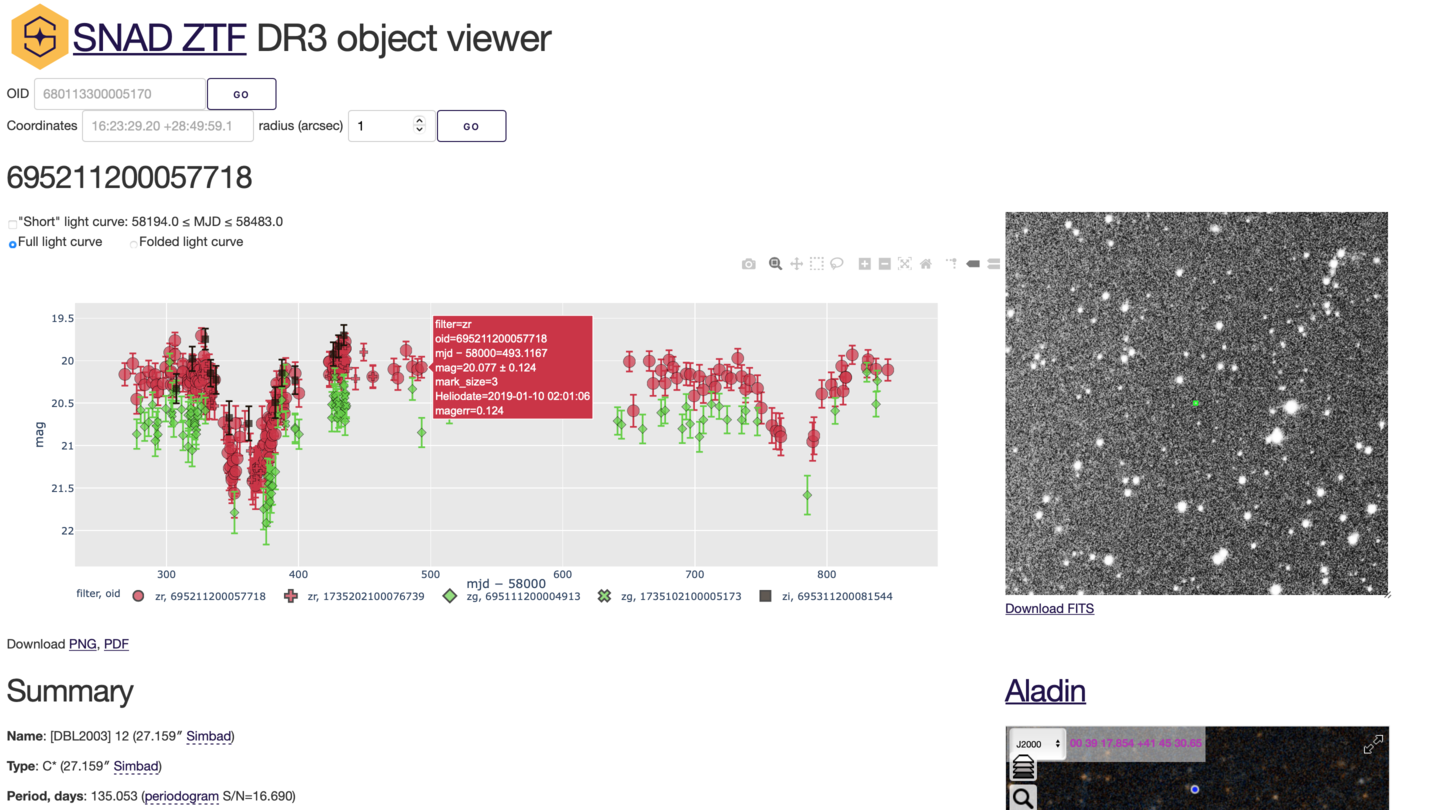}
    \caption{SNAD ZTF DR3 web-viewer page for object {\tt 695211200057718}.}
    \label{fig:viewer}
\end{figure*}

We developed web-based graphical user interface for expert-analysis of outliers, however it can be used by any researcher working with ZTF DRs as a light-curve viewer and cross-match tool.
We use ClickHouse\footnote{\url{https://clickhouse.tech}} column-based relation database management system as a storage for ZTF DR light-curves.
A column-based database is sub-optimal for querying a small number of objects which is required by a web-service, but in our case the main purpose of the database was selecting of batches of objects for feature extraction for anomaly detection, while an opportunity of getting of individual objects is a useful by-product.
The web-viewer is written on Dash\footnote{\url{https://dash.plotly.com}} Python framework which is suitable for interactive data visualisation.
An ZTF object page shows the main properties, light-curve and result of cross-matching with different catalogues.
Due to the fact that several ZTF DR3 objects (OIDs) can represent the same source, the viewer shows not only light-curve of the current object, but also finds all neighbour ZTF DR light-curves within some radius, the default value is one arcsecond.
A user can interact with the plot, changing this search radius, turn off and turn on light-curves, click on an observation to load image into embedded FITS viewer powered by JS9 library\footnote{\url{https://js9.si.edu}}.
Also a user can see folded light-curve which can be useful for periodic variable star analysis.
The object page contains cross-matching information with various variable star and transient catalogues, has embedded Aladin \citep{2014ASPC..485..277B} Sky Atlas where object position is marked.
Feature set which is presented in current paper (see Sec.~\ref{sec:appendix-features}) is listed on the object page too.
Web-viewer supports SIMBAD-powered \citep{2000A&AS..143....9W} object cone search which allows a user to find ZTF DR3 objects by common source name or by sky coordinates.
A screenshot of the web-viewer is presented in Fig.~\ref{fig:viewer} and its source code is available on Github\footnote{\url{https://github.com/snad-space/ztf-viewer}}.

\section{Light curve fit of supernova candidates}
\label{ap:lcmodels}

\begin{figure*}
\begin{center}
\includegraphics[width=2\columnwidth]{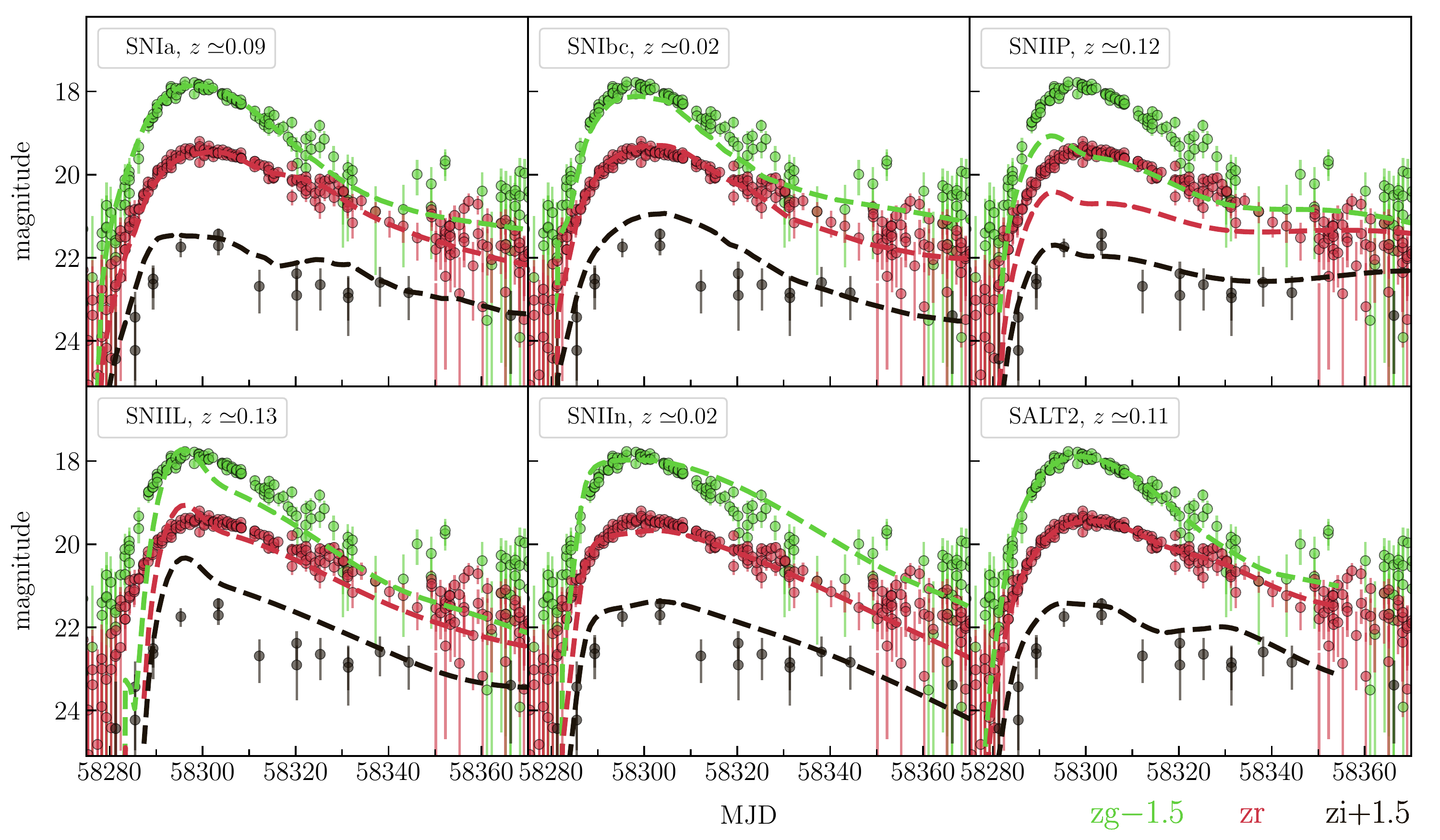}
\caption{Results of light curve fit of {\tt 795209200003484} by Nugent's supernova models and \textsc{SALT2} model. Observational data correspond to OIDs: {\tt 795109200001660} ($zg$), {\tt 795209200003484} ($zr$), {\tt 795309200004249} ($zi$), {\tt 796312200003601} ($zi$).}
\label{fig:795209200003484}
\end{center}
\end{figure*}

\begin{figure*}
\begin{center}
\includegraphics[width=2\columnwidth]{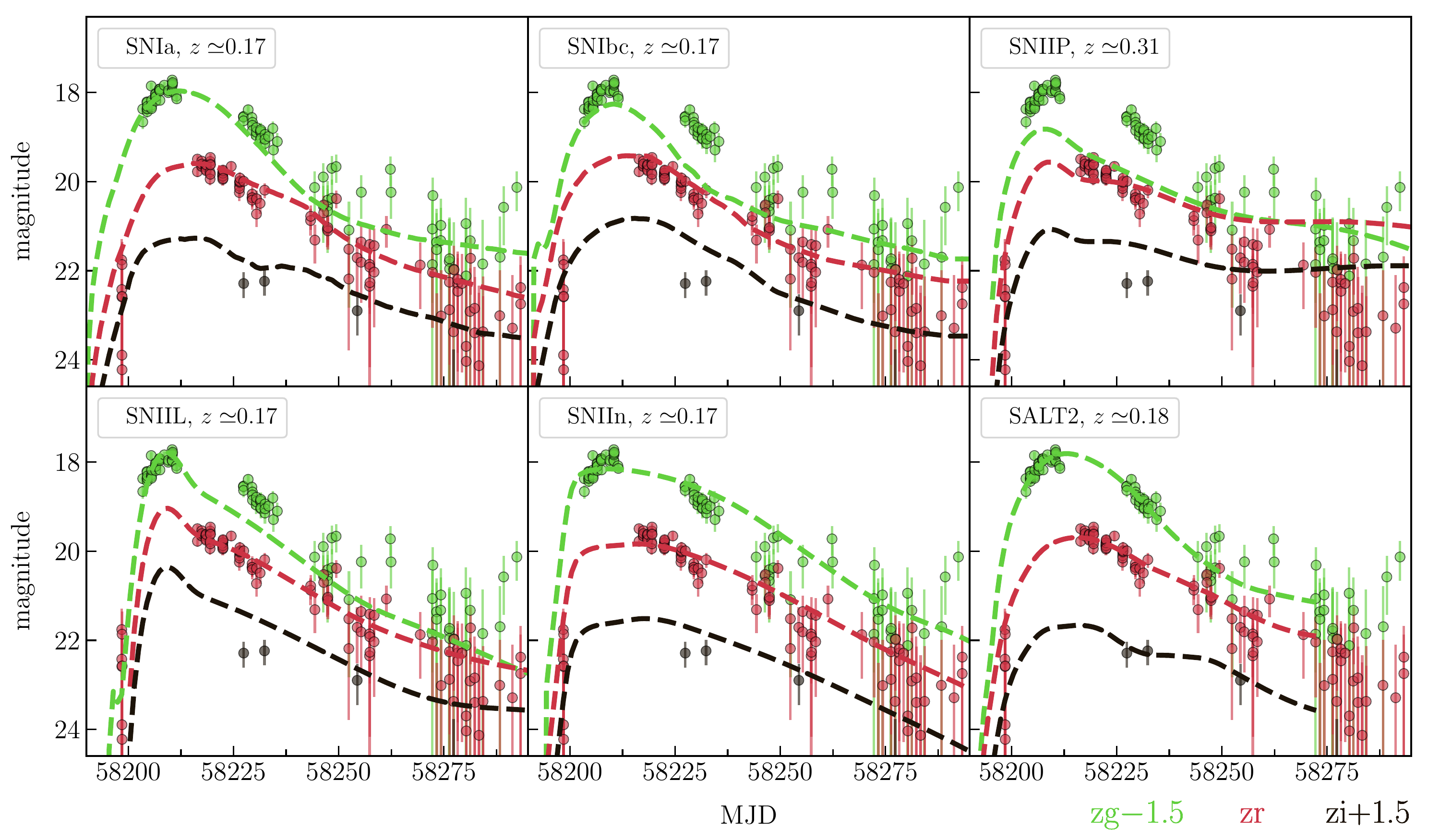}
\caption{Results of light curve fit of {\tt 795212100007964} by Nugent's supernova models and \textsc{SALT2} model. Observational data correspond to OIDs: {\tt 795112100014865} ($zg$), {\tt 795212100007964} ($zr$), {\tt 795312100008834} ($zi$).}
\label{fig:795212100007964}
\end{center}
\end{figure*}

\begin{figure*}
\begin{center}
\includegraphics[width=2\columnwidth]{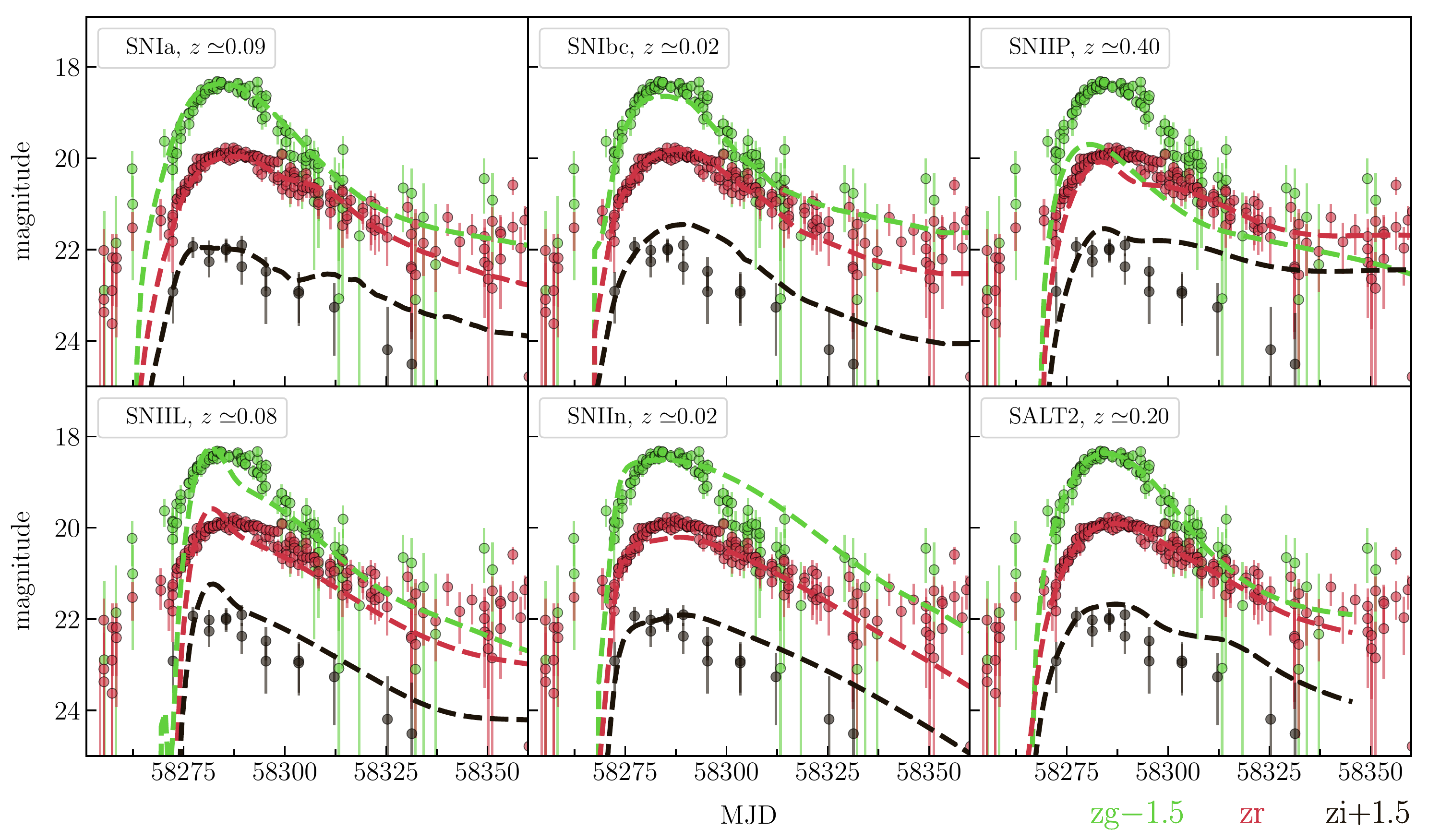}
\caption{Results of light curve fit of {\tt 795205100007271} by Nugent's supernova models and \textsc{SALT2} model. Observational data correspond to OIDs: {\tt 795105100002903} ($zg$), {\tt 795205100007271} ($zr$), {\tt 795305100008092} ($zi$), {\tt 796308200008245} ($zi$).}
\label{fig:795205100007271}
\end{center}
\end{figure*}

\begin{figure*}
\begin{center}
\includegraphics[width=2\columnwidth]{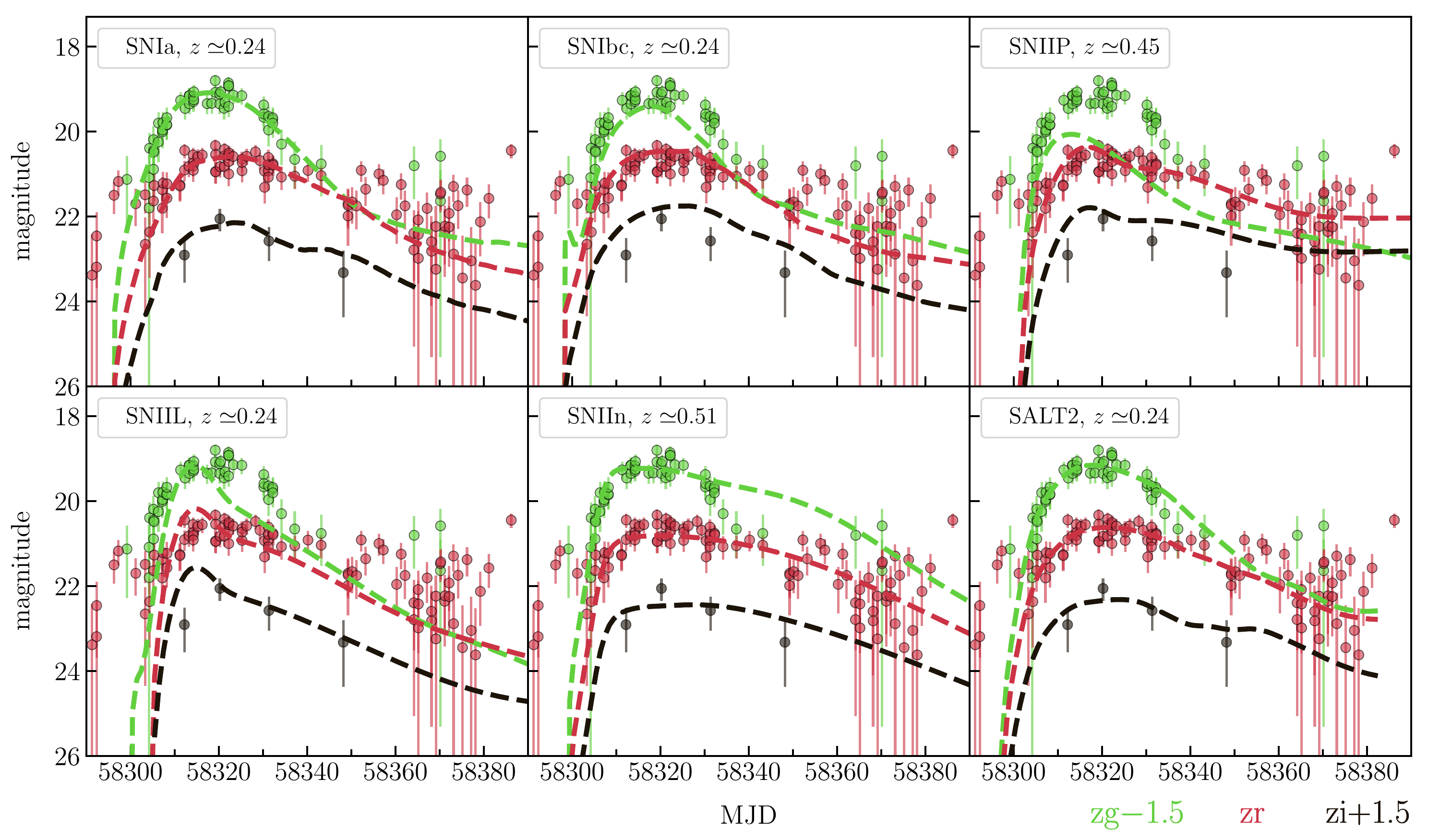}
\caption{Results of light curve fit of {\tt 795204100013041} by Nugent's supernova models and \textsc{SALT2} model. Observational data correspond to OIDs: {\tt 795104100011724} ($zg$), {\tt 795204100013041} ($zr$), {\tt 795304100015672} ($zi$).}
\label{fig:795204100013041}
\end{center}
\end{figure*}
\section{Anomaly candidates}
\label{sec:appendix-outliers}
\onecolumn
    \begin{landscape}
    \begin{longtable}{lllllrrrrlr}
        \caption{A complete list of anomaly candidates in the \andro{}, \deep{}, and \disk{}  fields.}
      \label{tab:disk_deep_m31}
      \\
\hline
OID	&	Other identifiers	&	$\alpha$, $\delta$ (deg) &	$m_{r, {\rm min}}$	&	$m_{r, {\rm max}}$	&	$E(B-V)$	& $D$ (pc) &	$M_{r, {\rm max}}$	&	$P_0^\dagger$ (d)	&	Type$^\star$ &	References	\\
\hline
\multicolumn{11}{c}{\andro{}} \\
\hline
{\tt 695211100003383}	&	2MASS J00452494+4207269	&	11.35388 42.12427	&	20.87$\pm$0.25	&	18.71$\pm$0.05	&	0.117	&M\,31 &$-$6.05	&	---	&	RSG	&	[14]	\\
{\tt 695211100015190}	&		&	11.79716 41.78059	&	21.45$\pm$0.31	&	19.49$\pm$0.09	&	0.078	&M\,31	&$-$5.17	&	---	&	---	&		\\
{\tt 695211100022045}	&	AT 2017ixs	&	10.98339 41.53641	&	21.27$\pm$0.29	&	18.66$\pm$0.05	&	0.268	&M\,31	&$-$6.50	&	---	&	PNV	&	[15]	\\
{\tt 695211100131796}	&	PSO J011.0457+41.5548	&	11.04581 41.55487	&	17.92$\pm$0.03	&	16.84$\pm$0.02	&	0.287	&M\,31	&$-$8.36	&	---	&	VAR	&	[16,17]	\\
{\tt 695211200018901}	&	2MASS J00421755+4135039	&	10.57313 41.58447	&	18.54$\pm$0.04	&	17.78$\pm$0.02	&	0.143	&M\,31	&$-$7.06	&	214.6	&	RSG	&	[14]	\\
\multirow{2}{*}{{\tt 695211200019653}}	&	2MASS J00415491+4133323, 	&	10.47879 41.55899	&	15.58$\pm$0.01	&	15.19$\pm$0.01	&	0.100	& 1642	&3.85	&	7.715	&	MSINE,	&	[2,3]	\\
	&	 ZTFJ004154.90+413332.3	&	&		&		&		&	&	&		&	RSCVN	&	\\
{\tt 695211200022958}	&	PSO J010.4744+41.4515	&	10.47441 41.45149	&	18.86$\pm$0.05	&	18.30$\pm$0.03	&	0.171	&M\,31	&$-$6.61	&	90.54	&	Delta Cep	&	[18]	\\
{\tt 695211200035023}	&		&	10.50601 41.81453	&	21.17$\pm$0.25	&	19.33$\pm$0.07	&	0.061	&M\,31	&$-$5.29	&	132.9	&	---	&		\\
{\tt 695211200046528}	&	[MAP97] 55	&	10.63267 41.48622	&	18.51$\pm$0.04	&	17.85$\pm$0.02	&	0.308	&M\,31	&$-$7.41	&	74.29	&	Delta Cep	&	[19]	\\
{\tt 695211200057718}	&		&	9.82489 41.76983	&	21.56$\pm$0.29	&	19.71$\pm$0.09	&	0.056	&M\,31	&$-$4.90	&	>200	&	---	&		\\
{\tt 695211200058391}	&		&	9.98496 41.74939	&	22.18$\pm$0.35	&	20.02$\pm$0.12	&	0.058	&M\,31	&$-$4.59	&	---	&	---	&		\\
\multirow{2}{*}{{\tt 695211200075348}}	&	M31N 2013-11b, 	&	10.35917 41.73029	&	21.61$\pm$0.29	&	19.40$\pm$0.07	&	0.060	&M\,31	&$-$5.22	&	168.4	&	VAR	&	[20,21]	\\
	&	 {\small MASTER OTJ004126.22+414350.0}	&	&		&		&		&	&	&		&		&	\\
{\tt 695211300004359}	&	PNV J00414894+4109173	&	10.45371 41.15395	&	20.61$\pm$0.21	&	17.56$\pm$0.03	&	0.261	&M\,31	& $-$7.58	&	---	&	PNV	&	[22]	\\
{\tt 695211300006331}	&	CSS\_J003827.1+410334	&	9.61336 41.05944	&	15.03$\pm$0.01	&	14.30$\pm$0.02	&	0.067	& 1546	& 3.18	&	3.138	&	EA	&	[1]	\\
{\tt 695211300007276}	&	SV* SON 10726	&	10.19874 41.03242	&	18.51$\pm$0.05	&	17.78$\pm$0.03	&	0.275	&M\,31	&$-$7.40	&	58.00	&	Delta Cep	&	[23]	\\
{\tt 695211400009049}	&		&	11.11946 40.99692	&	21.47$\pm$0.31	&	19.72$\pm$0.12	&	0.076	&M\,31	&$-$4.94	&	---	&	---	&		\\
{\tt 695211400025927}	&		&	11.44201 41.25969	&	21.82$\pm$0.34	&	20.07$\pm$0.16	&	0.072	&M\,31	&$-$4.58	&	---	&	---	&		\\
{\tt 695211400027347}	&	2MASS J00434749+4112585	&	10.94775 41.21573	&	21.50$\pm$0.31	&	19.57$\pm$0.11	&	0.332	&M\,31	&$-$5.76	&	>200	&	VAR	&	[24]	\\
{\tt 695211400046832}	&		&	11.26142 40.57456	&	22.17$\pm$0.37	&	20.21$\pm$0.18	&	0.063	&M\,31	&$-$4.41	&	---	&	---	&		\\
{\tt 695211400070144}	&	2MASS J00443180+4119083	&	11.13252 41.31898	&	19.63$\pm$0.12	&	18.10$\pm$0.04	&	0.275	&M\,31	&$-$7.07	&	---	&	VAR	&	[25]	\\
{\tt 695211400121607}	&	[JPN2003] V206	&	10.87915 41.09875	&	21.42$\pm$0.30	&	19.33$\pm$0.09	&	0.181	&M\,31	&$-$5.60	&	---	&	VAR	&	[24]	\\
\hline
\multicolumn{11}{c}{\deep{}} \\
\hline
\multirow{2}{*}{{\tt 795202100005941}}	&	MLS180307:163438+521642, 	&	248.65767 52.27841	&	21.90$\pm$0.27	&	19.45$\pm$0.07	&	0.026	&	---	&	---	&	---	&	SN candidate	&	[11]	\\
	&	ZTF18aanbnjh	&		&		&		&		& &		&		&		&		\\
{\tt 795202300001087}	&	NSVS 5270259	&	247.70051 51.96916	&	13.84$\pm$0.01	&	12.99$\pm$0.01	&	0.022	&	957	&	3.03	&	0.674	&	EA	&	[1]	\\
{\tt 795203200009604}	&	DDE 32	&	244.89910 52.77552	&	20.39$\pm$0.14	&	16.50$\pm$0.01	&	0.019	&	439	&	8.24	&	---	&	AM	&	[1]	\\
{\tt 795204100013041}	&	ZTF18abgvctp	&	242.30742 52.21426	&	21.87$\pm$0.27	&	20.02$\pm$0.11	&	0.016	&	---	&	---	&	---	&	---	&		\\
{\tt 795204200006882}	&	CSS\_J160450.1+520159	&	241.20883 52.03324	&	20.67$\pm$0.18	&	17.20$\pm$0.02	&	0.014	&	3909	&	4.20	&	316.8	&	EA	&	[1]	\\
{\tt 795205100007271}	&	ZTF18aayatjf	&	252.30216 54.11178	&	21.45$\pm$0.23	&	19.51$\pm$0.08	&	0.021	&	---	&	---	&	---	&	---	&		\\
\multirow{2}{*}{{\tt 795205400001697}}	&	SDSS J164749.77+534217.3, 	&	251.95740 53.70487	&	12.48$\pm$0.01	&	12.35$\pm$0.01	&	0.051	&	924	&	2.39	&	2.217	&	---	&		\\
	&	ZTF18aajsmrd	&		&		&		&		&		&	&		&		&		\\
{\tt 795205400022890}	&	IW Dra	&	251.56278 53.19847	&	14.69$\pm$0.01	&	11.57$\pm$0.01	&	0.047	&	4888	&	$-$1.99	&	193.2	&	Mira	&	[1]	\\
{\tt 795205400035251}	&	SDSS J164533.41+531522.6	&	251.38937 53.25627	&	16.37$\pm$0.01	&	15.99$\pm$0.01	&	0.046	&	1353	&	5.21	&	230.3	&	---	&		\\
{\tt 795209200003484}	&	ZTF18abbpebf	&	251.54866 56.33124	&	21.33$\pm$0.22	&	18.96$\pm$0.05	&	0.015	&	---	&	---	&	---	&	---	&		\\
{\tt 795209300012791}	&	CSS\_J164050.1+552654	&	250.20916 55.44835	&	16.27$\pm$0.01	&	15.17$\pm$0.01	&	0.015	&	5138	&	1.58	& 0.458	&	RRAB	&	[1]	\\
{\tt 795210400001565}	&	SDSS J163331.55+554406.1 &	248.38139 55.73513	&	12.50$\pm$0.01	&	12.43$\pm$0.01	&	0.006	&	240	&	5.52	&	3.731	&	---	&		\\
{\tt 795211200035931}	&	SN 2018coi	&	244.74136 56.71714	&	20.35$\pm$0.15	&	18.04$\pm$0.03	&	0.009	&	---	&	$-$19.07	&	---	&	SN Ia	&	[12]	\\
{\tt 795212100007964}	&	ZTF18aanbksg	&	242.93762 55.96133	&	21.72$\pm$0.27	&	19.28$\pm$0.07	&	0.007	&	---	&	---	&	---	&	---	&		\\
\multirow{2}{*}{{\tt 795213200000671}}	&	AT2018afr, Gaia18apj, 	&	251.93077 58.50559	&	20.46$\pm$0.17	&	18.86$\pm$0.05	&	0.012	&	---	&	---	&	---	&	SN candidate	&	[13]	\\
	&	ZTF18aaincjv	&		&		&		&		& &		&		&		&		\\
{\tt 795213300001714}	&	{\small ASASSN-V J164454.40+573232.3}	&	251.22667 57.54232	&	13.02$\pm$0.01	&	12.41$\pm$0.01	&	0.016	&	4974	&	$-$1.11	&	52.90	&	SR	&	[1]	\\
{\tt 795213400008053}	&	NSVS J1647409+565803	&	251.91816 56.96779	&	14.98$\pm$0.01	&	13.25$\pm$0.01	&	0.017	&	8363	&	$-$1.40	&	111.3	&	SR	&	[1]	\\
{\tt 795214300016648}	&	& 248.03423 57.75695	& 19.84$\pm$0.10	& 16.68$\pm$0.01&	0.010 &	403 &	8.62 &	--- &	---& \\
\hline
\multicolumn{11}{c}{\disk{}} \\
\hline
{\tt 807201100022915}	&	V0476 Cas	&	25.84085 59.36414	&	15.93$\pm$0.01	&	12.79$\pm$0.01	&	0.512	&	3256	&	$-$1.11	&	227.1	&	Mira	&	[1]	\\
\multirow{2}{*}{{\tt 807202100027080}}	&	J022.6771+60.0173, 	&	22.67713 60.01735	&	16.61$\pm$0.01	&	15.44$\pm$0.01	&	0.441	&	1875	&	2.93	&	0.824	&	DBF, EA	&	[2,3]	\\
	&	ZTFJ013042.51+600102.4	&		&		&		&	&		&		&		&		&	\\
{\tt 807202300038681}	&	MGAB-V574	&	20.22254 58.82538	&	20.74$\pm$0.22	&	17.24$\pm$0.02	&	0.474	&	2830	&	3.75	&	---	&	UG	&	[1]	\\
{\tt 807202400015654}	&	V0420 Cas	&	22.80860 58.79127	&	14.82$\pm$0.01	&	13.72$\pm$0.01	&	0.419	&	2717	&	0.46	&	13.78	&	EA	&	[1]	\\
\multirow{2}{*}{{\tt 807203200013118}}	&	J017.9869+59.7407, 	&	17.98693 59.74082	&	17.73$\pm$0.03	&	16.14$\pm$0.01	&	0.553	&	1026	&	4.65	&	2.185	&	---	&		\\
&	ZTF18abfyqzh	&		&	&		&		&	&		&	&		& 	\\
{\tt 807203300044912}	&	V0890 Cas	&	16.93576 59.05054	&	20.79$\pm$0.24	&	14.32$\pm$0.01	&	0.747	&	4536	&	$-$0.90	&	421.8	&	Mira	&	[1]	\\
{\tt 807203400031819}	&	Romanov V28	&	19.30153 58.46788	&	19.28$\pm$0.08	&	16.09$\pm$0.01	&	0.414	&	1485	&	4.15	&	>20	&	UGSS	&	[1]	\\
\multirow{2}{*}{{\tt 807203400037050}}	&	Dauban V254, &	19.20494 59.08340	&	16.74$\pm$0.02	&	14.17$\pm$0.01	&	0.491	&	2625	&	0.80	&	256.0	&	Mira	&	[3]	\\
&	ZTFJ011649.18+590500.2	&		&	&		&		&	&		&	&		& 	\\
{\tt 807204200004799}	&		&	14.29017 59.90208	&	20.19$\pm$0.17	&	18.05$\pm$0.03	&	0.435	&	4891	&	3.47	&	4.320	&	---	&		\\
{\tt 807204400014494}	&	ZTF18abhakel	&	16.16915 58.85506	&	14.88$\pm$0.01	&	13.86$\pm$0.01	&	0.607	&	1514	&	1.38	&	6.017	&	---	&		\\
{\tt 807205200017536}	&	V0367 Cas	&	25.29686 61.25982	&	16.30$\pm$0.01	&	13.48$\pm$0.01	&	0.964	&	1955	&	$-$0.48	&	3.709 &	EA	&	[1]	\\
\multirow{2}{*}{{\tt 807206200003542}}	&	{\small ASASSN-V J012356.42+615458.8}	&	20.98505 61.91633	&	17.18$\pm$0.02	&	14.74$\pm$0.01	&	10.394	&	793	&	$-$21.81	&	---	&	VAR, YSO,	&	[1,4,5]	\\
	&		&		&		&		&		& &		&		&	PMS	&		\\
{\tt 807206200004116}	&	MGAB-V1397	&	21.18448 61.89163	&	20.17$\pm$0.16	&	17.63$\pm$0.02	&	1.411	&	984	&	3.99	&	---	&	YSO, PMS	&	[1,5] 	\\
{\tt 807206200014645}	&	MGAB-V1388	&	20.49273 61.50519	&	18.45$\pm$0.04	&	16.28$\pm$0.01	&	0.887	&	935	&	4.11	&	---	&	YSO, PMS	&	[1,5] 	\\
{\tt 807206200023036}	&	NSVS J0122238+611352	&	20.59913 61.23097	&	16.53$\pm$0.01	&	12.84$\pm$0.01	&	0.626	&	4401	&	$-$2.01	&	325.1	&	Mira	&	[1]	\\
{\tt 807206300013468}	&	II Cas	&	21.67063 60.77933	&	15.62$\pm$0.01	&	12.32$\pm$0.01	&	0.560	&	2997	&	$-$1.52	&	282.0	&	Mira	&	[1]	\\
{\tt 807206400014916}	&	Mis V1358	&	22.34756 60.74668	&	17.36$\pm$0.02	&	12.92$\pm$0.01	&	0.489	&	2911	&	$-$0.67	&	310.0	&	Mira:	&	[1]	\\
\multirow{2}{*}{{\tt 807207100005929}}	&	IRAS 01145+6132, 	&	19.45793 61.80168	&	17.80$\pm$0.03	&	14.25$\pm$0.01	&	1.013	&	3247	&	$-$0.95	&	278.0	&	LPV, Mira	&	[3,6]	\\
&	ZTFJ011749.90+614805.9	&		&	&		&		&	&		&	&		&	\\
{\tt 807207300036103}	&	BMAM-V362  	&	16.98932 60.35615	&	20.46$\pm$0.18	&	17.79$\pm$0.03	&	0.540	&	2311	&	4.57	&	---	&	UGZ	&	[1]	\\
{\tt 807208200059506}	&	V0450 Cas	&	12.87116 61.64957	&	16.42$\pm$0.01	&	13.93$\pm$0.01	&	1.223	&	560	&	2.01	&	---	&	INS	&	[1]	\\
{\tt 807208300012891}	&	OU Cas	&	13.32034 60.69593	&	16.80$\pm$0.02	&	15.17$\pm$0.01	&	0.542	&	2434	&	1.82	&	1.265	&	EA	&	[1]	\\
{\tt 807208300016714}	&	OT Cas	&	13.26166 60.59657	&	16.10$\pm$0.01	&	12.49$\pm$0.01	&	0.496	&	3145	&	$-$1.29	&	292.0	&	Mira	&	[1]	\\
{\tt 807208400015695}	&	AV Cas	&	14.89163 60.72181	&	15.72$\pm$0.01	&	11.95$\pm$0.01	&	0.570	&	1695	&	$-$0.68	&	330.0	&	Mira	&	[1]	\\
{\tt 807208400036953}	&	ZTF18abflqpt	&	14.44257 60.25802	&	20.01$\pm$0.14	&	17.48$\pm$0.02	&	0.445	&	2553	&	4.29	&	1.544	&	---	&		\\
\multirow{2}{*}{{\tt 807209100042420}}	&	Dauban V252, 	&	27.07326 63.12797	&	18.25$\pm$0.04	&	13.44$\pm$0.01	&	1.031	&	3624	&	$-$2.03	&	327.6	&	Mira	&	[3]	\\
&	ZTFJ014817.58+630740.6	&		&	&		&		&	&		&	&		& 	\\
{\tt 807209300012026}	&	WISE J014237.7+623700	&	25.65734 62.61680	&	14.32$\pm$0.01	&	13.20$\pm$0.01	&	1.294	&	3020	&	$-$2.57	&	6.399	&	EA	&	[1]	\\
{\tt 807209300037143}	&	NSVS J0141549+623406	&	25.47716 62.56828	&	16.89$\pm$0.02 &	12.32$\pm$0.01	&	1.349	&	4302	&	$-$4.36	&	297.0	&	Mira	&	[1]	\\
{\tt 807209400037670}	&	IRAS 01438+6208	&	26.85197 62.38822	&	18.56$\pm$0.05	&	12.56$\pm$0.01	&	1.426	&	3518	&	$-$3.88	&	286.0	&	Mira	&	[7]	\\
\multirow{2}{*}{{\tt 807210100028861}}	&	NSVS J0129109+631249,	&	22.29801 63.21435	&	16.28$\pm$0.01	&	12.70$\pm$0.01	&	1.357	&	5972	&	$-$4.71	&	454.2	&	LPV, 	&	[1,7] 	\\
&	IRAS 01257+6257	&		&	&		&		&	&		&	&	SR+L, C	&	\\
{\tt 807210200004045}	&		&	21.24037 63.85572	&	19.48$\pm$0.09	&	17.84$\pm$0.03	&	1.550	&	3353	&	1.18	&	---	&	---	&		\\
\multirow{2}{*}{{\tt 807210200026027}}	&		&	21.96268 63.28808	&	18.81$\pm$0.06	&	16.56$\pm$0.01	&	1.451	&	1704	&	1.63	&	421.7	&	YSO, PMS, 	& [4,5,8]		\\
	&		&	&		&		& &		&		&		&	AGN	&	\\
{\tt 807211100059466}	&	V0724 Cas	&	18.72328 63.61240	&	15.55$\pm$0.01	&	12.54$\pm$0.01	&	1.488	&	5475	&	$-$5.03	&	280.0	&	Mira	&	[1]	\\
\multirow{2}{*}{{\tt 807211300006190}}	&	Mis V0818, 	&	16.17487 62.88210	&	17.33$\pm$0.02	&	13.02$\pm$0.01	&	1.332	&	4768	&	$-$3.85	&	315.0	&	Mira	&	[3]	\\
&	ZTFJ010441.96+625255.5	&		&		&		&		&		&		&		&		&		\\
{\tt 807211300012948}	&	ZTFJ010526.59+624324.4	&	16.36079 62.72345	&	16.09$\pm$0.01	&	14.31$\pm$0.01	&	1.463	&	2785	&	$-$1.72	&	302.9	&	Mira	&	[3]	\\
{\tt 807211400009493}	&	ZTF18abjcvcb	&	18.74697 62.83884	&	20.08$\pm$0.15	&	17.63$\pm$0.02	&	1.472	&	3334	&	1.18	&	2.546	&	---	&		\\
{\tt 807212100012737}	&	ZTF17aaaeesd	&	15.01403 63.61420	&	19.01$\pm$0.07	&	16.51$\pm$0.01	&	1.196	&	1087	&	3.22	&	---	&	---	&		\\
{\tt 807212100020830}	&	ZTFJ005605.95+632426.8	&	14.02475 63.40745	&	16.80$\pm$0.02	&	13.26$\pm$0.01	&	1.016	&	3188	&	$-$1.90	&	394.4	&	Mira	&	[3]	\\
{\tt 807212200034055}	&	[I81] M 633	&	13.10185 63.06180	&	16.83$\pm$0.02	&	13.64$\pm$0.01	&	1.039	&	2171	&	$-$0.75	&	364.0	&	Mira	&	[9]	\\
\multirow{2}{*}{{\tt 807212300038616}}	&	NSVS J0053173+623614, 	&	13.32484 62.60320	&	19.61$\pm$0.11	&	14.55$\pm$0.01	&	1.306	&	3530	&	$-$1.59	&	311.0	&	L, Mira	&	[1,9] 	\\
&[I81] M 644	&		&	&		&		&	&		&	&		& 	\\
{\tt 807214100007080}	&	{\small [WWV2004] J0129273+653917}	&	22.37209 65.65551	&	18.30$\pm$0.04	&	13.00$\pm$0.01	&	1.051	&	3655	&	$-$2.55	&	295.2	&	Mira	&	[7]	\\
\multirow{2}{*}{{\tt 807214100026393}}	&	Dauban V269, 	&	22.36480 65.19663	&	20.94$\pm$0.27	&	16.75$\pm$0.02	&	1.074	&	3884	&	1.00	&	454.2	&	Mira	&	[3]	\\
&	ZTFJ012927.55+651147.8	&		&	&		&		&	&		&	&		& 	\\
{\tt 807214100042070}	&	MGAB-V1402	&	24.06955 65.31108	&	20.11$\pm$0.16	&	18.24$\pm$0.04	&	1.125	&	3876	&	2.37	&	600.0	&	Mira	&	[1]	\\
{\tt 807214300010833}	&	NSVS 1695145	&	21.24271 64.77038	&	16.53$\pm$0.01	&	13.84$\pm$0.01	&	1.315	&	4633	&	$-$2.91	&	560.0	&	Mira:	&	[1]	\\
{\tt 807214300051120}	&	ZTFJ012430.53+641432.2	&	21.12717 64.24230	&	17.69$\pm$0.03	&	15.41$\pm$0.01	&	1.194	&	3852	&	$-$0.62	&	325.0	&	Mira	&	[3]	\\
{\tt 807215100019764}	&	VSX J011759.4+652153	&	19.49756 65.36491	&	19.37$\pm$0.09	&	14.03$\pm$0.01	&	1.593	&	4243	&	$-$3.26	&	320.0	&	Mira:	&	[1]	\\
{\tt 807215200006611}	&	NSVS J0108368+653701	&	17.15140 65.61674	&	16.78$\pm$0.02	&	12.71$\pm$0.01	&	1.580	&	5238	&	$-$5.00	&	309.0	&	Mira	&	[1]	\\
\multirow{2}{*}{{\tt 807216100038423}}	&	IPHAS J005934.24+651815.1, 	&	14.89278 65.30412	&	18.25$\pm$0.04	&	14.47$\pm$0.01	&	1.412	&	2598	&	$-$1.27	&	295.2	&	AGB, Mira	&	[3,10]	\\
&	ZTFJ005934.27+651814.9	&		&		&		&		&		&		&		&		&	\\
{\tt 807216200026731}	&	NSV 15193	&	13.35496 65.03225	&	16.29$\pm$0.01	&	12.10$\pm$0.01	&	1.305	&	2681	&	$-$3.44	&	507.0	&	Mira	&	[1]	\\
{\tt 807216200045725}	&	NSVS 1633482	&	11.72349 65.61549	&	21.30$\pm$0.34	&	15.52$\pm$0.01	&	1.411	&	3882	&	$-$1.10	&	390.0	&	Mira	&	[1]	\\
{\tt 807216400013229}	&		&	14.15960 64.56684	&	17.41$\pm$0.02	&	16.29$\pm$0.01	&	1.381	&	2530	&	0.69	&	2.160	&	---	&		\\
\hline
% }
\end{longtable}
\noindent
$^\dagger$ 
The best period extracted from either the Lomb–Scargle periodogram or one of the catalogues listed in the ZTF-viewer or determined by us.

\noindent
$^\star$ 
AM --- AM Herculis-type variable\\
AGB	--- Asymptotic Giant Branch Star\\
C --- Carbon star\\
DBF --- Distant binary, full period\\
Delta Cep --- Classical Cepheid\\
EA ---	$\beta$ Persei-type (Algol) eclipsing system\\
INS ---	Orion variable with rapid light variations\\
L --- Slow irregular variable, stars are often attributed to this type because of being insufficiently studied\\
LPV ---	Long Period Variable\\
Mira --- ${o}$ (omicron) Ceti-type variable\\
MSINE --- Star showing modulated sinusoids\\
PNV --- Possible nova\\
RRAB --- RR Lyrae variable with asymmetric light curves\\
RSCVN --- RS Canum Venaticorum-type binary system\\
RSG --- Red supergiant\\
SN --- Supernova\\
SR --- Semi-regular variable\\
UG	--- U Geminorum-type variable, quite often called dwarf nova\\
UGSS --- SS Cygni-type variable\\
UGZ	--- Z Camelopardalis-type star\\
VAR --- Variable of unspecified type\\
YSO	--- Young stellar Object of unspecified variable type\\

\noindent
[1]~---~\citealt{2006SASS...25...47W};
[2]~---~\citealt{2018AJ....156..241H};
[3]~---~\citealt{2020ApJS..249...18C};
[4]~---~\citealt{2016MNRAS.458.3479M};
[5]~---~\citealt{2020AA...638A..21V};
[6]~---~\citealt{2008OEJV...87....1U};
[7]~---~\citealt{2004AJ....128.2965W};
[8]~---~\citealt{2012ApJ...751...52E};
[9]~---~\citealt{2018RMxAA..54..341N};
[10]~---~\citealt{2009MNRAS.400.1413W};
[11]~---~\citealt{2009ApJ...696..870D};
[12]~---~\citealt{2018ATel11745....1G};
[13]~---~\citealt{2018TNSTR.345....1D};
[14]~---~\citealt{2009ApJ...703..420M};
[15]~---~\citealt{2017TNSTR1418....1C};
[16]~---~\citealt{2014ApJ...785...11L};
[17]~---~\citealt{2017ApJ...836...64H};
[18]~---~\citealt{2018AJ....156..130K};
[19]~---~\citealt{1997AAS..126..401M};
[20]~---~\citealt{2013ATel.5569....1O};
[21]~---~\citealt{2016ATel.9470....1S};
[22]~---~\citealt{2015ATel.7462....1H};
[23]~---~\citealt{2013AJ....145..106K};
[24]~---~\citealt{2003AA...402..113J};
[25]~---~\citealt{1999AJ....118..346K}.
\end{landscape}

\twocolumn

\clearpage
\newpage
%%%%%%%%%%%%%%%%%%%%%%%%%%%%%%%%%%%%%%%%%%%%%%%%%%%%

%\section{Links}

%\href{https://www.aavso.org/LCGv2/}{AAVSO Light Curve plotter}\\
%\href{http://ogledb.astrouw.edu.pl/~ogle/OCVS/catalog_query.php}{OGLE Catalogue of Variable Stars Database Query Page}\\
%Useful papers: \citealt{Rebbapragada2009,2014ApJ...793...23N}\\
%Master thesis \href{https://repositorio.uc.cl/bitstream/handle/11534/16560/000641231.pdf?sequence=1}{SUPERVISED DETECTION OF ANOMALOUS LIGHT-CURVES IN MASSIVE ASTRONOMICAL CATALOGS}

% \section*{Acknowledgements}

%%%%%%%%%%%%%%%%%%%%%%%%%%%%%%%%%%%%%%%%%%%%%%%%%%

%%%%%%%%%%%%%%%%%%%% REFERENCES %%%%%%%%%%%%%%%%%%

% The best way to enter references is to use BibTeX:

\bibliographystyle{mnras}
\bibliography{ztf_anomalies} % if your bibtex file is called example.bib

% Alternatively you could enter them by hand, like this:
% This method is tedious and prone to error if you have lots of references
%\begin{thebibliography}{99}
%\bibitem[\protect\citeauthoryear{Author}{2012}]{Author2012}
%Author A.~N., 2013, Journal of Improbable Astronomy, 1, 1
%\bibitem[\protect\citeauthoryear{Others}{2013}]{Others2013}
%Others S., 2012, Journal of Interesting Stuff, 17, 198
%\end{thebibliography}

%%%%%%%%%%%%%%%%%%%%%%%%%%%%%%%%%%%%%%%%%%%%%%%%%%

%%%%%%%%%%%%%%%%% APPENDICES %%%%%%%%%%%%%%%%%%%%%

% \appendix

% \section{Some extra material}

%%%%%%%%%%%%%%%%%%%%%%%%%%%%%%%%%%%%%%%%%%%%%%%%%%

% Don't change these lines
\bsp	% typesetting comment
\label{lastpage}
\end{document}